\documentclass[prd, onecolumn, superscriptaddress, nofootinbib, notitlepage, floatfix, preprintnumbers]{revtex4-1}
\usepackage{xcolor}
\usepackage{graphicx}
\usepackage{wrapfig}
\usepackage{amsmath}
\usepackage{amsfonts}
\usepackage{amssymb}
\usepackage{multirow}
\usepackage{slashed}
\usepackage{physics}
\usepackage{soul}

\usepackage{ulem}
\usepackage[colorlinks=true, linkcolor=blue, citecolor=blue, urlcolor=blue]{hyperref}
\newcommand{\msbar}{{\overline{\mathrm{MS}}}}
\newcommand\befs{\begin{figure*}}
\newcommand\eefs[1]{\label{fig:#1}\end{figure*}}
\newcommand\bef{\begin{figure}}
\newcommand\eef[1]{\label{fig:#1}\end{figure}}
\newcommand\beq{\begin{equation}}
\newcommand\eeq[1]{\label{#1}\end{equation}}
\newcommand\beqa{\begin{eqnarray}&&\quad\cr}
\newcommand\eeqa[1]{\cr &&\quad \label{#1}\end{eqnarray}}
\newcommand\bet{\begin{table}}
\newcommand\eet[1]{\label{tb:#1}\end{table}}
\newcommand\bets{\begin{table*}}
\newcommand\eets[1]{\label{tb:#1}\end{table*}}
\newcommand\fgn[1]{Fig.\ \ref{fig:#1}}
\newcommand\eqn[1]{Eq.\ (\ref{#1})}
\newcommand\scn[1]{Section \ref{sec:#1}}

\begin{document}
\widetext
\title{
   The transversity parton distribution function of the nucleon using the pseudo-distribution approach
}
\newcommand*{\WM}{Department of Physics, William and Mary, Williamsburg, Virginia, USA.}\affiliation{\WM}       
\newcommand*{\Jlab}{Thomas Jefferson National Accelerator Facility, Newport News, Virginia, USA.}\affiliation{\Jlab}    
\newcommand*{\CU}{Department of Physics, Columbia University, New York City, New York, USA.}\affiliation{\CU}   
\newcommand*{\ODU}{Department of Physics, Old Dominion University, Norfolk, Virginia, USA.}\affiliation{\ODU}    
\newcommand*{\CNRS}{Aix Marseille Univ, Universit\'e de Toulon, CNRS, CPT, Marseille, France.}\affiliation{\CNRS}    

\author{Colin Egerer}\affiliation{\WM}\affiliation{\Jlab}
\author{Christos Kallidonis}\affiliation{\Jlab}
\author{Joseph Karpie}\affiliation{\CU}
\author{Nikhil Karthik}\affiliation{\WM}\affiliation{\Jlab}
\author{Christopher J. Monahan}\affiliation{\WM}\affiliation{\Jlab}
\author{Wayne Morris}\affiliation{\ODU}\affiliation{\Jlab}
\author{Kostas Orginos}\affiliation{\WM}\affiliation{\Jlab}
\author{Anatoly  Radyushkin}\affiliation{\ODU}\affiliation{\Jlab}
\author{Eloy Romero}\affiliation{\Jlab}
\author{Raza Sabbir Sufian}\affiliation{\WM}\affiliation{\Jlab}
\author{Savvas Zafeiropoulos}\affiliation{\CNRS}
\collaboration{On behalf of the \textit{HadStruc Collaboration}}
\begin{abstract}
    We present a determination of the non-singlet transversity
    parton distribution function (PDF) of the nucleon, normalized
    with respect to the tensor charge at $\mu^2=2$ GeV$^2$ from
    lattice quantum chromodynamics. We apply the pseudo-distribution
    approach, using a gauge ensemble with a lattice spacing of 0.094
    fm and the light quark mass tuned to a pion mass of 358 MeV.
    We extract the transversity PDF from the analysis of the
    short-distance behavior of the Ioffe-time pseudo-distribution
    using the leading-twist next-to-leading order (NLO) matching
    coefficients calculated for transversity.  We reconstruct the
    $x$-dependence of the transversity PDF through an expansion in
    a basis of Jacobi polynomials in order to reduce the PDF ansatz
    dependence.  
    Within the limitations imposed by a heavier-than-physical pion mass and a fixed lattice spacing, we present a comparison of our estimate for the valence transversity PDF
    with the recent global fit results based on single transverse spin asymmetry. We find the intrinsic
    nucleon sea to be isospin symmetric with respect to transversity.
\end{abstract}

\preprint{JLAB-THY-21-3521}
\date{\today}
\maketitle

\section{Introduction}
\label{sec:intro}

The determination of the collinear quark and gluon structures of
polarized hadrons has been a vigorously pursued research program,
spurred by the abundant cross-section data from previous and ongoing
experiments, such as at HERA, Tevatron, JLab, RHIC and the LHC.
More exciting discoveries pertaining to hadron structure are to
come with the planned electron-ion collider (EIC)~\cite{Accardi:2012qut}
and the JLab 12 GeV~\cite{Dudek:2012vr,Chen:2014psa} upgrade.  The global-fit
analyses (for example,
see~\cite{Harland-Lang:2014zoa,Dulat:2015mca,Accardi:2016qay,NNPDF:2017mvq})
of the available fully-inclusive experimental data have led to a
high-precision extraction~\cite{Accardi:2016ndt} of the leading-twist,
unpolarized and polarized nucleon parton distribution functions
(PDFs) over a wide range of momentum fraction $x$, especially for the non-singlet
case, which has smaller experimental systematic uncertainties at
small $x$.  A complete understanding of the leading-twist collinear
structure of the proton, however, includes not only the unpolarized PDF
and polarized PDF of a longitudinally polarized  nucleon, but
also the transversity quark distribution that characterizes the
correlation of the transverse spin of a collinear parton with the
transverse polarization direction of the nucleon.

The transversity distribution, denoted by $h(x)$ or $\delta q(x)$
in the literature, measures the difference in the probabilities for
a hard virtual photon to scatter from a quark with spin aligned
parallel and antiparallel to the transverse polarization direction
of the nucleon.  The transversity distribution is the only chiral-odd
leading-twist collinear PDF.  This decouples the transversity PDF
from the inclusive deep-inelastic scattering (DIS) experiments, and hence, one has to rely on
other processes that can accommodate the required helicity-flip of
the scattered parton, such as those initially suggested
in~\cite{Ralston:1979ys,Artru:1989zv,Cortes:1991ja,Jaffe:1991kp,Ji:1992ev}.
The first determination of the nucleon transversity PDF resulted
from an analysis~\cite{Anselmino:2007fs} incorporating the experimental
data for the single spin asymmetry in semi-inclusive DIS (SIDIS)
process in HERMES~\cite{HERMES:2004mhh} and COMPASS~\cite{COMPASS:2006mkl}
experiments and chiral-odd TMD fragmentation functions from the
Belle data~\cite{Belle:2005dmx}.  The transversity distributions
for the valence $u$ and $d$ quarks were also extracted using the
data for dihadron production in
SIDIS~\cite{Radici:2018iag,Bacchetta:2011ip,Benel:2019mcq}.  Recently,
the first global analysis of the single spin asymmetry in SIDIS and
various other processes was presented by the JAM collaboration in
Ref.~\cite{Cammarota:2020qcw}, which demonstrated a universal
description of single spin asymmetry with a comparatively well
determined transversity PDF.  The scarcity of available data for
extracting the transversity PDF through a global analysis and the
non-conservation of the tensor charge make it less constrained, and
is therefore well-suited for an extraction from
first-principles lattice QCD.

Complementary to the global-fit determinations of the leading-twist
PDFs, {\sl in silico} lattice QCD computations of $x$-dependent
hadron structure are fast developing as a reliable framework. The
perturbative matching frameworks that use equal-time matrix elements
have proved particularly promising --- the large momentum effective
theory (LaMET)~\cite{Ji:2013dva,Ji:2014gla} and the perturbative QCD short-distance
factorization based approaches, the pseudo-distribution
approach~\cite{Radyushkin:2017cyf,Orginos:2017kos}, and the
factorizable lattice cross-section approach~\cite{Ma:2014jla,Ma:2017pxb}
as applied to the current-current
correlators~\cite{Braun:2007wv,Sufian:2019bol,Sufian:2020vzb}.  We
should, however, note that there are other methods to probe the
$x$-dependent hadron structure, such as through the direct computation
of the Mellin moments using leading-twist local
operators~\cite{Martinelli:1987zd}, the analytic continuation of
the hadronic tensor~\cite{Liu:1993cv}, operator product expansion (OPE) of the Compton
amplitude~\cite{Chambers:2017dov}, and the OPE of heavy-light current
correlators (HOPE method)~\cite{Detmold:2005gg,Detmold:2021uru}.
We refer the readers
to the recent
reviews~\cite{Ji:2020ect,Radyushkin:2019mye,Cichy:2018mum,Monahan:2018euv,Cichy:2021lih}
on these topics for technical discussions.

In this work, we apply the pseudo-distribution approach, for which
one uses a universal perturbative matching kernel ${\cal C}(u, z^2)$
to relate, in a short-distance regime at non-zero hadron momentum,
the invariant amplitudes associated with the renormalized matrix
elements of equal-time spacelike separated parton bilinears to the
$\nu$-Fourier transform of the $\msbar$ collinear PDF, or Ioffe-time
distribution $\mathcal{I}\left(\nu,\mu\right)$. Using the pseudo-distribution
and related approaches, lattice QCD computations of the unpolarized
and polarized quark
distributions~\cite{Egerer:2021ymv,Karpie:2021pap,Joo:2020spy,Joo:2019jct,Orginos:2017kos,Bhat:2020ktg,Fan:2020nzz,Alexandrou:2020qtt,Alexandrou:2018pbm,Lin:2020fsj},
and the valence distribution of the
pion~\cite{Sufian:2019bol,Sufian:2020vzb,Izubuchi:2019lyk,Gao:2020ito,Lin:2020ssv}
have been performed.  These studies demonstrate the ability of the
perturbative matching approaches to capture the expected behaviors
of the unpolarized and polarized PDFs from the global fits to a
reasonable degree, which one can consider in the experimentalists'
parlance as the {\sl controls} for the methodology.  With this
initial success, the lattice QCD investigations of some of the
experimentally less-constrained leading-twist quantities have begun
to appear; for example, the computations of the generalized parton
distribution functions~\cite{Alexandrou:2020zbe,Lin:2020rxa,Chen:2019lcm},
gluon PDFs~\cite{HadStruc:2021wmh,Fan:2018dxu,Fan:2020cpa,Fan:2021bcr},
and the topic of this paper, the transversity PDF.

Previous lattice QCD
studies~\cite{Bhattacharya:2016zcn,Bhattacharya:2015wna,Green:2012ej,Aoki:2010xg,Abdel-Rehim:2015owa,Bali:2014nma,Yamazaki:2008py}
based on the local operator approaches have computed the tensor
charge, $g_T(\mu)$, which is the first moment of the transversity
PDF, and the second
moments~\cite{Mondal:2020ela,Mondal:2020cmt,Alexandrou:2019ali,Harris:2019bih,Bali:2018zgl,Abdel-Rehim:2015owa}
of the transversity PDF. A study in Ref.~\cite{Lin:2017stx} found a
considerable impact of using the tensor charge $g_T$ from the lattice QCD
determinations as a constraint in the fits to the SIDIS data for the transversity PDF. Closely related to the present work, the $x$-dependence of the
transversity PDF has been computed before based on the perturbative
NLO $x$-space matching of the LaMET approach by two independent
groups in Refs.~\cite{Liu:2018hxv,Alexandrou:2018eet,Chen:2016utp}.
More recently, the first lattice QCD computation of the $x$-dependent
transversity generalized parton distribution function (GPD) based
on the LaMET approach was presented in Ref.~\cite{Alexandrou:2021bbo}.
The aim of this paper is to complement those previous studies with
an independent, first computation of the leading-twist transversity
PDF of the nucleon using the short-distance factorization based
pseudo-distribution approach. Independent computations of the
transversity PDF using different lattice quantities and factorization
approaches are crucial, because the different approaches suffer
from different systematic effects, such as those generated by power
corrections, renormalization prescriptions or perturbative truncation
effects.  The usage of the pseudo-distribution approach using
renormalization group invariant rations separate the computation
of the transversity PDF into two stages --- first, a computation
of the $x$-dependence of the PDF at a fixed normalization, and then
using standard lattice QCD methods to perform a computation of the
tensor charge $g_T$ to change the normalization from 1 to $g_T$.
Therefore, in this paper, we focus on the ratio $h(x,\mu)/g_T(\mu)$
that captures the $x$-dependence  and its corresponding perturbative
matching for the pseudo-distribution approach.

The structure of the paper is as follows. In \scn{matching}, we
present the definitions of the non-singlet valence and antiquark
transversity distributions, and then present the analytical results
for the NLO perturbative matching in real-space to match the
pseudo-distribution to the leading-twist $\msbar$ transversity PDF.
We discuss the details of the gauge ensemble and lattice measurements
in \scn{setup}.  In \scn{mel}, we present our determination of the
bare nucleon matrix elements that form the basis of our analysis
in the following sections. As a prelude to the extraction of the
transversity PDF, in \scn{opewoope} we present an analysis of the
efficacy of NLO leading-twist framework in explaining our lattice
data, and thereby deduce the necessary corrections we need to add
to the leading-twist framework. Finally, in \scn{results}, we present
our strategy for the reconstruction of the $x$-dependence of
transversity PDF using a Jacobi polynomial basis, and present a
comparison of our estimation with the available data on the
transversity PDF from the global fits.

\section{Theoretical framework: definitions and NLO matching}
\label{sec:matching}
In this work, we make use of the factorization of the pseudo-ITD
matrix element at the perturbatively small quark-antiquark separations,
$z$, into a hard perturbative matching kernel $C(u, \mu^2 z^2)$ and
the parton distribution function; in our case, the transversity
PDFs corresponding to the isotriplet flavor combinations at scale $\mu$. We first
explicitly define the relevant isovector combinations of the
transversity PDF and then discuss the NLO matching kernel that
relates the ratio of hadronic matrix elements, calculable on the
lattice, to the light-cone transversity PDF in the $\msbar$ scheme.

\subsection{Definition of non-singlet transversity distributions}
The transversity PDF of the nucleon with  spin $S^{\nu_\perp}$
polarized in a transverse direction $\rho_\perp$ and an on-shell
momentum $P$ can be defined within QCD in terms of the quark-fields
$\psi$ and $\bar\psi$ that are displaced along the light-cone as,
\beqa
&&h(x,\mu) = \int_{-\infty}^\infty \frac{d\nu}{2\pi} e^{-i x \nu} {\cal I}(\nu,\mu)\qquad\text{with}\qquad,\cr
&&2 P^+ S^{\rho_\perp} {\cal I}(P^+ z^-,\mu) = \left\langle P,S^{\rho_\perp}|\bar\psi(z^-)\gamma^+\gamma^{\rho_\perp} \gamma_5 W_+(z^-,0)\psi(0) | P, S^{\rho_\perp}\right\rangle,
\eeqa{tpdf}
with the straight Wilson-line $W_+(z^-,0)$ making the definition
gauge-invariant.  The non-singlet transversity PDF that we compute
can be succinctly written as
\beq
h_{u-d}(x)=h_u(x)-h_d(x),\quad x\in[-1,1].
\eeq{huminusd}
It is more useful to write the above quantity in terms of quark ($q$) and
antiquark ($\bar q$) distributions that have support from $[0,1]$
by identifying $h_q(-|x|)=-h_{\bar q}(|x|)$.
Following the conventions laid down in the community white
paper~\cite{Lin:2017snn}, the non-singlet transversity distributions in 
this paper are
\beqa
h_-(x)\equiv h_{u^--d^-}(x) &=& h_{u}(x)-h_{\bar u}(x) - h_{d}(x)+h_{\bar d}(x),\cr
h_+(x) \equiv h_{u^+-d^+}(x) &=& h_{u}(x)+h_{\bar u}(x) - h_{d}(x)-h_{\bar d}(x),
\eeqa{pdfdef1}
for $x\in[0,1]$, and their Mellin moments given as 
\beq
\langle x^n \rangle_\pm \equiv \langle x^n \rangle_{u^\pm - d^\pm} = \int_0^1 dx x^n h_\pm(x).
\eeq{mellindef}
The factorization scale $\mu$ is implicit in the above equations,
and the evolution of $h(x,\mu)$ and their moments with the scale
is given in~\cite{Vogelsang:1997ak}.  By defining $h_-(x)$ as the
valence quark distribution, $h_{\rm v}(x)$, and $h_{\bar u - \bar
d}(x) = h_{\bar u} - h_{\bar d}$ as the isotriplet antiquark
distribution that characterizes the intrinsic sea, we see that,
\beqa
h_{\rm v}(x) &\equiv& h_{-}(x),\cr
h_{\rm v}(x) + 2 h_{\bar u - \bar d}(x)  &\equiv& h_+(x).
\eeqa{pdfdef2}
In contrast to the unpolarized quark distribution, which corresponds to the
distribution of the conserved charge amongst the partons, the
underlying tensor charge,
\beq
g_T(\mu) = \langle x^0 \rangle_-,
\eeq{gtfrompdf}
is not conserved, and hence, it depends on the renormalization
scheme and it runs with the renormalization scale $\mu$. We express
the tensor charge and the transversity distribution in the $\msbar$
scheme.  A global fit to the lattice QCD results for the tensor
charge gives $g_T(\mu)=1.00(5)$ at $\mu^2 = 2 {\ \rm
GeV}^2$~\cite{Lin:2017stx}.  In this work, we focus on the shape
of the $x$-dependent transversity distribution, and defer a dedicated
computation of $g_T(\mu)$ to the future. Therefore, the aim of this
work is to compute $h_{\rm v}(x,\mu)/g_T(\mu)$ and $h_{\bar u -
\bar d}(x,\mu)/g_T(\mu)$ as a function of $x$ from the appropriately
defined pseudo-PDF matrix element.

\subsection{NLO matching from the pseudo-ITD to $\msbar$ transversity PDF}
Let us consider an on-shell proton with a momentum four-vector
$P=\left(E(\mathbf{P}), \mathbf{P}\right)$ and spin vector $S^{\perp}$
satisfying $\left(S^{\perp}\right)^2=-1, S^{\perp}\cdot P=0$, and
such that it points in a spatial direction that is transverse to
spatial momentum $\mathbf{P}$; the relativistically normalized 
quantum state  is denoted as $|P, S^{\perp}\rangle$.  Within both the short-distance 
factorization and the LaMET approaches, the expectation value of 
an appropriately chosen bilocal quark operator is evaluated in the
boosted hadron state.
Such a flavor non-singlet Wilson-line 
connected bilocal quark bilinear operator that is appropriate for 
obtaining the transversity PDF is
\beq
O_{\gamma_5 \gamma_\lambda\gamma_\rho}(z) \equiv \bar\psi \gamma_5\gamma_\lambda\gamma_\rho W(0,z) \tau_3 \psi,
\eeq{transop}
where $\psi=(u,d)$, and $W(0,z)$ is the straight Wilson-line
connecting the quark and antiquark separated by $z$.
The Lorentz decomposition~\cite{Musch:2010ka}
of its forward nucleon matrix element is
\beqa
&&\langle P,S^\perp|O_{\gamma_5 \gamma_\lambda\gamma_\rho}(z)| P, S^\perp \rangle =\cr &&\quad
2(P_\lambda S^\perp_\rho-P_\rho S^\perp_\lambda){\cal M}(z\cdot P,z^2)+
2i m_N^2 (z_\lambda S^\perp_\rho-z_\rho S^\perp_\lambda) {\cal N}(z\cdot P,z^2)+
2 m_N^2 (z_\lambda P_\rho-z_\rho P_\lambda) (z\cdot S^\perp) {\cal R}(z\cdot P,z^2).
\eeqa{lordec}
As is
conventional, in this work, we choose $z=(0,0,0,z_3)$ and $P=(E(P_3),0,0,P_3)$,
thereby making $\nu= -z\cdot P= z_3 P_3$ and $-z^2= z_3^2$.  The quantity $\nu=-z\cdot P$ 
is referred to as the Ioffe-time~\cite{Ioffe:1969kf,Braun:1994jq}.
Of the
three independent form-factors ${\cal M}, {\cal N}$ and ${\cal R}$,
only ${\cal M}$ gives the leading-twist contribution. Hence, by a
good choice of directions $\rho$ and $\lambda$, we can project onto
${\cal M}$; such a choice is $\lambda=0$ (that is, along the temporal
direction) and $\rho=1,2$ (that is, either of the two spatial
directions transverse to the nucleon momentum). Coincidentally, it
is precisely this choice that is purely multiplicatively
renormalizable without any mixing~\cite{Constantinou:2017sej}.  
For these choices of directions $\lambda=0$ and $\rho=1,2$,
the spin vectors are $S^{\perp}=(0,1,0,0)$ and $(0,0,1,0)$ respectively.
Using these choices in \eqn{lordec}, and by using the rotational
invariance, we find
\beq
{\cal M}(z_3, P_3) =\frac{1}{4 E(P_3)} \sum_{\rho=1}^2 \langle P,S^{\perp}|O_{\gamma_5 \gamma_0\gamma_{\rho}}(z)| P, S^{\perp} \rangle.
\eeq{itddef}
For convenience in what follows, we have written the arguments of
${\cal M}$ as $(z_3,P_3)$ without making use of the Lorentz structure.
The above matrix element is not renormalized due to the self-energy
divergence of the Wilson-line, the logarithmic end-point divergences,
and standard field renormalizations for
$\psi$~\cite{Ji:2017oey,Ishikawa:2017faj,Green:2017xeu}. Due to the
multiplicative renormalizability for the choices of directions as
made above, we can define the reduced pseudo-ITD
(rpITD)~\cite{Radyushkin:2017cyf,Orginos:2017kos} for the transversity
PDF as
\beq
\mathfrak{M}(\nu, z_3^2) \equiv \frac{{\cal M}(z_3, P_3)}{{\cal M}(z_3, 0)} \frac{{\cal M}(0, 0)}{{\cal M}(0, P_3)}.
\eeq{ritddef}
The first factor on the right-hand side above removes the self-energy
divergence of the Wilson-line, and the second factor above ensures
that in the local operator limit, $z_3\to 0$, the rpITD becomes
$\mathfrak{M} \to 1$ independent of renormalization scale. Thus,
it is clear that by using the above definition of rpITD, we have
forsaken the information on the tensor charge, $g_T(\mu)$, that
would have been otherwise obtained in the limit $z_3\to 0$ at fixed
$P_3$. Hence, we expect that $\mathfrak{M}$ matches onto the
transversity PDF that is normalized to unity, that is $h(x,\mu)/g_T(\mu)$;
this expectation indeed gets borne out of an actual perturbative
calculation to compute the rpITD-to-$\msbar$ PDF matching kernel
using on-shell quark external states.  The renormalization choice
of setting the $z_3=0$ matrix element to 1 has further advantage
of reducing the statistical errors for the matrix elements at other
smaller $z_3$ due to correlations in the data. From our experience
with the rpITD for the unpolarized PDF, we expect it might
help in the cancellation of higher-twist effects and finite volume
effects (through the complete removal of all corrections at ${\cal O}(\nu^0)$) 
for the transversity rpITD as well--- however, this
expectation needs to be checked through further studies.
\bef
\centering
\includegraphics[scale=0.9]{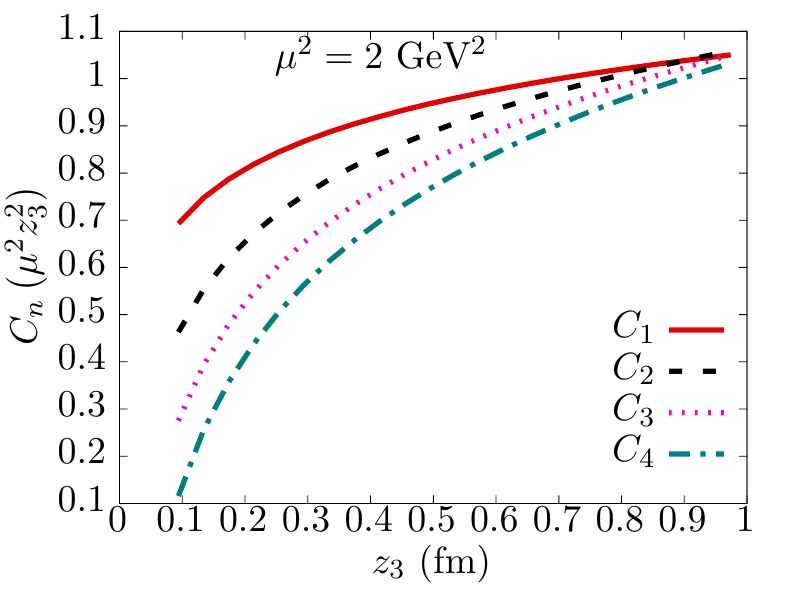}
\caption{The $z_3$ dependence of the Wilson coefficients, $C_n(\mu^2
z^2_3)$, in the leading-twist OPE for transversity for $n=1,2,3$
and $4$. The value of $\mu=\sqrt{2}$ GeV.}
\eef{cn}

The
matching relation involving the perturbative kernel ${\cal C}$ has the general form 
of the lightcone OPE \cite{Balitsky:1987bk}
\beq
\mathfrak{M}^{\rm twist-2}(\nu, z_3^2) = \int_0^1 du\, {\cal C}\left(u, \mu^2 z_3^2\right) {\cal I}(u \nu, \mu),
\eeq{transmatching}
where the normalized $\msbar$ light-cone transversity ITD $ {\cal I}(u \nu, \mu)$ is related to the transversity PDF by 
\beq
{\cal I}(\nu, \mu) = \int_{-1}^1 dx\, e^{i x \nu}\, \frac{h_{u-d}(x,\mu)}{g_T(\mu)}.
\eeq{lcitd}
The expression for the matching kernel at NLO was found to be given by\footnote{During the preparation of this paper we have learned that the  equivalent result has been obtained by Braun et.al. ~\cite{Braun:2021gvv}.}
\beq
{\cal C}(u,\mu^2 z_3^2) = \delta(1-u)-\frac{\alpha_s C_F}{2\pi}\bigg{\{}
\left[\frac{2u}{1-u}\right]_+ \ln\left(\frac{z_3^2 \mu^2 e^{2\gamma_E+1}}{4}\right)
+4\left[\frac{\ln(1-u)}{1-u}\right]_+\bigg{\}}.
\eeq{nlokernel}
Here we use  the standard definition of the plus-prescription at $u=1$.
The matching formula may also be rewritten~\cite{Braun:2007wv,Izubuchi:2018srq} in the 
form of the leading-twist local OPE
\beq
\mathfrak{M}^{\rm twist-2}(\nu, z_3^2) = \sum_{n=0}^{2 N_{\rm max}}  a_{n+1}(\mu) C_n\left(\mu^2 z_3^2\right) \frac{\left(i \nu\right)^n}{n!} \ ,
\eeq{tw2ope}
which is nothing but the Taylor expansion 
in $\nu$ of the lightcone OPE  to
an order $N_{\rm max}$.
The accuracy of the leading-twist local OPE improves as $N_{\rm max}\to\infty$,
but a {\sl large-enough} value of $N_{\rm max}$ is sufficient given
the statistical precision of the lattice data, as well as the finite
range of $\nu$ and $z_3$ that the lattice data spans.  The Mellin moments
normalized by $g_T(\mu)$ are given by
\beq
a_{n+1}(\mu)  = \begin{cases} 
{\langle x^n \rangle_-}/{g_T}, \quad\text{even\ } n,\cr 
{\langle x^n \rangle_+}/{g_T}, \quad\text{odd\ } n, \end{cases}
\eeq{mellinmom}
with $a_1(\mu)=1$. The leading-twist NLO Wilson coefficients,
$C_n(\mu^2 z_3^2) = \int_0^1 du\, {\cal C}(u, \mu^2 z_3^2) u^n$, for
transversity are given by
\beq
C_n\left(\mu^2 z_3^2\right) = 
1+\frac{\alpha_s C_F}{\pi}\bigg{\{}   \ln\left(\frac{z_3^2 \mu^2 e^{2\gamma_E+1}}{4}\right)\sum_{k=2}^{n+1} \frac{1}{k}
-\left (\sum_{k=1}^{n}\frac{1}{k} \right )^2 -\sum_{k=1}^{n}\frac{1}{k^2} \bigg \} \ .
\eeq{wilcoeff}
By fitting the lattice data for ${\rm Re}\, \mathfrak{M}$ using the
above expression for ${\rm Re}\, \mathfrak{M}^{\rm twist-2}$, we
can obtain $h_-(x,\mu)$.  Similarly, we can obtain  $h_+(x,\mu)$
from ${\rm Im}\, \mathfrak{M}$. We use the value of $\alpha_s$ from
the PDG~\cite{Tanabashi:2018oca} at the same scale $\mu$ used to
determine the PDF.

In \fgn{cn}, we show the variation of the Wilson coefficients $C_n$
with $z_3^2$ at a scale of $\mu=\sqrt{2}$ GeV.  As the Mellin moments
typically decrease rapidly with the order $n$, and also due to the
$n!$ suppression of higher-orders in \eqn{tw2ope}, only the few
lowest $n$ mainly contribute in \eqn{tw2ope} given a finite range
in $\nu$.  Therefore, \fgn{cn} shows the effect of $O(\alpha_s)$
corrections to $C_n$ for the lowest four $n$. The 1-loop effect on
$C_1$ and $C_2$ at intermediate $z_3\approx 0.4$ fm is about 10\%
and 20\% respectively, whereas the effect is about 35\% on $C_3$
and $C_4$. For even smaller $z_3$ where the effect of $\ln(\mu^2z_3^2)$
increases, typically only the $n=1$ and 2 dominate \eqn{tw2ope},
for which the 1-loop effect is about 20\% and 40\% respectively at
$z_3=0.2$ fm which is about two lattice units in the ensemble we
use for this work. Practically, such ${\cal O}(\alpha_s)$ corrections
could have an even smaller effect when convoluted with realistic
PDFs.  Thus, at the level of matching, we are working in a region
of $z_3$ where the 1-loop corrections at a fixed $\alpha_s(\sqrt{2} {\ \rm
GeV})$ are small.

\section{Lattice setup}
\label{sec:setup}
The computation presented in this paper was performed using a lattice
ensemble generated by the JLab/W\&M/LANL collaboration~\cite{edwardsetal}
with a
lattice spacing $a=0.094$ fm and the pion mass tuned to $M_\pi=358$
MeV with a physical strange quark mass. The computation
is unitary using 2+1 flavor isotropic Wilson-clover fermion action
in both the sea and the valence quark sectors.  We used a fixed
lattice size of $L^3\times  L_t = 32^3\times 64$. Further details
of the ensemble are presented in Refs.~\cite{Yoon:2016jzj,Yoon:2016dij}.

In order to project onto the nucleon ground-state $|P,S_\perp\rangle$
with spatial momentum $\mathbf{P}=(0,0,P_3)$ and with the spin
polarization $S^\perp$ that is in a spatial direction $\nu$,
perpendicular to $\mathbf{P}$, we insert the nucleon interpolating
operator ${\cal N}(t',P_3,S^\perp)$ in time-slices $t'=t$ and $t'=0$.
The key features of this computation are the usages of distillation~\cite{HadronSpectrum:2009krc} 
and its modification using phases~\cite{Egerer:2020hnc} that make determination of 
high-momentum matrix elements possible. The details related to the
implementation of distillation, that is pertinent to the ensemble used here, is given in
our previous publication~\cite{Egerer:2021ymv}.  The spin projection is achieved
via the projectors ${\cal P}^\perp = \frac{1}{2}(1+\gamma_5
\slashed{S}^\perp)=\frac{1}{2} (1+\gamma_5\gamma_\nu)$.  In the
Pauli-Dirac representation we use in our computations, the spin
projector for the positive parity state reduces to a more familiar
$2\times 2$ matrix, ${\cal P}^\perp = \frac{1}{2}(1+\sigma_\nu)$.
We computed the set of spatial momenta,
\beq
P_3 = n_3 \Delta;\qquad \Delta = \frac{2\pi}{L a} = 0.41 {\rm\ GeV},
\eeq{p3}
for $n_3=0,1,2,3,4,5,6$. In physical units, these momenta correspond
to $P_3 = 0, 0.41, 0.82, 1.23, 1.64, 2.06$ and 2.47 GeV respectively.
For the sake of lattice corrections, the pertinent scale is $a^{-1}$,
in units of which these momenta correspond to $0.196 n_z$; that is,
the lowest four momenta are well below $a^{-1}$, where as the highest
two momenta are comparable to $a^{-1}$.

We extracted the bare matrix element ${\cal M}(z_3,P_3)$ by computing
the two-point function,
\beq
C_{\rm 2pt}(t_s; P_3)  = \left\langle {\cal N}(t_s, -P_3, S^\perp) \overline{\cal N}(0,P_3, S^\perp)\right\rangle,
\eeq{2pt}
and the three-point function,
\beq
C_{\rm 3pt}(t_s,\tau; z_3, P_3)  =\frac{1}{2} \sum_{\rho=1}^2  \left\langle {\cal N}(t_s, -P_3, S^\perp) O_{\gamma_5\gamma_0 \gamma_\rho}(z_3; \tau) \overline{\cal N}(0,P_3, S^\perp)\right\rangle,
\eeq{3pt}
where the operator $O_{\gamma_5\gamma_0 \gamma_\rho}(z_3;\tau)$ is
inserted at a time-slice $\tau$, for $0 < \tau < t_s$.  We used
$t_s = 4a, 6a, 8a, 10a, 12a, 14a$ in our computation. In physical
units, the source-sink separation ranges from 0.388 fm to 1.358 fm.
As we will see, at the three highest momenta, reasonable signal was
obtained up to $t_s=10a$ corresponding to 0.97 fm.  Our values of
quark-antiquark separations $z_3$ ranged from $0$ to $16a$ for
momenta $n_3 < 4$, and ranged from $0$ to $8a$ for the higher three
momenta. Since, we performed fits in shorter $z_3<1$ fm, only the
values of $z_3 \le 10a$ were actually usable in the analysis. In
\eqn{3pt}, we have averaged over the two spatial directions that
are transverse to $P_3$, but we checked to ensure that the two
individual three point functions are consistent with each other
well within 1-$\sigma$ errors.

\section{Extraction of bare matrix element}
\label{sec:mel}

\bef
\centering
\includegraphics[scale=0.82]{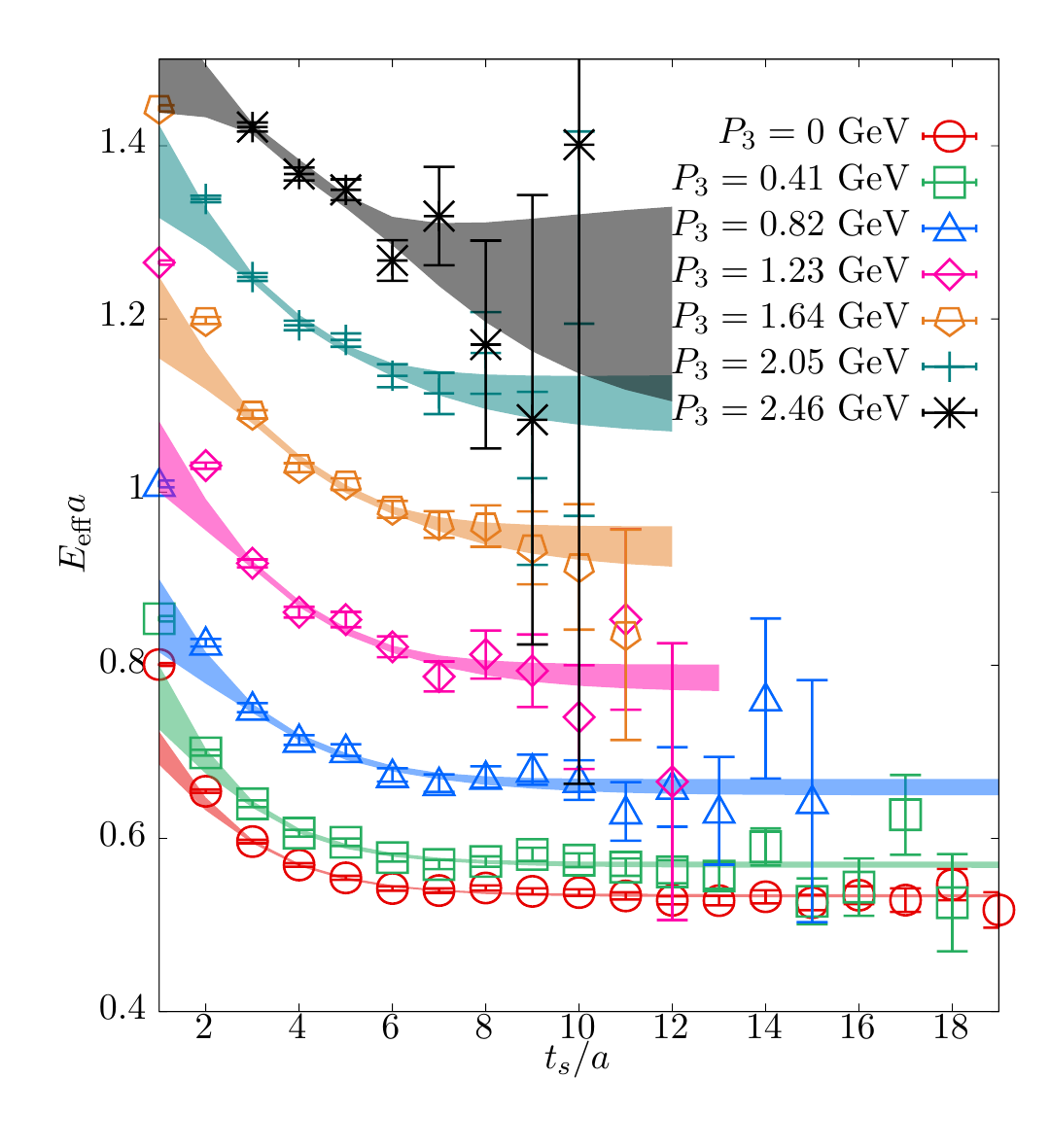}
\caption{
    The effective masses $E_{\rm eff}(t_s)$ determined from the two-point
    function of transversely polarized nucleon at different momenta
    $P_3$ along the $z$-direction are shown as a function of
    source-sink separation $t_s/a$.  The filled bands are the
    expectations for $E_{\rm eff}(t_s)$ based on the two-state fits to
    the nucleon correlator over a fit range $t_s\in[3a,18a]$.
}
\eef{eff}

\bef
\centering
\includegraphics[scale=0.82]{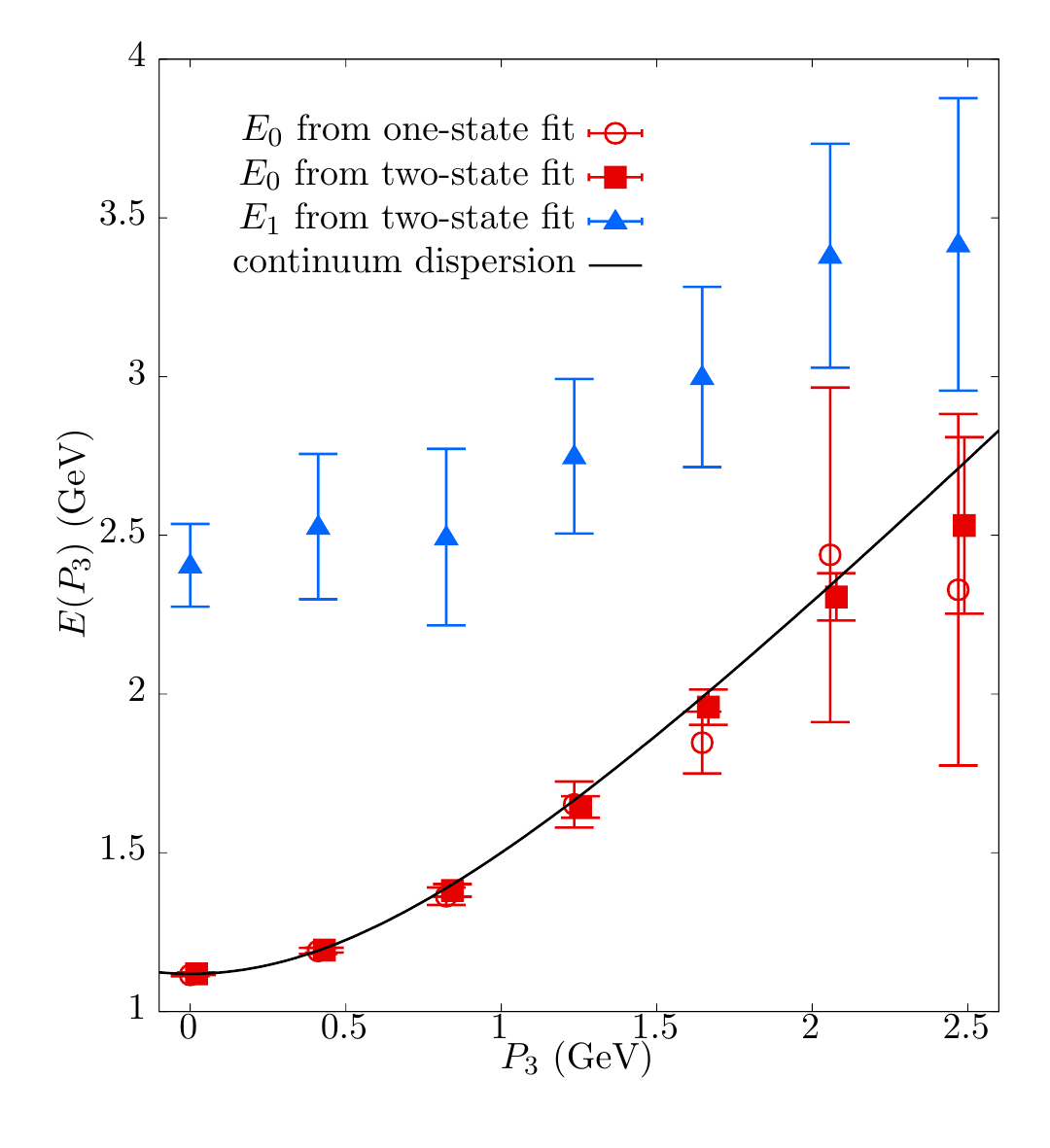}
\caption{
    The plot shows the ground state nucleon energy, $E_0$, (filled
    red squares) and the first excited state, $E_1$, (filled blue
    triangles) as extracted from the nucleon two-point function using
    the two-state fits over a range $t_s\in [3a,18a]$, at
    different nucleon momenta $P_3$. For comparison, the nucleon
    ground state masses obtained from the one-state fits over a range of
    $t_s\in[10a,18a]$ are shown using the open red circle symbols.
}
\eef{disp}

We follow the standard ways to obtain the bare matrix element from
the three-point and two-point functions in \eqn{2pt} and \eqn{3pt};
namely, two-state fits to the ratio of three-point to two-point
functions and via summation method~\cite{Maiani:1987by,Capitani:2012gj}.
In the end, we will primarily use the summation method to cross-check
the consistency of the extrapolations from the two-state fits of
the ratio, and input the extrapolated matrix elements from the
three-point to two-point ratio in the analysis of transversity PDF
in the rest of the paper.

For the fits, we use the spectral decomposition of the two-point
and three-point functions in terms of the excited-state energies
$E_n$ and their amplitudes $Z_n$, namely,
\beq
C_{\rm 2pt}(t_s; P_3) = \sum_{n=0}^{N-1} |Z_n|^2 e^{-E_n t_s}; \qquad Z_n = \frac{1}{\sqrt{2E_n}} \langle 0|{\cal N}|n\rangle,
\eeq{2ptspectral}
and 
\beq
C_{\rm 3pt}(t_s,\tau; z_3,P_3) =\sum_{n,m=0}^{N-1} \frac{Z_n^* Z_m}{2\sqrt{E_n E_m}}\langle n|O(z_3)|m\rangle  e^{-E_n (t_s-\tau)-E_m \tau}.
\eeq{3ptspectral}
It is clear that the leading ground-state contribution in $C_{\rm
3pt}$ is the desired ${\cal M}(z_3,P_3)$. Given the statistical
error in the data, we truncated the above spectral decomposition
at $N=2$ in both \eqn{2ptspectral} and \eqn{3ptspectral}; we refer
to fits performed with this $N=2$ truncation as the two-state fits.
Our methodology is to use the two-state fits using \eqn{2ptspectral} to obtain the energies and
amplitudes of the nucleon and the first excited state from the two-point
function data. Using the
jackknife samples of fitted values as the input, we then performed
two-state fits to the $t_s$- and $\tau$-dependencies of the three-point
function data using the matrix elements, $\langle n|O(z_3)|m\rangle$,
as the fit parameters. The resultant jackknife samples of the
fitted values of the ground-state matrix element, ${\cal M}(z_3,P_3)$,
were then used in the analysis of transversity PDF that we will
discuss in the following sections. It is convenient to implement this
excited-state analysis scheme by defining the ratio,
\beq
R(t_s,\tau; z_3,P_3) \equiv \frac{C_{\rm 3pt}(t_s,\tau;z_3,P_3)}{C_{\rm 2pt}(t_s;P_3)},
\eeq{ratioR}
so that the leading term in its corresponding spectral decomposition
that follows from \eqn{2ptspectral} and \eqn{3ptspectral} is simply
the bare matrix element ${\cal M}(z_3,P_3)$. A related technique
is the summation method, which uses the quantity,
\beq
R^{\rm sum}(t_s;z_3,P_3) = \sum_{\tau = \tau_0}^{t_s-\tau_0} R(t_s,\tau;z_3,P_3),
\eeq{rsum}
where one can skip $\tau_0$ data points closer to the source and
the sink. From the spectral decomposition, it is clear that the
leading $t_s$ dependence is a straight-line,
\beq
R^{\rm sum}(t_s;z_3,P_3) = t_s {\cal M}(z_3,P_3) + R_0 + {\cal O}\left(e^{-(E_1-E_0) t_s}\right).
\eeq{sumspec}
In the $t_s\to\infty$ limit, one would expect $t^{-1}_s R^{\rm
sum}(t_s)$ to approach ${\cal M}$.

We used ($N=1$) one-state and ($N=2$) two-state fits to the nucleon
two-point function to extract the ground-state energy $E_0(P_3)$.
We varied the fit range $t_s\in [ t_{\rm min}, t_{\rm max} ]$ to
check for the robustness of the fit parameters. For the one-state
fits, we found using a fit range $[10a,18a]$ to be optimal and be
consistent with the larger $t_{\rm min}$. For $P_3=0$, we found the
nucleon mass in the ensemble to be 1.115(5) GeV. As a cross-check, the estimate for nucleon mass here using a single interpolating operator is
consistent with an earlier estimate~\cite{Khan:2020ahz} on the same ensemble using an extensive GEVP basis. Since, we use
values of $t_s$ and $\tau$ which are smaller than $10a$, a single-state
fit is not a feasible approach to obtain the matrix elements, and
therefore, we performed two-state fits to the nucleon correlator
with smaller values of $t_{\rm min}=2a,3a$ and $4a$, and $t_{\rm
max}=18a$.
At all the momenta, we found that such two-state fits
resulted in $E_0$ that were consistent with those obtained using
one-state fits with $t_{\rm min} \ge 10a$. It was also encouraging
that the central values of $E_0$ and $E_1$ obtained from the two-state
fits, showed only small variations ($< 1\%$ for $E_0$ and $<10\%$
for $E_1$)  when $t_{\rm min}$ was changed and such variations were
within the statistical errors. Therefore, we used the results of
two-state fits over a range $[3a, 18a]$ in the extrapolation of
three-state fits to be discussed next. In \fgn{eff}, we show the
effective mass, $E_{\rm eff}(t_s)$, as a function of source-sink
separation, $t_s$. In the figure, we have differentiated the data
at different $P_3$ using different colored symbols as specified in
the legend.  We have compared the expectation for $E_{\rm eff}(t_s)$
from the two-state fits, shown as the bands of different colors for
different $P_3$, with the actual data for $E_{\rm eff}$. The
goodness of the two-state fits is evident in the agreement with 
the data for $t_s\ge 3a$.

\befs
\centering
\includegraphics[scale=0.54]{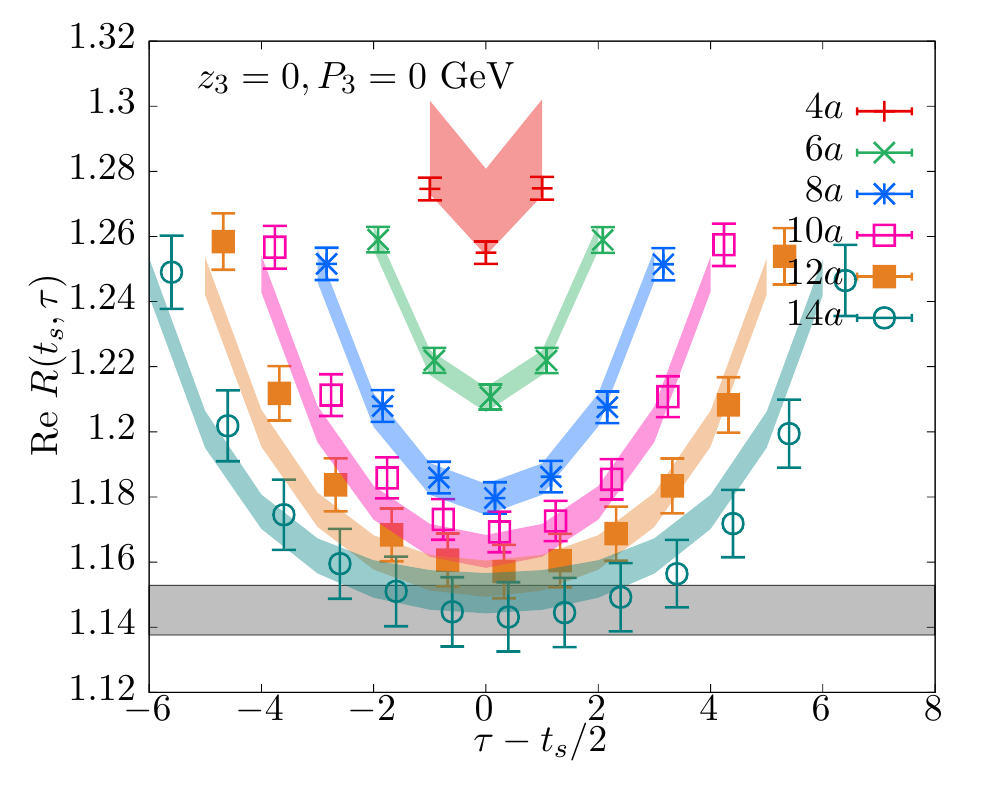}
\includegraphics[scale=0.54]{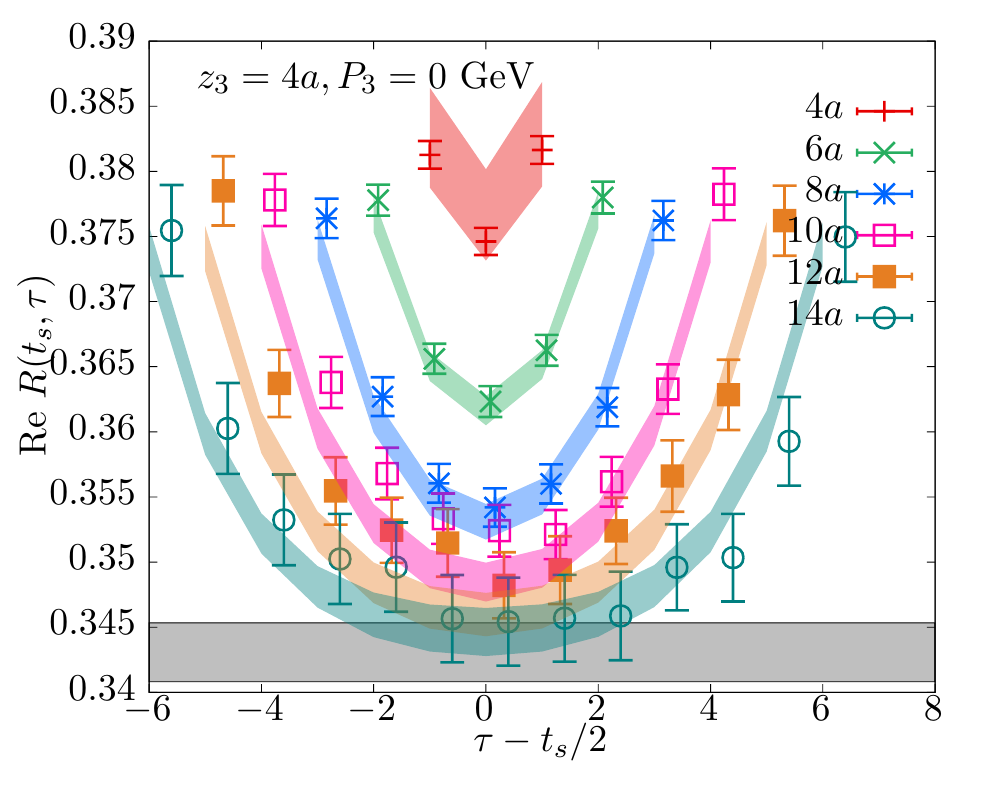}
\includegraphics[scale=0.54]{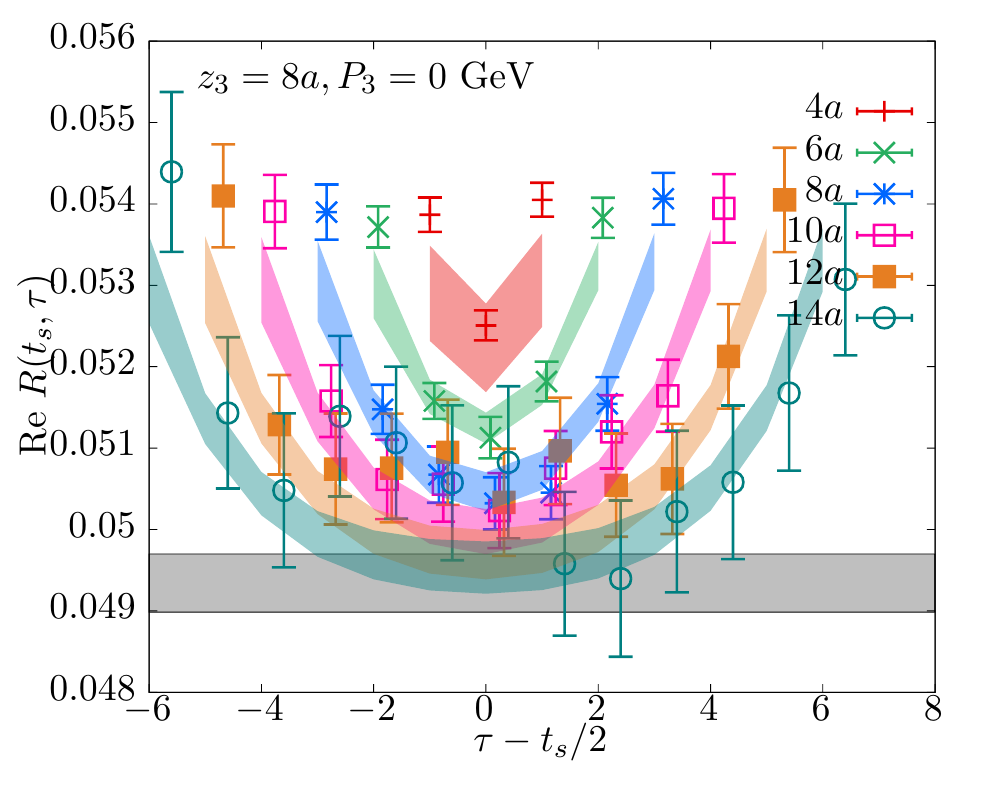}

\includegraphics[scale=0.54]{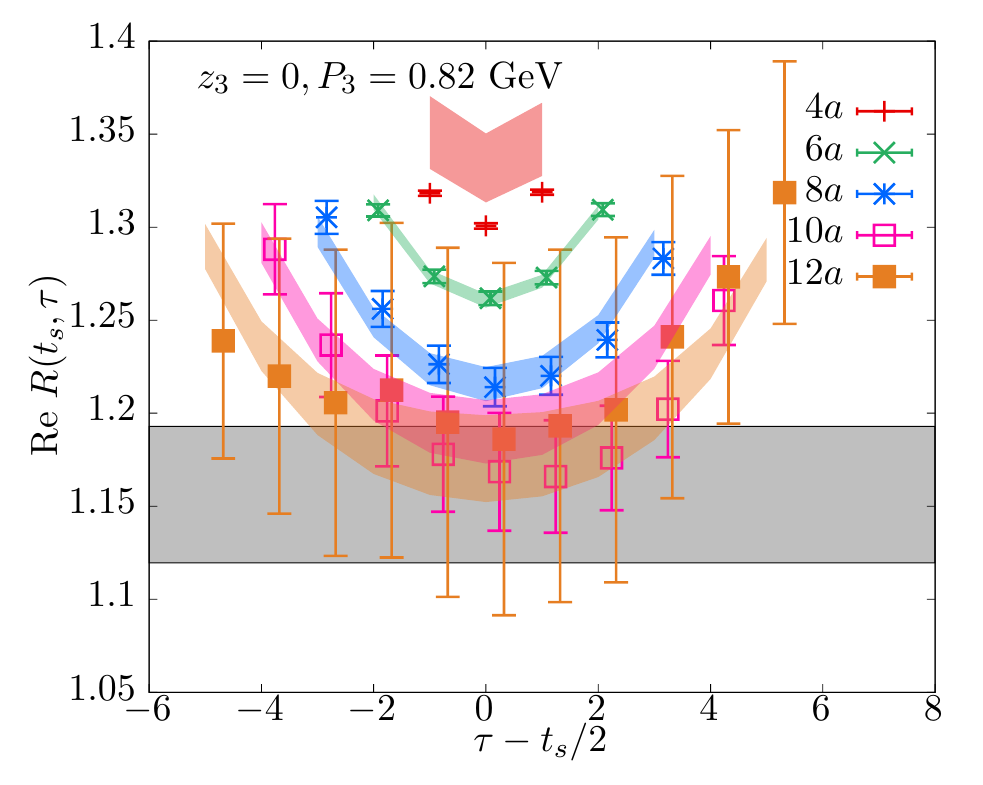}
\includegraphics[scale=0.54]{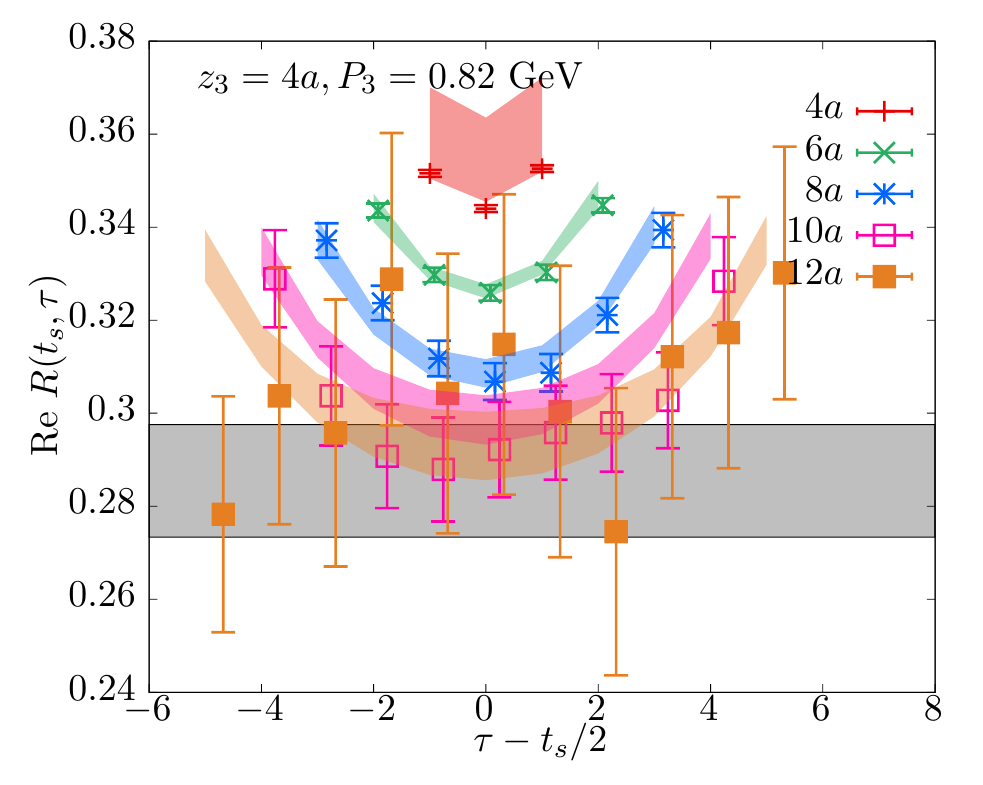}
\includegraphics[scale=0.54]{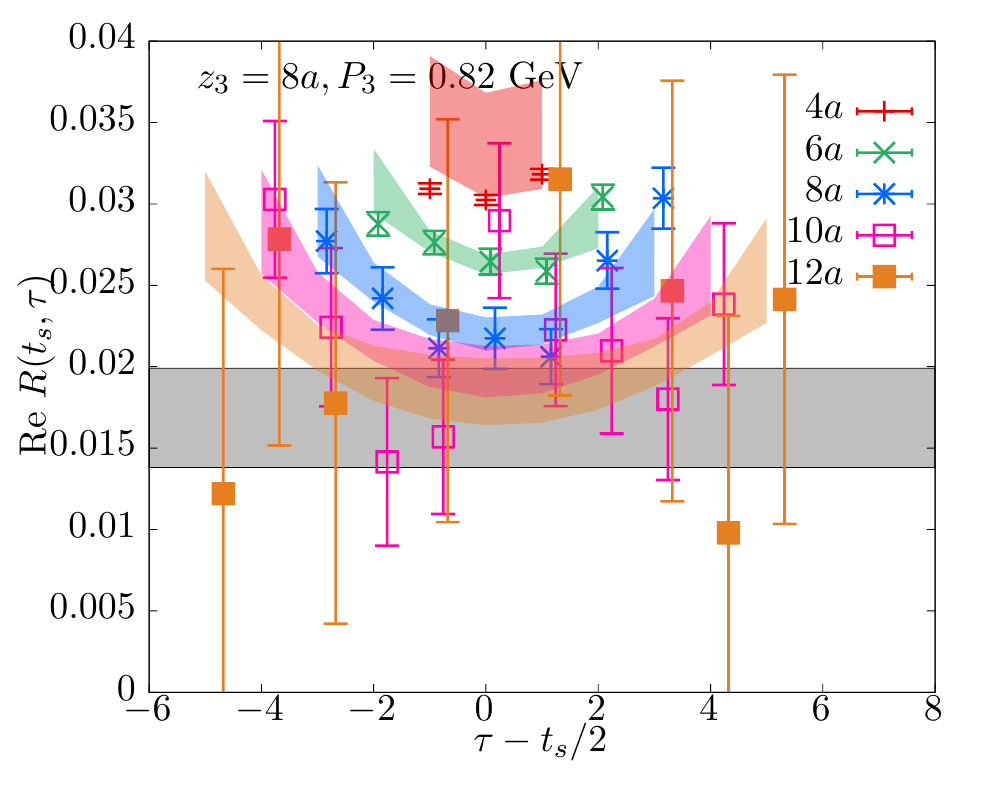}

\includegraphics[scale=0.54]{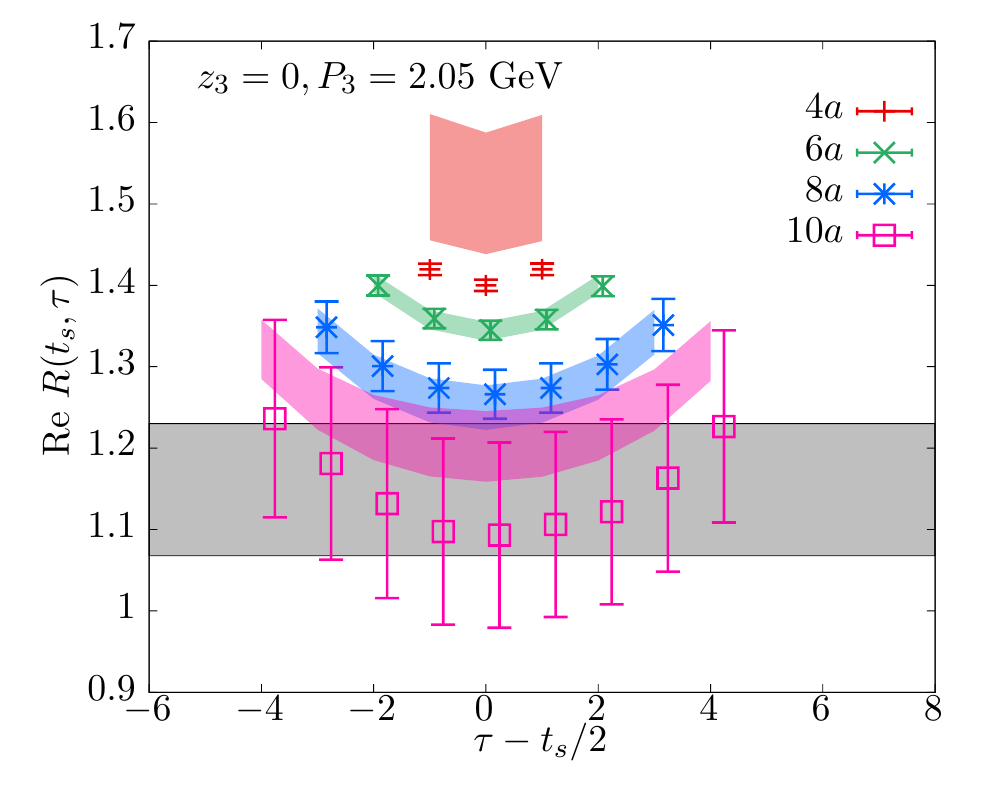}
\includegraphics[scale=0.54]{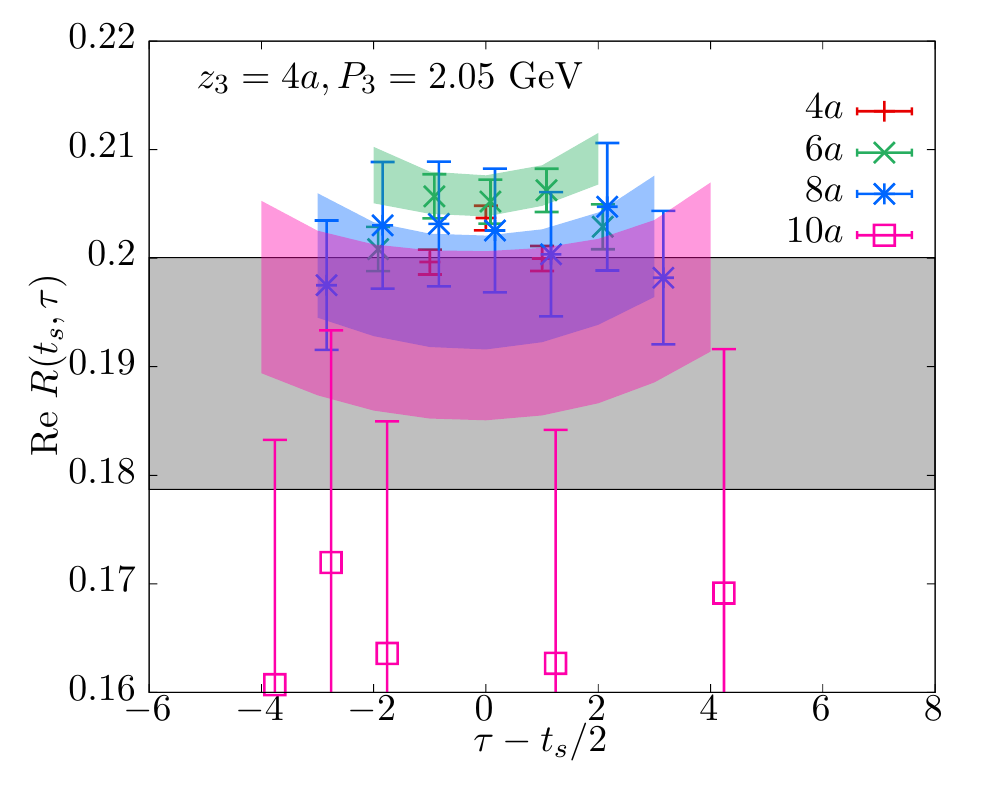}
\includegraphics[scale=0.54]{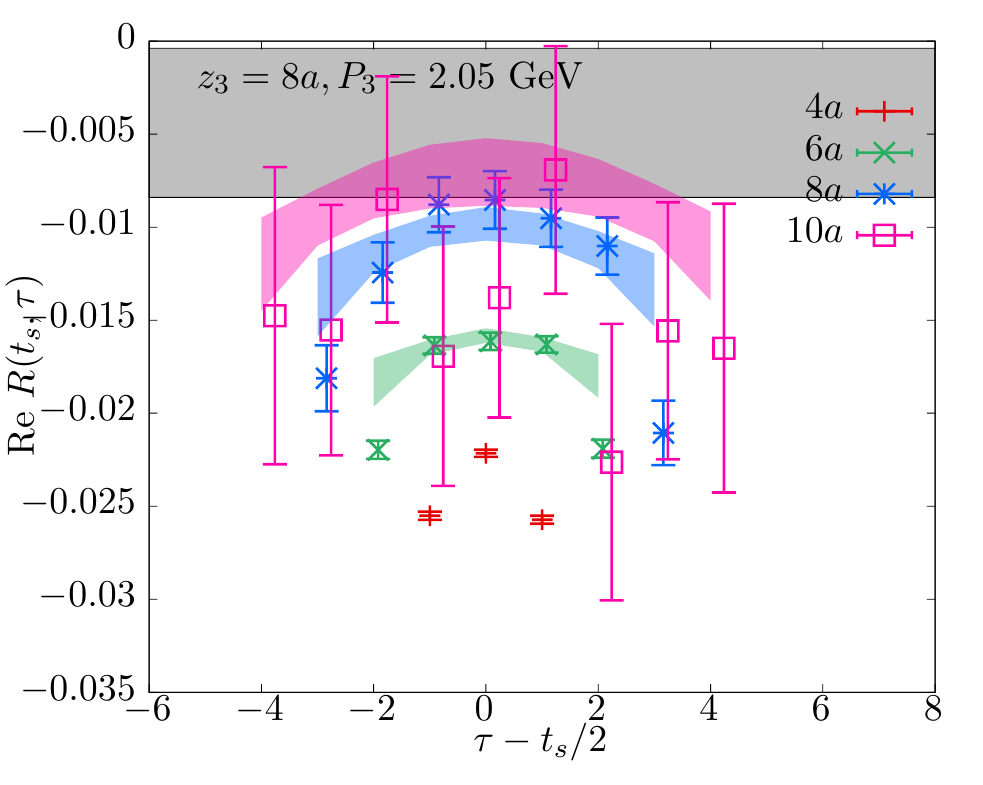}
\caption{
    The plot shows a sample of the excited-state extrapolations of the ratio $R(t_s,\tau)$ of the 
    three-point function
    to two-point function ratio to obtain the bare
    ground-state nucleon matrix element, ${\cal M}(z_3,P_3)$.
    As the real and imaginary parts of $R$ displayed similar behaviors, only the real part of $R$ is shown. In each panel,
    ${\rm Re\ }R(t_s,\tau)$ is shown as a function of $t_s-\tau/2$,
    where $t_s$ is the source-sink separation and $\tau$ is the
    operator insertion time. The points are the lattice measurements
    and the bands are the expectations based on the two-state fits to
    ${\rm Re\ }R(t_s,\tau)$ over a range of $t_s\in[6a,14a]$. The ratios at different fixed $t_s$,
    as specified in the plot legends, are distinguished by the colored
    symbols and bands used. 
    The horizontal gray band is the extrapolated value.
    The matrix of panels are such that the
    three rows from the top to bottom show the results at momenta $P_3=0, 0.82$
    and 2.05 GeV, and the three columns from the left to
    right are for quark-antiquark separations $z_3=0, 4a$ and $8a$
    respectively. 
}
\eefs{extrapoltwostate}

In \fgn{disp}, we show the dispersion relation for the ground state
and the excited state. For the ground state, we have shown the
consistency between the results for $E_0(P_3)$ from the two-state
fits with those from the one-state fits. The black curve is the
expected continuum single particle dispersion $E_0=\sqrt{M_N^2 +
P_3^2}$ with $M_N = 1.115$ GeV.  The fitted data for $E_0(P_3)$
agrees with the continuum dispersion over the entire range of $P_3$,
with only a slight tendency for the central values of $E_0(P_3)$
to be smaller than the continuum values at the largest three momenta,
which could be an effect of a small lattice correction, specifically an $\mathcal{O}\left(a^2 P_z^2\right)$ error. 
We have
also shown the dispersion of the first excited state as the blue
triangles. At $P_3=0$, the gap $E_1-E_0 = 1.3(2)$ GeV is larger
than the expectation that the leading excitation are $N\pi\pi$
multi-particle state, for which the gap is about 0.7 GeV. This
suggests that the first excited state from our two-state fits only
effectively captures the tower of excited states above the ground-state
nucleon.

Using the spectral content data from the two-state analysis of the
nucleon two-point function, we performed the extrapolation of the
real and imaginary parts of $R(t_s,\tau)$ using two-state fits to
obtain ${\cal M}(z_3,P_3)$. For the two-states, there are four
independent parameters (i.e., the matrix elements) as the fit
parameters for each of the real and imaginary parts of $R$. For the
fits, we skipped the shortest $t_s=4a$ and used only $t_s \in [6a,
14a]$, and for each $t_s$, we used only the operator insertion time
values $2a \le \tau \le t_s - 2a$ to reduce any end-point effects.
Thus, the number of data points being fit is 35 for the choice of fit
range using 4 parameters, albeit with correlated data points and
with larger $t_s > 10a$ being noisy for the largest two momenta
effectively reduces the number of data points being
fit. In our fits, we included the correlations between the data points at a 
given $t_s$ and also the cross-correlations at different $t_s$. 
We found the correlated $\chi^2/{\rm dof}$ to vary in the
acceptable range around 1 for all the cases studied here.  In
\fgn{extrapoltwostate}, we show some sample two-state fits to ${\rm
Re\ }R(t_s,\tau; z_3, P_3)$ at $P_3=0, 0.82$ and 2.05 GeV, and for
$z_3=0, 4a$ and $8a$. In addition to the data for $R$ and the bands
resulting from the two-state fits, we also show the extrapolated
value for ${\cal M}(z_3,P_3)$ as the grey band in the different
panels.  From the figure, it is clear that at the lower momenta
where the data at all $t_s$ are well-determined, the two-state
extrapolation describes the ratio data well. For the intermediate
momenta around 0.82 GeV, the data for $t_s > 10a$ become noisy and
do not contribute to the fits. For the largest two momenta, as seen
in the example $P_3=2.05$ GeV data shown in the figure, the fits
are constrained mainly by the $t_s=6a$ and $8a$ source-sink
separations. 

\bef
\centering
\includegraphics[scale=0.8]{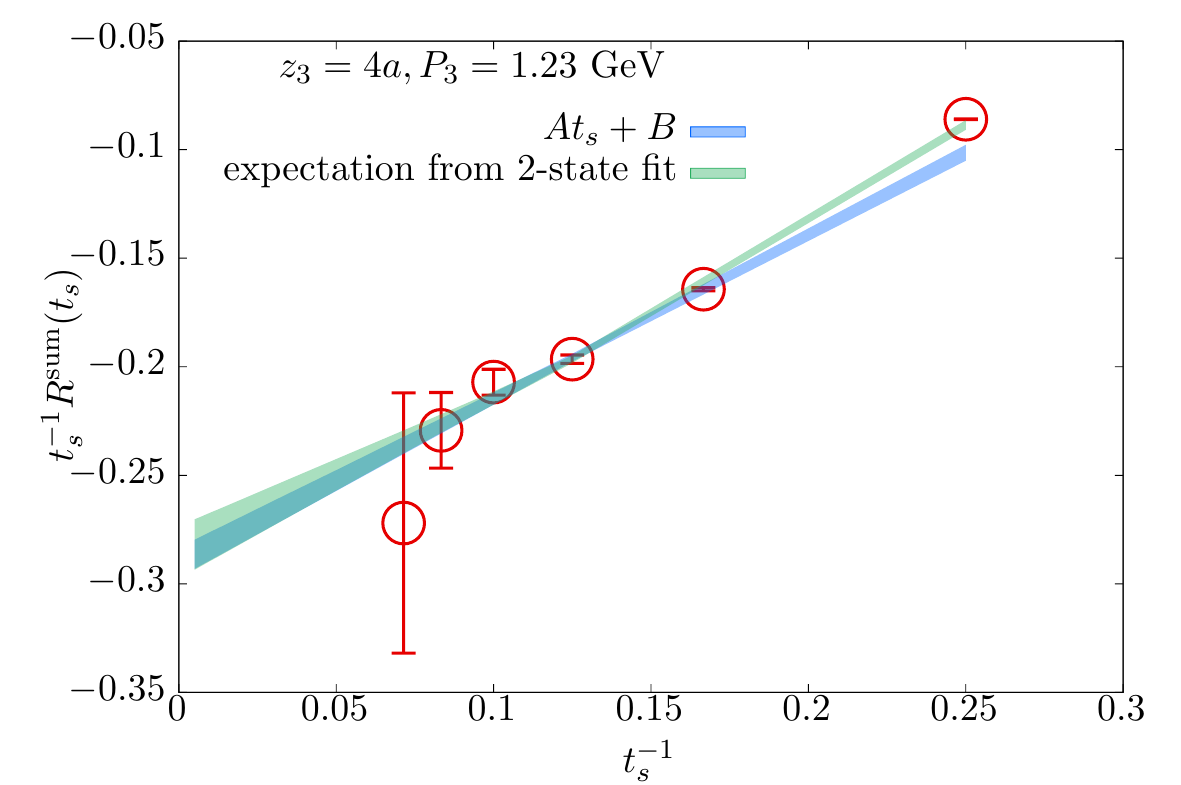}
\caption{
Extraction of the ground state matrix element by using the summation
method via fits to $R^{\rm sum}(t_s)$.
The plot shows $t_s^{-1} R^{\rm sum}(t_s)$
as a function of $t_s^{-1}$, defined conveniently such that the
$y$-intercept at $t_s\to\infty$ is the ground-state matrix element, ${\cal M}$.
The data points for $R^{\rm sum}(t_s)$ at a fixed
$P_3=1.23$ GeV and $z_3=4a$ are shown as the red circles.  The resulting
curve from the straight-line fit to $R^{\rm sum}(t_s)={\cal M} t_s
+ B$ over a range $t_s\in [6a,16a]$ is shown as the blue band. The
expectation for $R^{\rm sum}(t_s)$ from the two-state fits to the ratio
$R(t_s,\tau)$ over a range $t_s\in[6a,16a]$ is shown as the green
band.
}
\eef{sumfit}

We performed further consistency check on our two-state extrapolations
by using summation method to determine ${\cal M}(z_3,P_3)$. For
this, we fitted the straight-line in \eqn{sumspec} to the $t_s$
dependence of the lattice data for $R^{\rm sum}(t_s;z_3,P_3)$. We
used $\tau_0 = 2a$ to skip the end-points to find $R^{\rm sum}$,
but changing its value was not crucial. We did the straight-line
fits over the range of $t_s \in [6a,14a]$; the deterioration of
signal for $R^{\rm sum}(t_s)$ at larger $t_s$ with increasing $P_3$
followed the same trend as we explained above for the ratio $R$.
In \fgn{sumfit}, we have shown a sample straight-line fit to $R^{\rm
sum}(t_s)$ at $P_3=1.23$ GeV and $z_3=4a$. The $y$-axis in \fgn{sumfit}
is $t_s^{-1} R^{\rm sum}(t_s)$ and the $x$-axis is $t_s^{-1}$, such
that when $t_s^{-1}=0$, the $y$-intercept will give the value of
ground-state matrix element. The blue-band is the result from the
straight-line fit over $t_s\in[6a,14a]$, which passes through all
the data points satisfactorily, and not surprisingly, misses the
data point at the smallest $t_s=4a$ which did not enter the fit.
For comparison, we also show the expectation for $t_s^{-1} R^{\rm
sum}(t_s)$ from the two-state fits to the ratio $R(t_s,\tau)$ over
$t_s\in[6a,14a]$, that we discussed previously, as the green band.
The two estimates for ${\cal M}$ are consistent within error-bars
as seen from the $y$-intercepts of the two bands, validating the
extrapolations at least for the specific $(z_3,P_3)$ shown in the
figure. However, the surprising feature in \fgn{sumfit} (and also
for other $(z_3,P_3)$ as well), is that the expected curve for
$R^{\rm sum}(t_s)$ from the two-state fit always passes through the
$t_s=4a$ data point as well, unlike the summation fit curve. This
seems to suggest that the two-state fit has a slight advantage from
the sensitivity to the tower of higher excited states captured
through the effective first excited state $E_1$. We attempted to
test the robustness of summation fits by supplementing the straight-line
fit form with a term proportional to $e^{-(E_1-E_0)t_s}$, but it
however resulted in unstable fits with large errors in the fit
parameters. In the different panels of \fgn{mecompare}, we show the
results of ${\cal M}(z_3,P_3)$ as a function of $z_3$ that were
obtained from the two-state fit extrapolations (shown using circles)
and the summation fit extrapolations (shown using squares), at
different $P_3$. In each panel, ${\rm Re}{\cal M}$ and ${\rm Im}{\cal
M}$ are shown using red and blue symbols respectively. The comparison
nicely demonstrates the consistency between the two different ways
of extrapolations to get ${\cal M}$, thereby indirectly, justifying
a good estimation of the ground state matrix element. Therefore,
we will use the bare matrix element
obtained from the two-state fit in the rest of the paper, due to its usage of more data points
in its fits, especially at the larger $P_3$ where the summation fit
essentially uses only two data points, as well as due to its good
ability to describe even the smaller $t_s$ that did not even enter
the fits.

\befs
\centering
\includegraphics[scale=0.9]{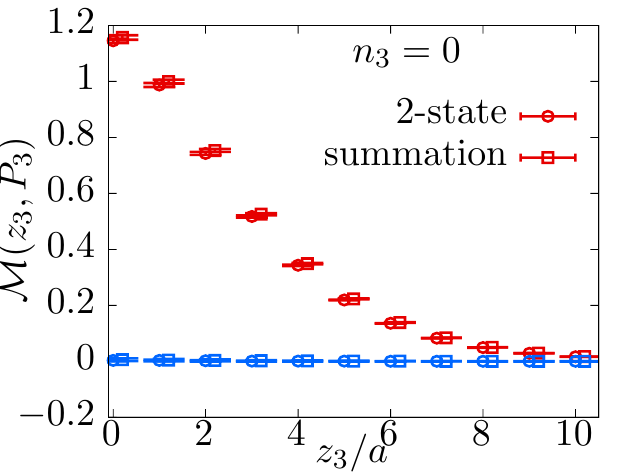}
\includegraphics[scale=0.9]{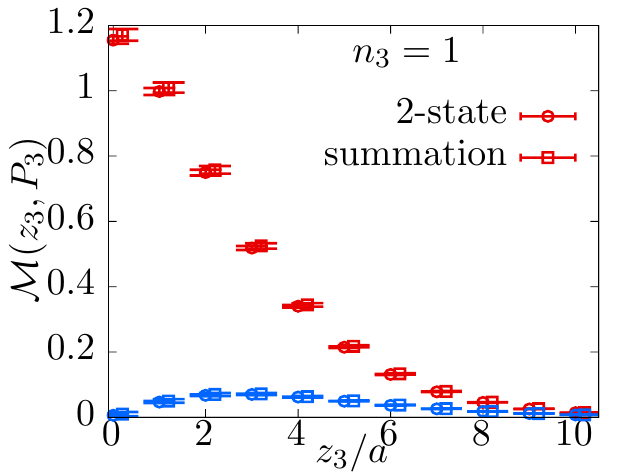}
\includegraphics[scale=0.9]{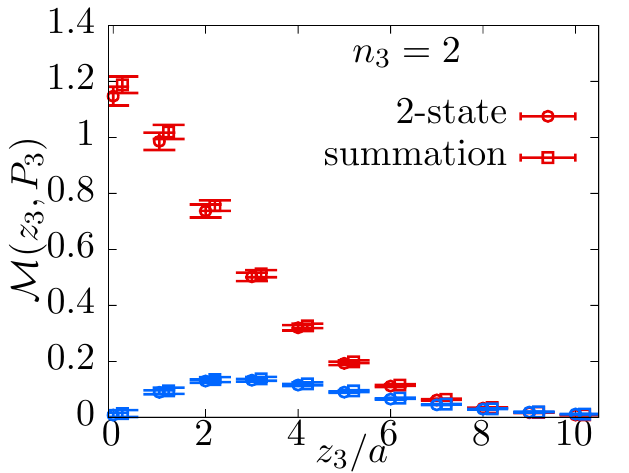}

\includegraphics[scale=0.9]{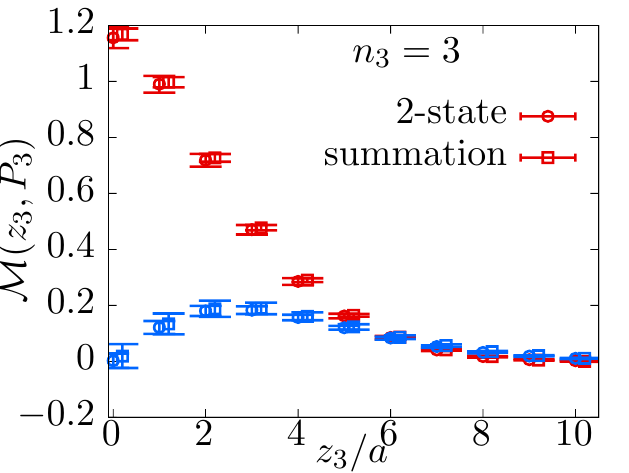}
\includegraphics[scale=0.9]{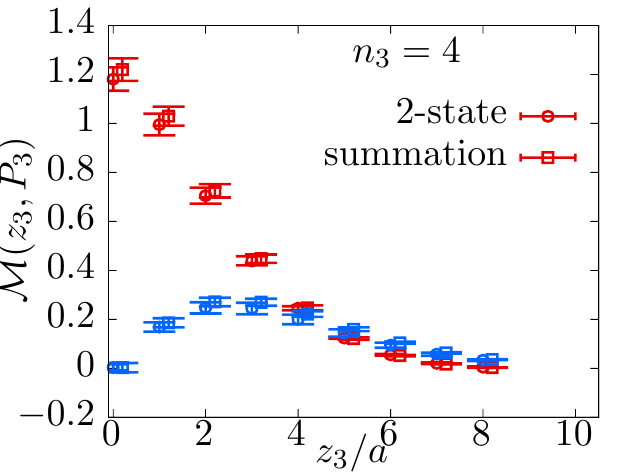}
\includegraphics[scale=0.9]{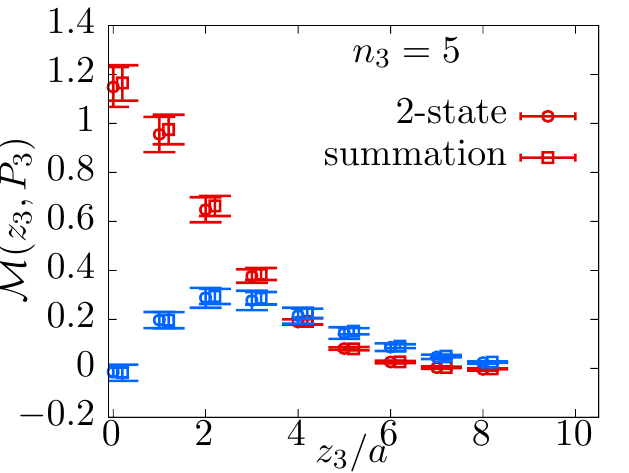}
\caption{
The plot demonstrates the consistency in the extracted bare
matrix elements ${\cal M}(z_3,P_3)$ by means of comparison between
the extrapolated values from two-states fits to $R(t_s,\tau)$
(circles) and straight-line fits to $R^{\rm sum}(t_s)$ (squares).  The red and blue symbols are for ${\rm Re\ } {\cal M}$ and ${\rm Im\ } {\cal M}$ respectively. The different
panels show this comparison at various momenta $P_3= 0.41 n_3$ GeV for $n_3=0,1,2,3,4,5$.
}
\eefs{mecompare}

\section{A numerical analysis of corrections to continuum leading-twist formalism}
\label{sec:opewoope}

The simplest analysis of the lattice pseudo-ITD data, without
incorporating any ansatz for the PDF is to use the Mellin moments as the fit parameters, as 
first introduced in Ref.~\cite{Karpie:2018zaz}. The premise
of the calculation is to find the best fit values of the Mellin
moments by fitting the Ioffe-time, $\nu$, dependence of
the real and imaginary parts of $\mathfrak{M}(\nu, z_3^2)$ using
the leading-twist OPE given in \eqn{tw2ope} at various fixed values
of $z_3$. In this way, we can obtain the Mellin moments $\langle
x^n\rangle_{\pm}$ as a function of $z_3$. If the leading-twist
OPE at a given perturbative order by itself is sufficient to describe
the lattice data in a given range of $z_3$ and $\nu$, then we should
find no $z_3$-dependence in the fitted values of $\langle
x^n\rangle_{\pm}$. By turning the argument around, by
assuming that the NLO leading-twist OPE is sufficient except that
it needs to be supplemented by small additional $\nu$ and $z_3$
dependent lattice corrections as well as higher-twist corrections,
then the moments analysis at fixed $z_3$ is a nice way to query the nature of these
small corrections. The idea is the following --- 
if the lattice pseudo-ITD data is an admixture of the 
leading-twist part  $\mathfrak{M}^{\rm twist-2}$
and some leading corrections in $1/|z_3|$ and $|z_3|$, such as,
\beq
\mathfrak{M}(\nu,z_3^2) = \mathfrak{M}^{\rm twist-2}(\nu,z_3^2) +
\sum_{k,n} \left(L_{k,n} \left(\frac{a}{|z_3|}\right)^k + H_{k,n} \left(\Lambda^2_{\rm QCD} z_3^2\right)^k \right)\frac{(i\nu)^n}{n!},
\eeq{tw2correc}
for some numerical coefficients $L_{k,n}$ and $H_{p,n}$,
then we can absorb the corrections into the leading-twist OPE, which effectively results in
a $z_3$-dependent $n$-th moment given by,
\beqa
a^{\rm eff}_{n+1}(z_3) =
\langle x^n \rangle + \frac{1}{C_n(\mu^2 z_3^2)}\sum_{k} \left(L_{k,n} \left(\frac{a}{|z_3|}\right)^k + H_{k,n} \left(\Lambda^2_{\rm QCD} z_3^2\right)^k \right).
\eeqa{effectivemom}
In practice, since the Wilson coefficients depend on $z_3$
logarithmically, one will see some power-law corrections in $1/|z_3|$
and $|z_3|$ to the moments extracted from OPE-without-OPE analysis,
thereby allowing us to deduce what the leading corrections are from
the lattice data itself. Also, only corrections with $n>0$ can appear
as $a_1=1$ by construction. Such an approach was also considered
previously in~\cite{Gao:2020ito} to deduce the nature of lattice corrections
for $z_3\sim {\cal O}(a)$.  Here, we take a similar stance and ask
whether there are corrections to the leading-twist OPE as seen in
the Mellin moments, and if so, what is the simplest correction that
we need to add to the leading-twist OPE in order to extract the
PDF?

At any given $z_3$, we only have six data points from the different
$P_3$. Therefore, we needed to truncate the leading-twist OPE in
\eqn{tw2ope} at modest values of $N_{\rm max}$ for this
analysis of moments; we used $N_{\rm max}=2,3,4$ and checked for the convergence
of the results.  In the top panels of \fgn{opewope}, we show the
results from the moments analysis using $N_{\rm max}=4$ truncation.
We used $\mu=\sqrt{2}$ GeV to do the matching and used $\alpha_s(\sqrt{2}
{\rm GeV})=0.36$ in the Wilson coefficients.  The data points in
the top-left and top-right panels are our data for ${\rm Im}\mathfrak{M}$
and ${\rm Re}\mathfrak{M}$ respectively. We have differentiated the
data at fixed values of $z_3$ ranging from $2a$ to $7a$ using
different colored symbols.  Along with the data points, we show the
resulting bands from the fits at each values of $z_3$. The fits
indeed describe the Ioffe-time dependence as well as the $z_3$
dependencies of the lattice data well, albeit at the expense of
allowing for  $z_3$ dependence of the moments as seen in the two
bottom panels.

\befs
\centering
\includegraphics[scale=0.99]{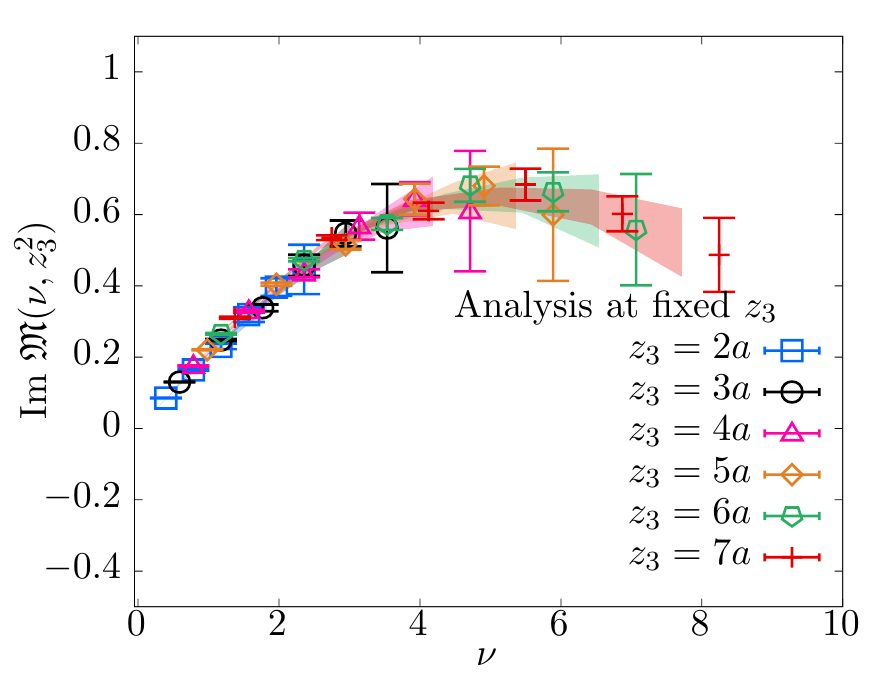}
\includegraphics[scale=0.99]{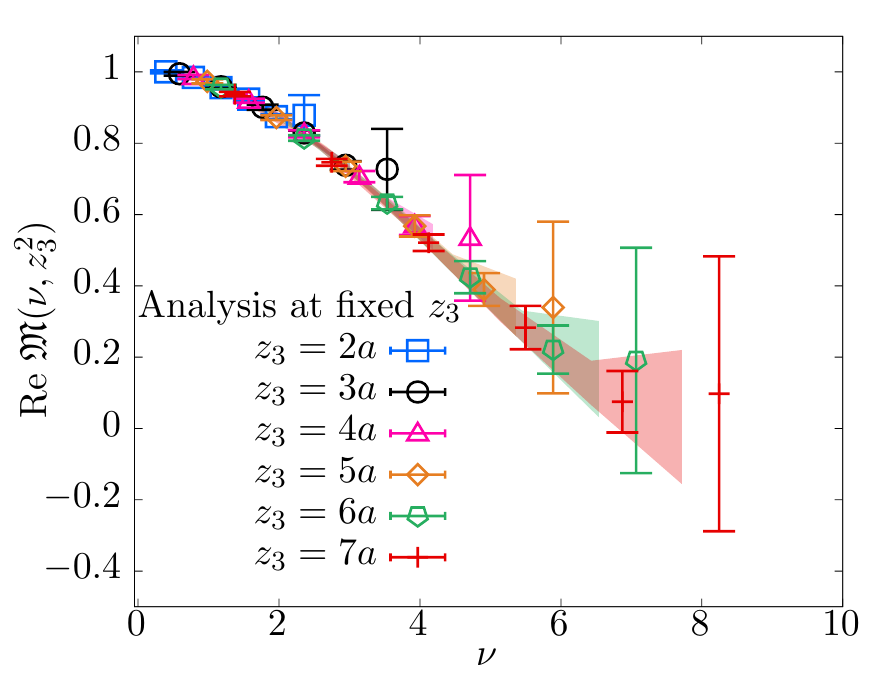}

\includegraphics[scale=0.99]{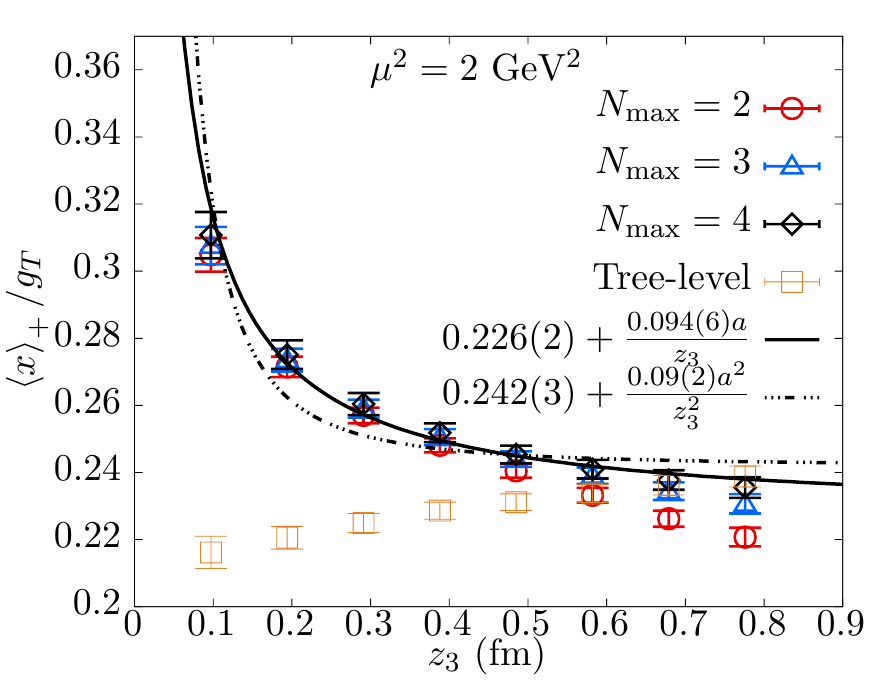}
\includegraphics[scale=0.99]{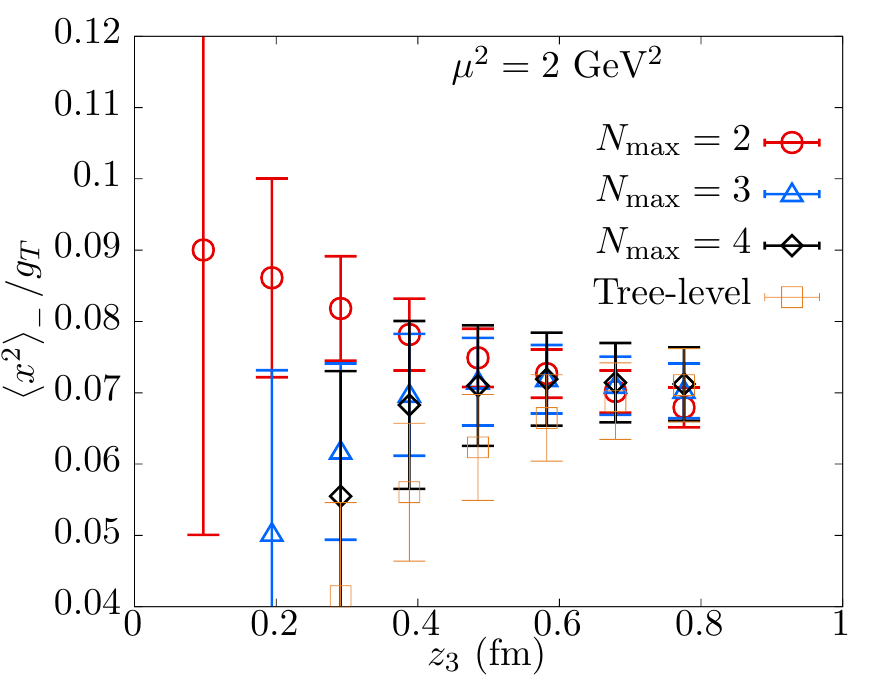}
\caption{
The analysis of $\nu$ dependence at different fixed $z_3$.  The
top-left and top-right panels show ${\rm Im}\mathfrak{M}$ and ${\rm
Re}\mathfrak{M}$ as a function of the Ioffe-time, $\nu$. The data
points for $\mathfrak{M}(\nu,z_3^2)$ at different fixed $z_3$ are
differentiated by the colored symbols. The corresponding fits to
the leading-twist OPE with $N_{\rm max}$  Mellin moments as the fit
parameters, at various fixed values of $z_3^2$ are the colored
bands. In the fits shown in the top panels, $N_{\rm max}=4$ Mellin moments were used.
The bottom-left panel shows the corresponding $z_3$-dependence of
$\langle x\rangle_+(\mu)/g_T(\mu)$ as obtained from fits to ${\rm
Im}\mathfrak{M}(\nu,z^2)$ at the different fixed $z_3^2$. The results
using the truncation order $N_{\rm max}=2,3,4$ are shown.  The black
curve is the expectation for the observed residual $z_3$ dependence
based on a short-distance lattice artifact of the type
$L_{1,1}(a/|z_3|)\nu$ (refer text).  The results from fits
using tree-level Wilson coefficients (i.e., $C_n=1$) and $N_{\rm
max}=4$ are also shown to see the effect of 1-loop matching.  A
similar $z_3$-dependence of $\langle x^2\rangle_{-}/g_T$ at
$\mu=\sqrt{2}$ GeV as obtained from the leading-twist OPE fits to
${\rm Re}\mathfrak{M}(\nu,z_3^2)$ is shown in the bottom-right
panel. }
\eefs{opewope}

The bottom-left and right panels of \fgn{opewope} show the $z_3$ dependencies of the
dominant fit parameters in the OPE of ${\rm Re}\mathfrak{M}$ and
${\rm Im}\mathfrak{M}$, namely, the normalized Mellin moments
$\langle x\rangle_{+}/g_T$ and $\langle x^2 \rangle_{-}/g_T$.  Let
us first focus on the bottom-left panel in  \fgn{opewope} --- the results from
the analyses of ${\rm Im}\mathfrak{M}$ using $N_{\rm max}=2,3,4$ are
the different colored symbols in the plot.  We can infer that by
$N_{\rm max}=4$, the fits have more or less converged. The $z_3$
dependence of $\langle x\rangle_{+}/g_T$ is striking, without any
region in the perturbative range of $z_3$ that can be identified
as a plateau. Thus, it is important to take care of the corrections
to the leading twist framework. Since we have analyzed only one
ensemble, we have to rely on previous works to deduce the origin of
the corrections seen here.  We observe that the corrections are
larger at shorter $z_3$, and hence, suggests that the dominant source
of the correction could be due to the lattice corrections when $z_3$
is comparable to the lattice cut-off itself.  Indeed, a similar
observation has been made in previous
works~\cite{Gao:2020ito,Karpie:2021pap} that used more than one
lattice spacing. Therefore, in this work, we will proceed  under the
hypothesis that the leading correction is a lattice spacing correction
of the type $ L_{k,m} \left(a/|z_3|\right)^k (i\nu)^{m}$ that we
discussed above. The solid black curve in the bottom-left panel is
a fit using the form $a_2^{\rm eff}(z_3) = \langle x\rangle_{+}/g_T
+ L_{1,1} \left(a/|z_3|\right)$, with $ \langle x\rangle_{+}/g_T =
0.226(2)$ and $L_{1,1}=0.094(6)$.  On the other hand, a fit to an
alternate correction of the type $L_{2,1}\left(a/|z_3|\right)^2$
performs poorly as seen from the dashed curve shown in the bottom-left
panel of~\fgn{opewope}.  Thus, we infer that the leading correction
to ${\rm Im}\mathfrak{M}$ is a correction of the form $L_{1,1}
\left( a/|z_3|\right) \nu$.  In addition to the lattice correction,
we do not find any perceptible higher twist corrections of the form
$(\Lambda_{\rm QCD} |z_3|)^2 \nu$ present in our data for ${\rm
Im}\mathfrak{M}$ up to $z_3 = 0.8$ fm, indicating that most of the
higher twist effects have presumably canceled between the bare
matrix elements ${\cal M}(z_3,P_3)$ and ${\cal M}(z_3,0)$ in their
ratio.  A similar plot of the effective $a_3^{\rm eff}(z_3)$ as
extracted from ${\rm Re}\mathfrak{M}$ is shown on the bottom right
panel. Unlike the results on the bottom-left panel, the fitted
values of $a_3^{\rm eff}(z_3)$ are comparatively noisier, especially
at the shorter $z_3<0.4$ fm. For $z_3 > 0.4$ fm up to 0.8 fm, a
plateau is seen.  Thus, to the precision of the data, we found no
indications of small-distance lattice correction nor any higher-twist
corrections in ${\rm Re}\mathfrak{M}$. To see the effect of DGLAP
as enshrined in the $\ln(\mu^2 z_3^2)$ in the NLO Wilson coefficients,
we also performed the above analysis using the tree-level matching
as obtained using $\alpha_s=0$ and therefore lacks the logarithmic
part as well as some finite $\alpha_s$ corrections (the resulting
tree-level moments can also be inferred as the moments of the
pseudo-ITD). From the bottom panels, we see that the effect of
1-loop is quite important for the ${\rm Im}\mathfrak{M}$ compared
to ${\rm Re}\mathfrak{M}$. From the $z_3$ behavior for $\langle
x\rangle_+$, we see that the effect of DGLAP and the effect of the
$a/z_3$ lattice correction have opposing behaviors, and taking care
of the them together is important in lattice studies at finite
lattice spacings.

Based on the above analysis, the explicit functional forms for the
leading-twist OPE along with the simplest leading lattice-spacing
correction and higher-twist correction, that we will use in the
extraction of the $x$-dependent PDF in the remaining part of the
paper is
\beq
{\rm Re}\left( \mathfrak{M}\left(\nu, z_3^2\right)\right) = \left( 1 + \sum_{n=1}^{N_{\rm max}} C_{2n}\left(z_3^2 \mu^2\right) \frac{(-1)^n \nu^{2n}}{(2n)!}\int_0^1 x^{2n} \frac{h_{-}(x,\mu)}{g_T(\mu)} dx \right) + L_{1,2} \frac{a}{|z_3|} \frac{\nu^2}{2} + H_{1,2} \left(\Lambda_{\rm QCD} z_3\right)^2 \frac{\nu^2}{2},
\eeq{refitform}
and for the imaginary part is,
\beq
{\rm Im}\left( \mathfrak{M}\left(\nu, z_3^2\right)\right) =
\left(\sum_{n=1}^{N_{\rm max}} C_{2n-1}\left(z_3^2 \mu^2\right) \frac{(-1)^{n-1} \nu^{2n-1}}{(2n-1)!}\int_0^1 x^{2n-1} \frac{h_{+}(x,\mu)}{g_T(\mu)} dx\right) +L_{1,1} \frac{a}{|z_3|} \nu + H_{1,1} \left(\Lambda_{\rm QCD} z_3\right)^2 \nu,
\eeq{imfitform}
with the terms within the larger parentheses in the above expression are simply the 
convolution term in \eqn{transmatching} expanded in $\nu$ for convenience in implementation.
We will use a value ${\Lambda_{\rm QCD}}=0.286$ GeV as a typical scale 
simply to get dimensionless
values above.
We found actual evidence in the data only for a non-zero $L_{1,1}$
in the imaginary part, and whereas, we have added the other correction
terms, namely $L_{1,2}$, $H_{1,1}$ and $H_{1,2}$, in order to be
conservative in our fits and also because there is no {\sl a priori}
reason for the absence of such leading lattice correction terms or
the higher-twist correction terms. In the end, we found such terms
to come out with values close to zero, which we will take as an
empirical fact. We should also note that the lattice correction
term $L_{1,2}$ in the real part is proportional to the modulus
$|z_3|^{-1}$. This is in contrast to Ref.~\cite{Gao:2020ito} for the analysis
of pion valence PDF, where an analytic correction term $(a/z_3)^2\nu^2$
was used for ${\rm Re}\mathfrak{M}$ due to visible evidence for
such a term in the data. In our case, there is no such visible
evidence for ${\rm Re}\mathfrak{M}$ data, and therefore, we add a
term with lesser power of $|z_3|^{-1}$, which in principle could
be present.  Such an approach was also taken for the case of the
analysis for the nucleon unpolarized PDF~\cite{Karpie:2021pap,Egerer:2021ymv}.  We also cross-checked
by adding $(a/z_3)^2 \nu^2$ as a correction term instead of $(a/|z_3|)
\nu^2$ term to the real part in our studies, but it did not make
any statistically significant variation.  Therefore, we will present
the results with $L_{1,2}$ term in the real part.

We should note that one may parameterize the $\nu$
dependence of both the higher-twist as well as the lattice spacing errors with a more general form. 
In computations of the nucleon unpolarized PDFs~\cite{Karpie:2021pap,Egerer:2021ymv,HadStruc:2021wmh},  
Jacobi polynomials were used to describe the general $\nu$ dependence of correction terms. 
However, for the transversity rpITD data presented here, we found the results from an 
analysis using the Jacobi polynomial parametrization of corrections to leading-twist OPE 
to be consistent with the results using a simpler parametrization using leading correction terms
given above, and hence, we resort to this simpler parametrization in the rest of the paper.
Perhaps, an increased precision in the future might necessitate
more elaborate terms in the corrections. 

\section{Reconstruction of transversity PDF with reduced model-dependence}
\label{sec:results}

Having set up the required elements for the PDF analysis, we present
the results on the extraction of the PDF from the transversity
pseudo-ITD in this section.  Our approach is to reconstruct the
$x$-dependent transversity PDFs, $h_{\pm}(x)$ by assuming a
functional form for them, say $h_{\pm}(x; \{\alpha,\ldots\})$,
and then perform a combined fit of the parameters $\{\alpha,\ldots\}$
to the $\nu$ and $z_3$ dependencies of the pseudo-ITD lattice data
over a range of $P_3\in [ P_3^{\rm min}, P_3^{\rm max}]$ and $z_3
\in [z_3^{\rm min}, z_3^{\rm max}]$ using \eqn{refitform} and
\eqn{imfitform}. As we will explain, through a step by step
generalization of the functional form of the PDF ansatz, we reduce
the model-dependence. Our fitting method is by using the standard
$\chi^2$ minimization,
\beqa
&&\chi^2 = \sum_{p,p'}\Delta_{p} \Sigma^{-1}_{p,p'} \Delta_{p'};\quad p = [z_3,P_3],\cr
&& z_3\in[z_3^{\rm min},z_3^{\rm max}],\quad P_3\in[P_3^{\rm min},P_3^{\rm max}],
\eeqa{chisq}
where $\Delta_{p}={\rm Re}/{\rm Im}\mathfrak{M}_{\rm data}(p) -
{\rm Re}/{\rm Im}\mathfrak{M}_{\rm fit}(p;\{\alpha,\ldots\})$, and
$\Sigma_{p,p'}$ is the covariance between the different data points
$p,p'$. The covariance matrix uses its standard definition using
only statistical fluctuations, without folding any of the systematic
errors into it. We will take care of the systematic variations in
the fits in the end.

We used all the six available values of $P_3 \in [0, 2.46]$ GeV in
our analysis. However, we were cautious about the range of $z_3$
to use; too small values of $z_3$ will suffer from larger lattice
spacing corrections as we discussed in the last section, whereas
for $z_3\sim O(1)$ fm, we naively expect higher-twist effects and
higher-order perturbative terms could become important.  For this,
we skipped $z_3=0,a$ from our analysis and used only ranges with
$z_3^{\rm min}=2a, 3a$.  To see the variations due to the choice
of $z_3^{\rm max}$, we used $z_3^{\rm max}=8a, 10a= 0.75, 0.94$ fm.
We used the fixed order expressions for the Wilson coefficients in
\eqn{wilcoeff} at a factorization scale of $\mu=\sqrt{2}$ GeV in
our PDF analysis, that is comparable to $1/z_3$ that enters our
computation.

\befs
\centering
\includegraphics[scale=0.99]{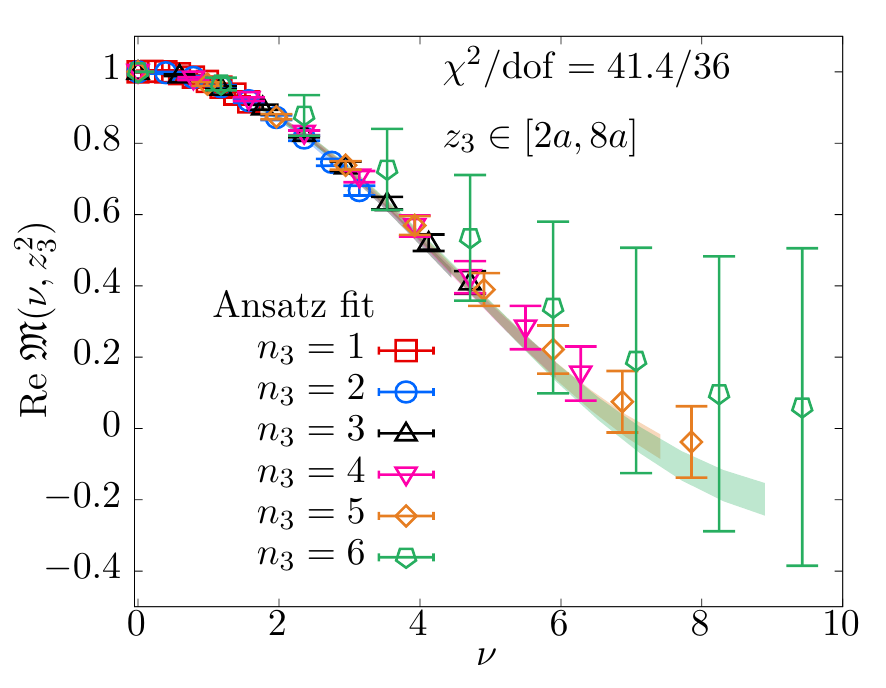}
\includegraphics[scale=0.99]{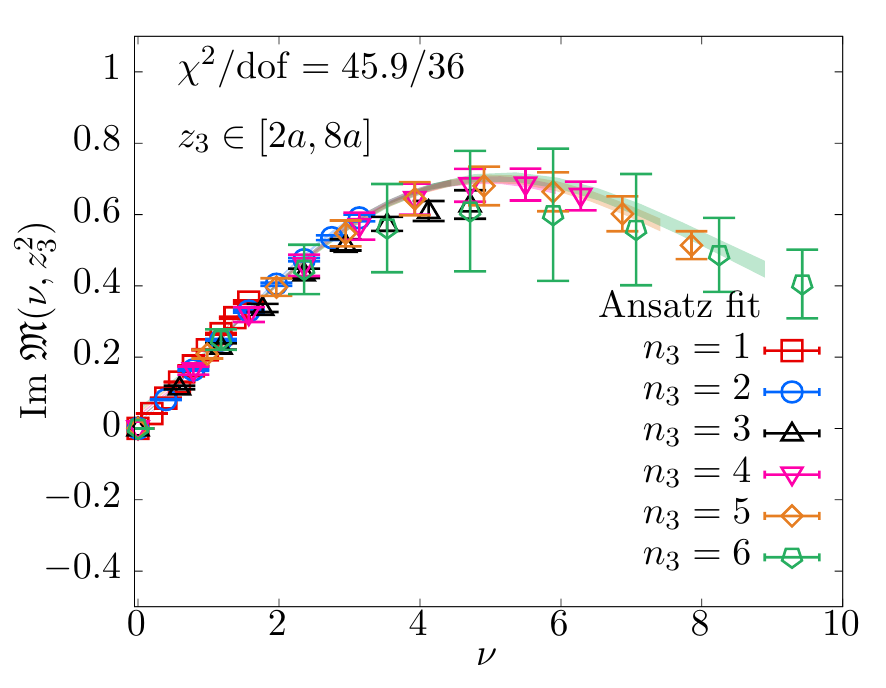}

\includegraphics[scale=0.99]{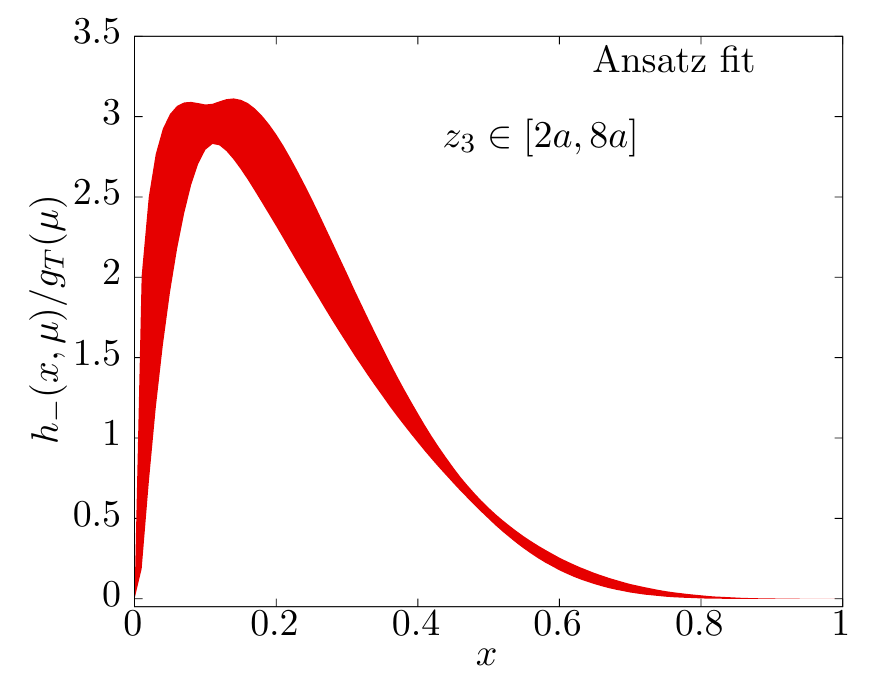}
\includegraphics[scale=0.99]{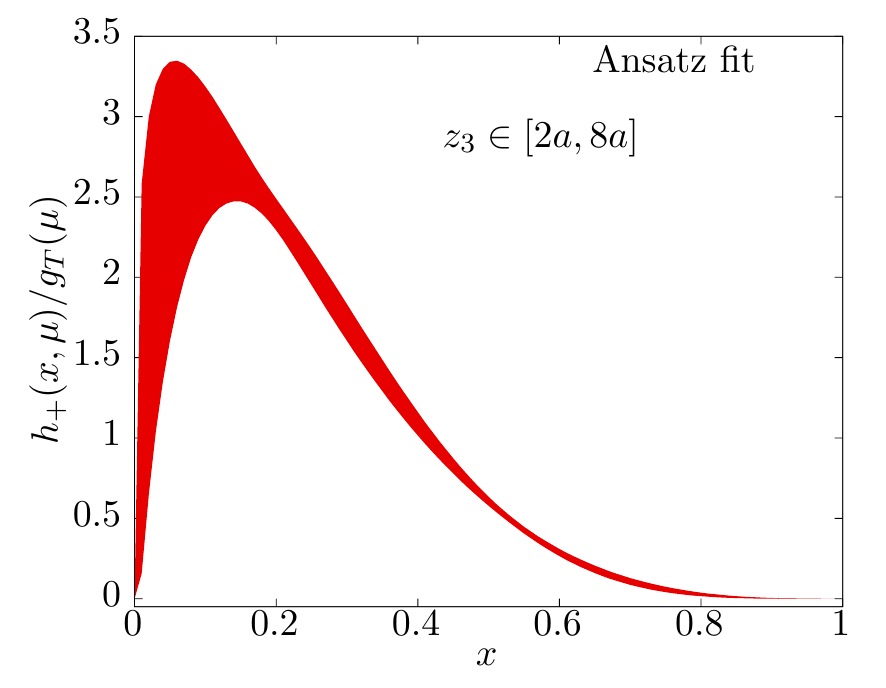}
\caption{ Reconstruction of transversity PDF based on the PDF ansatz
in \eqn{jamansatz}.  The top-left and top-right panels show the
real and imaginary parts of $\mathfrak{M}$ as a function of $\nu$.
The two panels show the best fit bands resulting from an analysis
assuming the PDF ansatz.  The fits shown in the figure incorporated
the data points at all momenta with $z_3\in[2a,8a]$. The color of
the bands and the data points distinguish the fixed value of momenta
$P_3=0.41 n_3$ GeV used.  The bottom-left and bottom-right panels
show the resultant transversity PDFs, $h_{-}(x)$ and  $h_{+}(x)$
respectively.}
\eefs{jamfit}

In the first step of the PDF reconstruction, we assumed a
functional form that is known to work well in the global fits to the PDFs from
experimental cross-sections data, namely,
\beq
\frac{h_{\pm}(x)}{g_T} = N_\pm x^{\alpha_\pm}(1-x)^{\beta_\pm}\left(1+\gamma_\pm \sqrt{x} + \delta_\pm x\right),
\eeq{jamansatz}
with $(\alpha_\pm, \beta_\pm, \gamma_\pm, \delta_\pm)$ as independent
fit parameters. The parameter $N_\pm$ is the normalizing constant.
We will simply refer this method as {\sl PDF ansatz fits}.  For the
valence case, $\int_0^1 dx \frac{h_{-}(x)}{g_T}=1$, which thereby
fixes $N_-=N_-(\alpha_-,\beta_-,\gamma_-,\delta_-)$ as a function
of the other independent parameters.  On the other hand, for $N_+$ there
is no such condition and therefore, we keep it as an additional fit
parameter in $h_{+}/g_T$.  We used the above functional form in
\eqn{refitform} and \eqn{imfitform} to fit our transversity
pseudo-ITD data. 
We evaluated the convolution integral for the leading-twist matching 
using the Taylor series in $\nu$ (see \eqn{refitform} and \eqn{imfitform}) using an expansion up to
order $N_{\rm max}=40$. This truncation achieves a machine precision 
approximation of the convolution kernel within the range of $\nu$ we use.

In the top left and right panels of \fgn{jamfit}, we compare our
PDF ansatz fit with the real and imaginary parts of our lattice
pseudo-ITD data in the left and right panels respectively. For the
fits shown in the two panels, we used the fitting ranges $z_3\in[2a,
8a (0.75 {\ \rm fm})]$ and $P_3 \in [0, 2.46 {\ \rm GeV}]$.  We
have represented the data points and fitted bands at a fixed $P_3$
by the same set of colors. The data points at different $P_3$ and
$z_3$ quite nicely fall on near universal curves as a function of
$\nu$, which means that the scaling violations to the
tree-level universality could be described by small perturbative
logarithmic terms. Indeed, it is clear from the two panels that the
corresponding fitted bands describe the data at different $P_3$
well over the range of $z_3$ we used.  Taking this range of fitted
data as a representative point for the sake of discussion, we found
the following set of parameters that enter the PDF ansatz:
\beqa
&& \alpha_+ = 0.49(42); \qquad \alpha_-=0.63(50)\cr
&& \beta_+ = 3.38(1.15); \qquad \beta_-= 4.37(1.75)\cr
&& \gamma_+ = -0.28(1.92); \qquad \gamma_- = -0.16(2.29)\cr
&& \delta_+ = -0.30(1.33); \qquad \delta_- = -0.17(1.75)\cr
&& N_+ = 10.85(92); \qquad  \cr
&& L_{1,1} = 0.0648(38) ; \qquad L_{1,2} = -0.038(20)\cr
&& H_{1,1} = -2.50(2.62)10^{-3}; \qquad H_{1,2} = 1.86(1.34)10^{-3}\cr
&& \chi^2/{\rm dof} = 45.9/35; \qquad \chi^2/{\rm dof} = 41.5/36\ .
\eeqa{paramsjam}
In the bottom-left and right panels of \fgn{jamfit}, we show the
corresponding best fit transversity PDFs, $h_{-}$ and $h_{+}$,
respectively for the representative values of fit ranges; in the
last half of this section, we will discuss more on the variability
of the fits as a systematic effect.  The quality of the fits are
acceptable as seen from the $\chi^2/{\rm dof}\approx 1.2$.  For
both $h_{\pm}$, a simpler two-parameter ansatz using only the
exponents $(\alpha_\pm,\beta_\pm)$ was also sufficient to capture
the shape of the transversity PDF, as one can see by the nearly
vanishing values of the small-$x$ corrections $\gamma_\pm$ and
$\delta_\pm$.  The role of the lattice correction $L_{1,1}$ in ${\rm
Im}\mathfrak{M}$ is not negligible as we discussed in the last
section, and such a term is necessary to obtain acceptable $\chi^2$.
Its real counterpart $L_{1,2}$ is comparatively smaller and consistent
with zero at 2-$\sigma$ level. The additive higher-twist corrections
$H_{1,1}$ and $H_{1,2}$ come out unimportant, and supports an
explanation that there are cancellations of higher-twist corrections
due to the ratio of nucleon matrix elements in \eqn{ritddef}.  The
fact that $\alpha_\pm > 0$ results in the transversity PDFs vanishing
at $x=0$ in the bottom panels.  The region of $x\in[0.1,0.8]$ where
the transversity PDF is significantly non-zero could perhaps help
their lattice determinations with lesser higher-twist contamination,
which is suggested~\cite{Braun:2018brg} to affect the $x\approx 0$ and $x\approx 1$ parts
of the extracted PDF.

At this point, we are concerned about the robustness of the
reconstructed transversity PDFs; 
by assuming a PDF ansatz, have we inadvertently restricted
the set of allowed PDFs severely and ruled out a wider possibility
of solutions? The answer to this question can only be found
by an actual inversion of the matching relation, and equivalently an
inversion of \eqn{refitform} and \eqn{imfitform}, to determine
$h_{\pm}(x)$ using a discrete set of data points that span
a finite range of $\nu$ and $z_3$.  This is well known to be an
ill-posed problem~\cite{Karpie:2019eiq}. 
Given the assumption (that is, our prior) that the
transversity PDF can be described using the PDF ansatz in
\eqn{jamansatz} to a good accuracy, and allowing for {\sl
small} fluctuations around this prior, we ask whether we can reconstruct the
transversity PDFs using a more flexible PDF parametrization that covers all such possible {\sl small} fluctuations.
We describe our method to answer this question
in our ensuing discussion on the reconstruction of the
transversity PDF using Jacobi polynomials that form a compete basis of functions of $x$ for $x\in[0,1]$.

The effectiveness of a Jacobi polynomial basis as an easy-to-implement
and complete set of functions for $x\in[0,1]$ was first investigated
in Ref.~\cite{Karpie:2021pap}.  The reader can refer to
Ref.~\cite{Egerer:2021ymv} for a more detailed description of a
related procedure as applied to the unpolarized PDF. The essential
properties of the Jacobi polynomials that we need for this paper
are as follows. Any pair of parameters $(\alpha,\beta)$ defines a
family of Jacobi polynomials, which we represent as $P_n^{\alpha,\beta}(u)$
for $u\in[-1,1]$ which are orthogonal with respect to a weight
function $W^{\alpha,\beta}(u)=(1-u)^\alpha (1+u)^\beta$.  We can
conveniently rewrite the polynomials as $P_n^{\alpha,\beta}(1-2x)$
which span the interval $x\in[0,1]$ that our PDFs are defined in,
and with the weight-function as $W^{\alpha,\beta}(x)=x^\alpha
(1-x)^\beta$. That is
\beq
\int_0^1 P_n^{\alpha,\beta}(1-2x) P_m^{\alpha,\beta}(1-2x) W^{\alpha,\beta}(x) dx = {\cal K}_n(\alpha,\beta) \delta_{m,n},
\eeq{ortho}
where ${\cal K}_n$ is a normalizing constant.  Due to this orthogonality
of the Jacobi polynomials, we can write the most general functional
form for our PDFs as,
\beqa
&& h_{\pm}\left(x; \{N_\pm,\alpha_\pm,\beta_\pm,s_{n\pm}\}\right) = N_\pm x^{\alpha_\pm} (1-x)^{\beta_\pm}\left(1 + \sum_{n=1}^{N_J} s_{n\pm} P^{\alpha_\pm,\beta_\pm}_n(1-2x)\right),\cr
&&\quad{\rm with}\quad s_n \equiv \frac{1}{{\cal K}_n(\alpha,\beta)} \int_0^1 h_{\pm}(x)  P^{\alpha_\pm,\beta_\pm}_n(1-2x) dx.\cr&&\quad
\eeqa{jacobiexpand}
The above expansion is exact for $N_J\to\infty$.  While it is
tempting to identify $(\alpha_\pm,\beta_\pm)$ with the small-$x$
and large-$x$ exponents due to the similarity of the above equation
with \eqn{jamansatz}, such an identification in general is not
correct --- for this, we note again that $(\alpha_\pm,\beta_\pm)$
can be any pair of real numbers, greater that -1, and due to the
completeness of the corresponding Jacobi polynomials $P_n^{\alpha,\beta}$,
the above expansion of $h_{\pm}(x)$ is always exact in the $N_J\to\infty$ limit. However, not
all choices of $(\alpha_\pm,\beta_\pm)$ are numerically optimal
when finite $N_J$ has to be used, as the above series in $n$ might
only slowly converge with $n$, or worse, it might not be uniformly
convergent as $n$ is increased unlike, for example, a series in the
Chebyshev polynomials. In Refs.~\cite{Karpie:2021pap,Egerer:2021ymv},
this convergence problem was approached by finding the best fit
values of $(\alpha_\pm,\beta_\pm)$ along with the coefficients
$s_{n\pm}$ by using the VarPro algorithm~\cite{doi:10.1137/0710036}.

\bef
\centering
\includegraphics[scale=0.9]{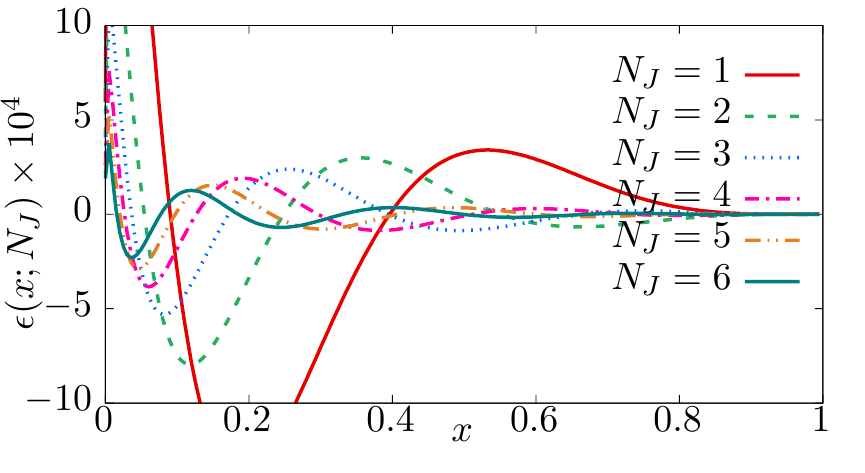}
\caption{Convergence of the Jacobi polynomial expansion for PDFs
that can be described by \eqn{jamansatz}. The error $\epsilon(x;N_J)$
due to truncation at order $N_J$ (see text) is plotted as a function
of $x$.  In the example shown,
$(\alpha,\beta,\gamma,\delta)=(0.49,3.38,-0.28, -0.3)$ in \eqn{jamansatz}
for $h_+(x)$.  The truncation order of the Jacobi polynomial expansion
is gradually increased from $N_J=1$ to 6 as seen from the different
curves.}
\eef{jacobiconverge}

In this work, we explore another possibility that makes full use
of the completeness of \eqn{jacobiexpand} and the empirically known
effectiveness of the PDF ansatz in \eqn{jamansatz}.  For
this, we specialize the above discussion from a generic $(\alpha,\beta)$
that define $P^{\alpha,\beta}_n$ to the case where we identify them
with the small-$x$ and large-$x$ exponents.  We generalize
\eqn{jamansatz} and assume that the PDF can be written as
\beq
h_{\pm}(x) = x^{\alpha_\pm}(1-x)^{\beta_\pm}{\cal G}_\pm(x), 
\eeq{jamtojacobi}
where $\alpha_\pm$ and $\beta_\pm$ are the actual small-$x$ and
large-$x$ exponents, in which case, it is justified to assume that
${\cal G}_\pm(x)$ is a slowly-varying function that can be expanded
linearly in $P^{\alpha_\pm,\beta_\pm}_n$ as
\beq
{\cal G}_\pm(x;N_J) = 1 + \sum_{n=1}^{N_J} s_{n\pm} P^{\alpha_\pm,\beta_\pm}_n(1-2x),
\eeq{gfunc}
with a good convergent behavior as the order of truncation $N_J$
in increased. In order to see if this is true, let us consider the
central values of $(\alpha_+, \beta_+, \gamma_+,\delta_+)=(0.49,3.38,-0.28,
-0.3)$ from the PDF ansatz in \eqn{jamansatz}.  
In this specific example, we would like to see if ${\cal
G}_+(x) = 1-0.28 \sqrt{x} -0.3 x$ exhibits a convergent behavior with
respect to
$n$, when it is expanded in the basis $P_n^{0.49,3.38}$. Let us define the
error committed by the truncation at $N_J$ polynomials, $\epsilon(x;N_J)
\equiv x^{\alpha_+}(1-x)^{\beta_+}\left({\cal G}_+(x;\infty) - {\cal
G}_+(x;N_J)\right)$.  In \fgn{jacobiconverge}, we show $\epsilon(x;N_J)$
as a function of $x$,
as $N_J$ is increased from 1 to 6 for the example ${\cal G}_+(x)$
considered. For the example shown and for similar
such four-parameter ansatz parameterizations of the PDF, we found the
convergence with $N_J$ was uniform over a range $x \in [x_{\rm
min},1]$ with $x_{\rm min}$ monotonically becoming smaller with
increasing $N_J$. Thus, to summarize our observational study of the
Jacobi polynomial expansion, at least for PDFs that closely resemble
the typical $x^\alpha (1-x)^\beta$ functional forms, we can consider
the Jacobi polynomial expansion of the PDF, using the same values
of $\alpha$ and $\beta$ as the small-$x$ and large-$x$ exponents,
to be uniformly convergent with $N_J$, and it is sufficient to
consider only the first few $P_n^{\alpha,\beta}$ in the expansion.

\befs
\centering
\includegraphics[scale=0.67]{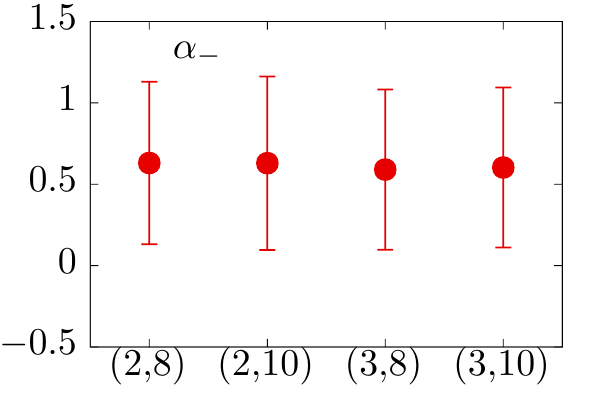}
\includegraphics[scale=0.67]{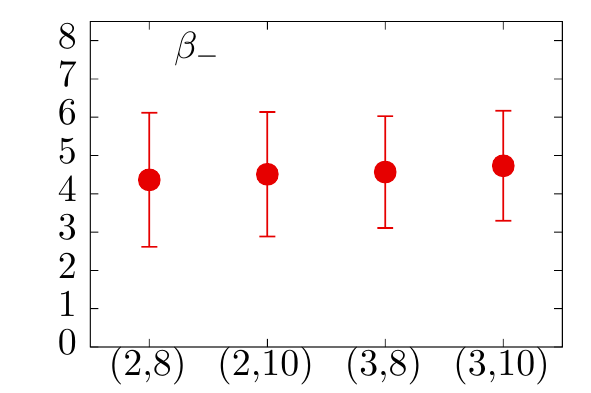}
\includegraphics[scale=0.67]{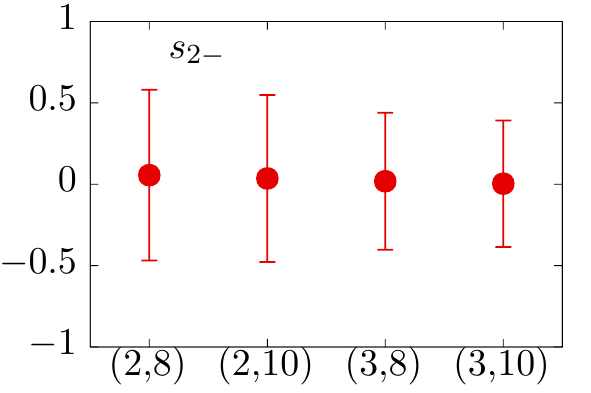}
\includegraphics[scale=0.67]{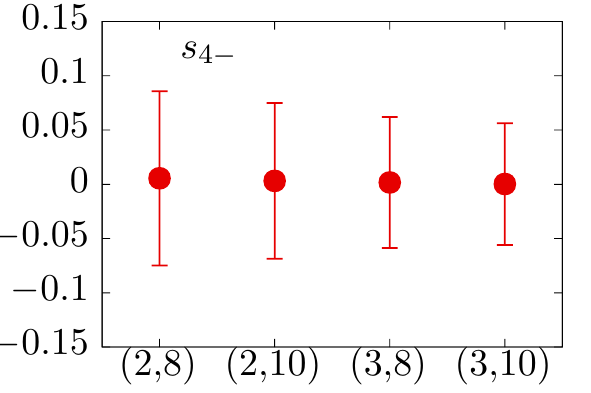}

\includegraphics[scale=0.67]{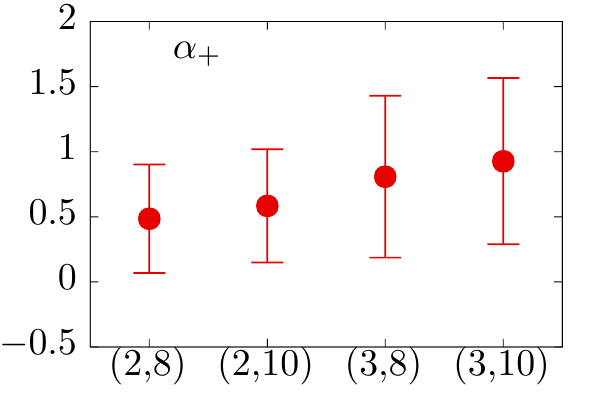}
\includegraphics[scale=0.67]{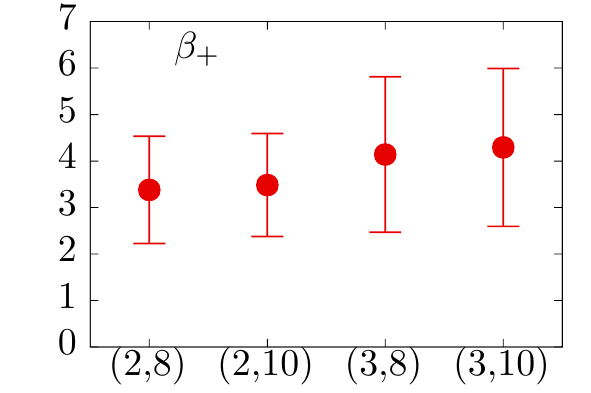}
\includegraphics[scale=0.67]{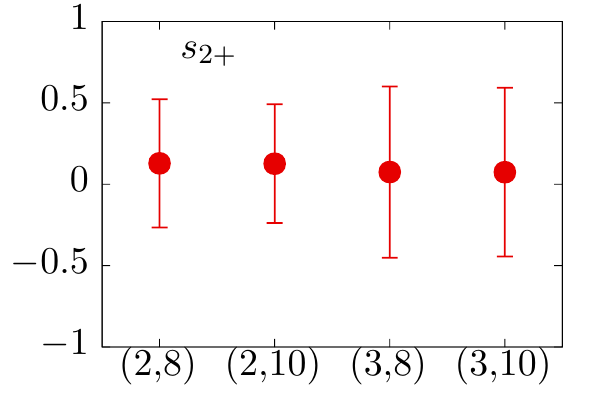}
\includegraphics[scale=0.67]{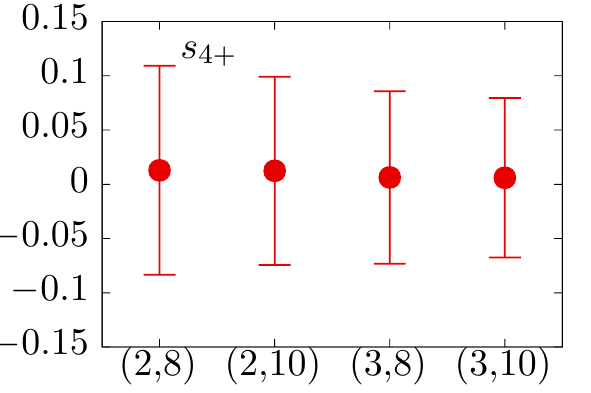}
\caption{
    The small-$x$ exponents, $\alpha_\pm$, and the large-$x$
    exponents, $\beta_\pm$, as inferred from the PDF ansatz fits
    are shown as a function of fit ranges $(z^{\rm min}_3, z^{\rm
    max}_3)$ in the left-half of the figure. The Jacobi polynomial
    expansion coefficients $s_{2\pm}, s_{4\pm}$ obtained from the
    decomposition of the PDF ansatz fits in a basis
    $P_n^{\alpha_\pm,\beta_\pm}$ are shown as a function of fit
    ranges in the right-half of the figure.  The central values and
    errors on the inferred expansion coefficients were then fed as
    prior and prior widths for the fits using the Jacobi polynomial
    parametrization (see text).
}
\eefs{jacprior}

Based on the above discussion, 
we improved upon our PDF ansatz reconstruction
in the following way. Let us denote the parameters and PDFs extracted from the 
PDF ansatz in \eqn{jamansatz} using ``ans" in the superscript in the discussion below.
\begin{enumerate}
    \item For each fit range $z_3\in[z_3^{\rm min},z_3^{\rm max}]$ and
	$P_3\in [P_3^{\rm min}, P_3^{\rm max}]$, we read off the
	small-$x$ and large-$x$ exponents, $(\alpha_\pm^{\rm ans},
	\beta_\pm^{\rm ans})$,  from the four-parameter ansatz
	reconstruction analysis we presented previously. We decomposed
	the four-parameter PDF that depends on $(\alpha_\pm^{\rm
	ans}, \beta_\pm^{\rm ans}, \gamma_\pm^{\rm ans}, \delta_\pm^{\rm
	ans})$ into a basis of Jacobi polynomials $P_n^{\alpha_\pm^{\rm
	ans},\beta_\pm^{\rm ans}}$ using \eqn{jacobiexpand}.
	The output of this decomposition were the expansion
	coefficients $s_{n\pm}^{\rm ans}$ for any order $n$.  By
	iterating this over jackknife samples of the four-parameter
	ansatz fits, we estimated the mean $\bar s_{n\pm}^{\rm ans}$
	and its error $\sigma_{s\pm}^{\rm ans}$ of the expansion
	coefficients.
    \item In the second step, with the same set of fit ranges as
        in the Step-1, we used the Jacobi polynomial expansion, \eqn{jacobiexpand},
	truncated at a chosen truncation order $N_J$ in \eqn{refitform}
	and \eqn{imfitform} with the expansion coefficient $s_{n\pm}$
	and the other correction parameters $L_{m,n}$, $H_{m,n}$
	as the fit parameters. One should note that the fit is
	linear in the expansion coefficients $s_{n\pm}$. We imposed
	our prior that the allowed PDFs are small fluctuations
	about the PDF ansatz fit, by using the log-likelihood
	function,
        \beq
        {\cal L} = \chi^2 + \sum_{n=1}^{N_J} \left(\frac{s_{n\pm}-\bar s^{\rm ans}_{n\pm}}{\sigma^{\rm prior}_{s_{n\pm}}}\right)^2,
        \eeq{likelihood}
	with $\chi^2$ defined in \eqn{chisq} and the second term
	is the negative logarithm of the Bayesian prior.  We took the central
	value of the prior from step-1 above.  The prior width,
	$\sigma_{s_{n\pm}}$, gives the handle to impose how {\sl
	small} the fluctuation around our prior ansatz based PDF
	can be. We chose a conservative, $\sigma^{\rm
	prior}_{s_{n\pm}}=3 \sigma_{s\pm}^{\rm ans}$ with
	$\sigma_{s\pm}^{\rm ans}$ taken from Step-1.  The sensitivity
	to $\sigma^{\rm prior}_{s_{n\pm}}$ was minimal as long as
        it was ${\cal O}(\sigma_{s\pm}^{\rm ans})$, 
        with even wider widths resulting in oscillatory,
	unphysical reconstructions of the PDF when $N_J$ was made
	larger than 4.  By minimizing ${\cal L}$, we obtained the
	maximum a posteriori estimates of $s_{n\pm}$ and their
	confidence intervals.  This step immediately resulted in
	the Jacobi polynomial based reconstruction of the transversity
	PDFs for a given specification of fit ranges for the lattice
	data. 
    We found the errors of $s_{n\pm}$ and the resulting PDF through
        a jackknife procedure.
      \item In the last step, we took care of the systematic error
      due to choices we made in the analysis steps-1 and -2 above,
      namely, the set
	  ${\cal R}_{\rm fit}$ of the fit choices uniquely labeled
	  by $\left(N_J, z_{\rm min}, z_{\rm max}, {\rm LC}, {\rm
	  HT}\right)$. We always made use of all six available
	  momenta in our analysis.  The term ${\rm LC}$ is Boolean
	  valued, denoting whether we included the lattice correction
	  term $L_{1,2}$ for ${\rm Re}\mathfrak{M}$ and $L_{1,1}$
	  for ${\rm Im}\mathfrak{M}$. Similarly, the Boolean term
	  ${\rm HT}$ denotes whether we added the terms $H_{1,1}$
	  and $H_{1,2}$ in the fits. We changed $z_{\rm min}$ from
	  $2a$ to $3a$, and changed $z_{\rm max}=8a$ to $10a$
	  corresponding to 0.75 fm to 0.94 fm. We successively
	  changed $N_J$ from 4 to 10 in our fits.  After collecting
	  together the analysis variations into the set ${\cal
	  R}_{\rm fit}$ per jackknife block, we used the Akaike
	  information criterion (AIC) model averaging to obtain a
	  single estimator $h^{\rm AIC}(x)$ per jackknife block,
	  and a single estimator $\Delta^{\rm AIC}(x)$ to capture
	  the systematic spread in PDFs per jackknife block:
          \beqa
          h^{\rm AIC}_{\pm}(x) &=&  \sum_{m \in {\rm R}_{\rm fit}} w^{(m)} h^{(m)}_{\pm}(x),\cr
          \Delta^{\rm AIC}_{\pm}(x) &=& \sqrt{\sum_{m \in {\rm R}_{\rm fit}} w^{(m)} \left( h^{(m)}_{\pm}(x) - h^{\rm AIC}_{\pm}(x)\right)^2},\cr
          \quad{\rm using\  weights}&&\  w^{(m)} \equiv \frac{ e^{-\frac{1}{2}{\rm AIC}(m)}}{\sum_{n \in {\rm R}_{\rm fit}} e^{-\frac{1}{2}{\rm AIC}(n)} },\cr &&\qquad
          \eeqa{aicmodelaveraging}
	  where ${\rm AIC}(n)$ is the (corrected) AIC value for the
	  $n$-th fit, namely, ${\rm AIC}(n) = {\cal L}_n + 2 p_n +
	  2 p_n(p_n+1)/(d_n-p_n-1)$, with $d_n$ being the number of
	  lattice data points being fitted in $n$-th fit and $p_n$
	  being the number of fit parameters, which is $N_J$ for
	  $h_{u^--d^-}$ and $N_J+1$ for $h_{u^+-d^+}$.
          
      \item Finally, we summarize our fits as $\bar h \pm \sigma
      \pm \Delta$,   where the central value $\bar h$, statistical error $\sigma$,
	  and systematic error $\Delta$ are defined as
          \beqa 
          \bar h_{\pm}(x) &=& {\rm Jack\-knife\ mean\ of\ }  h^{\rm AIC}_{\pm}(x),\cr
          \sigma_{\pm}(x) &=&  {\rm Jack\-knife\ error\ of\ }  h^{\rm AIC}_{\pm}(x),\cr
          \Delta_{\pm}(x) &=&  {\rm Jack\-knife\ mean\ of\ } \Delta^{\rm AIC}_{\pm}(x).
          \eeqa{statsys}
\end{enumerate}
The above choice which helps us separate the total error into
statistical and systematic parts is slightly different from another
choice~\cite{symonds2011brief,Karpie:2021pap} of adding $\sigma$
and $\Delta$ in quadrature to define a total error.  Below, we
discuss the results based on the above analysis methodology.

\befs
\centering
\includegraphics[scale=0.8]{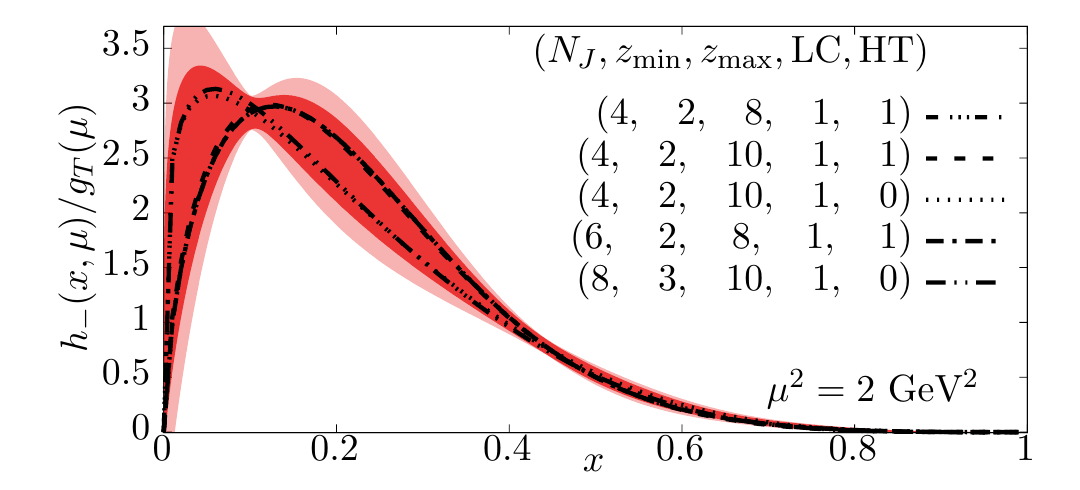}
\includegraphics[scale=0.8]{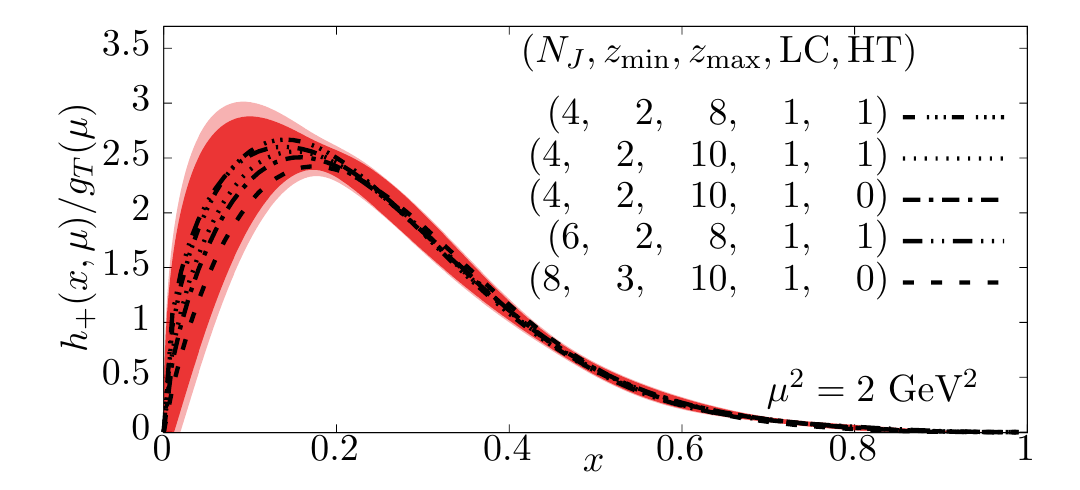}
\caption{
    The left and the right panels show the unit-normalized transversity
    PDFs $h_{-}(x)/g_T$ and $h_{+}(x)/g_T$ respectively, at
    $\mu=\sqrt{2}$ MeV as obtained from ${\rm Re}\mathfrak{M}(\nu,z_3^2)$
    and ${\rm Im}\mathfrak{M}(\nu,z_3^2)$ using the Jacobi polynomial
    reconstruction method (see text).  The legend specifies the
    maximum order of Jacobi polynomial used ($N_J$), the range of
    $z_3$-values $(z^{\rm min}_3, z^{\rm max}_3)$ used, whether the
    leading higher-twist (HT) correction term was used in the fit
    (1 or 0), and whether the leading short-distance lattice
    correction (LC) was used in the fit (1 or 0).  The different
    dashed curves are the central values of the PDFs reconstructed
    for some samples specifications of $(N_J, z^{\rm min}_3, z^{\rm
    max}_3, {\rm HT}, {\rm LC})$.  The inner red band is the
    1-$\sigma$ statistical error band and the outer red band is the
    combined statistical and systematic error (see text).
    }
\eefs{jacobisys}

In \fgn{jacprior}, we show the results for $s_{2\pm}^{\rm ans}$
and $s_{4\pm}^{\rm ans}$ from the Jacobi polynomial decomposition
of the PDF ansatz based fits. Along with the coefficients
$s_{n\pm}$, we have also shown the results for the small-$x$ and
large-$x$ exponents $\alpha_\pm$ and $\beta_\pm$ as inferred from
the fits. In each panel, we have shown the estimates for
$(\alpha_\pm,\beta_\pm, s_{2\pm}^{\rm ans}, s_{4\pm}^{\rm ans})$
at different $(z_3^{\rm min}, z_3^{\rm max})$ for the fit ranges.
The variability of the fitted parameters with $z_3$ range is rather
small and within the errors. These values of the exponents were
then used to form the family of $P_n^{\alpha_\pm,\beta_\pm}$
corresponding to each of the fit ranges. The central values and
statistical errors of $s_{n\pm}^{\rm ans}$ for $n$ up to 10 were
used as priors and the prior widths in the fits using \eqn{jacobiexpand}
as discussed in the step-2 above. It is at once clear from the consistency
of $s_{n\pm}$ with zero that the effect of the addition of Jacobi
polynomials with $n>0$ on the primary $x^{\alpha_\pm} (1-x)^{\beta_\pm}$
behavior is rather minimal. This is expected also from the observation
that the effect of ${\cal G}=1+\gamma_\pm \sqrt{x} +\delta_\pm x$,
was also minimal, and the transversity PDF could be described to a
good accuracy using a simpler $x^{\alpha_\pm} (1-x)^{\beta_\pm}$
two-parameter ansatz. However, these conclusions are made after the
fact and it is important to proceed with the Jacobi basis fits
in order to remove the slightest ansatz dependence and estimate the
systematic error in a more rigorous  manner.

\befs
\centering
\includegraphics[scale=0.8]{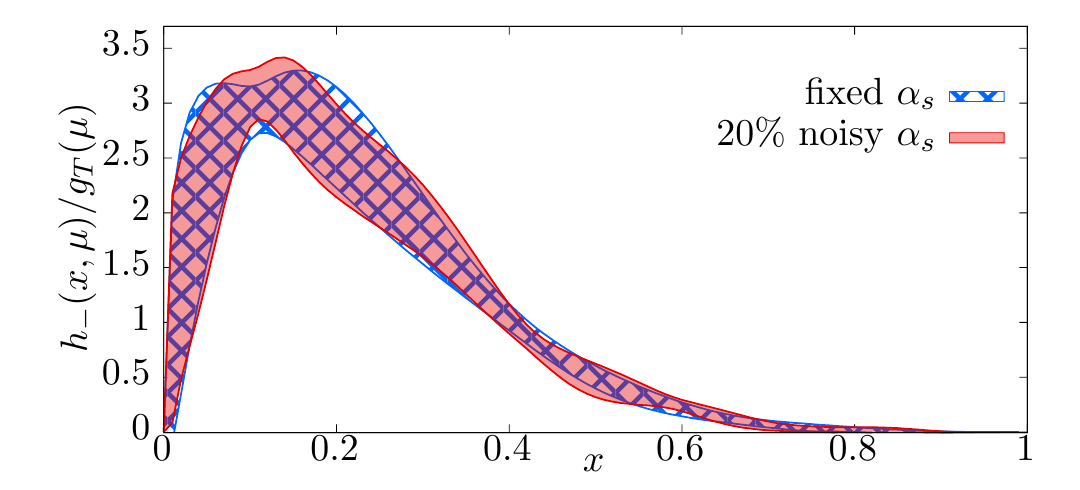}
\includegraphics[scale=0.8]{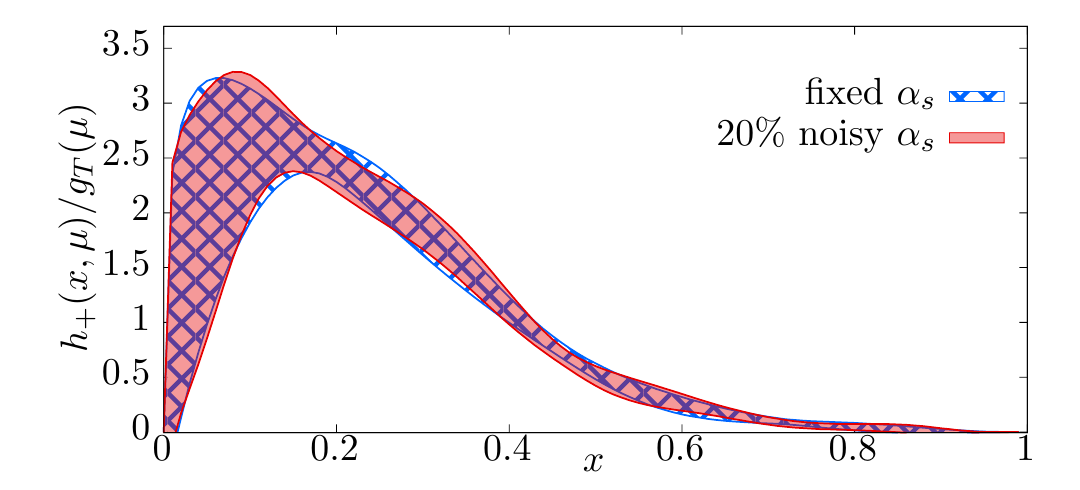}
\caption{
    The two panels show the transversity PDFs $h_{\pm}(x)/g_T$
    reconstructed using $(N_J=8, z^{\rm min}_3=2a, z^{\rm max}_3=8a, {\rm HT
    = 1}, {\rm LC=1})$ at a fixed value of the strong-coupling
    constant $\alpha_s(\mu=\sqrt{2}{\ \rm GeV})=0.36$ (shown as the
    patterned band) are compared with those using the same setup
    for the fits but $\alpha_s(\mu=\sqrt{2}{\ \rm GeV})$ is randomly
    picked from  the normal distribution with the central value of
    0.36 and with a width of $20\%$ (shown as the red band).
}
\eefs{pertsys}

In \fgn{jacobisys}, we show the results for $h_{\pm}(x)$
at $\mu=\sqrt{2}$ GeV from the fits using the Jacobi polynomial basis
obtained by minimizing the likelihood function ${\cal L}$ in
\eqn{likelihood}. In the figure, we have shown the central values
of $h_{\pm}(x)$ from some representative fitting choices,
$\left(N_J, z_{\rm min}, z_{\rm max}, {\rm LC}, {\rm HT}\right)$.
For $h_{-}(x)$, there is less scatter from changes to the fit
ranges than for  $h_{+}(x)$. For $h_{-}$, there is a tendency for central values 
with or without the higher-twist term to lie closer together, but such dependences were
well within statistical error and taken as part of systematic error. The AIC estimates of the central
values and their errors based on
\eqn{aicmodelaveraging} and \eqn{statsys} are shown as the red bands
in the two panels --- the darker red inner band includes only the
statistical error, whereas the lighter red outer band includes both
statistical and systematical errors. The AIC estimators nicely
envelope the PDFs resulting from sample individual fit choices. As
expected, the systematical error is not negligible in the case of
$h_{-}(x)$, whereas the systematic error committed in
$h_{+}(x)$ is small compared to the statistical one. The
results in \fgn{jacobisys} can be seen to be more or less the same
as our ansatz based estimation of the transversity PDFs in \fgn{jamfit}.
From the fits, we can also estimate the Mellin moments. This
is useful for making connection with the earlier estimates of $\langle x\rangle_+$
obtained via the leading-twist local operator approach, as well as 
with the possible estimates of $\langle x^2\rangle_-$ in the future.
Focusing on the first two Mellin moments, we find that at $\mu=\sqrt{2}$ GeV,
\beqa
&&\frac{\langle x\rangle_+}{g_T} = 0.2285(22)(17);\qquad \frac{\langle x\rangle_-}{g_T} = 0.2199(108)(101),\cr
&&\frac{\langle x^2\rangle_+}{g_T} = 0.0787(15)(08); \qquad \frac{\langle x^2\rangle_-}{g_T} = 0.0714(27)(12),
\eeqa{momfromjacobi}
where the errors in the first and second parenthesis are the
statistical and systematic errors using the procedure described
above. In Refs.~\cite{Yoon:2016jzj,Mondal:2020ela}, the values of
$g_T$ and $\langle x\rangle_+$ were computed using the ensembles
from the JLab/W\&M/LANL collaboration as used in this paper.
Unfortunately, the computations in those papers did not include the
ensemble used here, and therefore, for the sake of comparison we
take the results in~\cite{Yoon:2016jzj,Mondal:2020ela} that have
the same lattice spacing $a=0.094$ fm as in this paper, but a
slightly lighter pion mass of 270 MeV (which is the ensemble a094m270
as specified in those papers).  In these works, the value of tensor charge at 2 GeV
scale was found as $g_T=0.973(36)$ and $\langle x \rangle_+= 0.236(11)$, with a systematic
variation of about 0.02 around this value~\footnote{ In Ref.~\cite{Mondal:2020ela}, 
the results for $\langle x \rangle_+$ in the ensemble a094m270 shows variability with the 
excited state extrapolation methods and renormalization procedures. Therefore,
we consider a specific value from their
determination as $\langle x \rangle_+= 0.236(11)$, with
a variability of about 0.02 around this value.}.
From this, we find their
estimate for $\langle x \rangle_+/g_T = 0.242(14)$ at $\mu=2$ GeV
(with a systematic variation of about 0.02). In comparison, we find
our estimate for $\langle x\rangle_+/g_T$ to be 5\% smaller, which
is within a reasonable criteria for tolerance given both the
statistical and systematic errors, and the slight mismatch in
$\msbar$ renormalization scale $\mu$ in the two studies.

A remaining systematic error is the perturbative uncertainty
originating from the transversity matching kernel and the corresponding
Wilson coefficients due to the finite perturbative order used.  As
such, we only know the NLO matching kernel for transversity PDF at
this point, and therefore, we do not have a direct way to estimate
what the corrections from higher-order terms in the perturbative
series would be. This is unlike the unpolarized PDF case, where
there are recent results on the two-loop
matching~\cite{Chen:2020arf,Chen:2020iqi,Li:2020xml}, as well as
suggestions to estimate the higher-loop
uncertainties~\cite{Gao:2021hxl,Karthik:2021sbj}.  Instead, here we
tried to estimate the perturbative uncertainty in a simpler manner through the
sensitivity of the results to the value of $\alpha_s$ used in the
NLO coefficients in \eqn{nlokernel} and \eqn{wilcoeff} --- at NLO,
the scale $\mu$ at which we need to determine $\alpha_s(\mu)$ is not
specified and we implicitly assumed $\mu=\sqrt{2}$ GeV, same as the
factorization scale of the transversity PDF. Instead of fixing the
value of $\alpha_s=0.36$ at $\mu=\sqrt{2}$ GeV as done in all the
analysis presented above, we tried using a ``noisy" $\alpha_s$ by
randomly sampling $\alpha_s \sim {\cal N}(0.36, 0.072)$ and use
them in the fits.  We chose a Gaussian noise width of $20\%$ of
$\alpha_s=0.36$  as it is approximately the variation resulting in
$\alpha_s$ by changing the scale from $\mu/2$ to $2\mu$ for $\mu=\sqrt{2}$ GeV, a
variation that is traditionally used to evaluate the perturbative
uncertainties.  The results for  $h_{\pm}(x)$ using fixed $\alpha_s$
and 20\% noisy $\alpha_s$ are compared in \fgn{pertsys}.  We only
show a sample case for the fit choice using $N_J=8$
Jacobi polynomial reconstruction and $[z_{\rm
min},z_{\rm max}]=[2a,8a]$ in the figure, but the comparisons were
similar at other choices as well. One can see that the PDF
reconstruction is quite robust and only develops slight wiggles
when $\alpha_s$ is randomly varied, and such variations are masked
at the level of precision we are working at. This leads
us to think that the perturbative uncertainty of
our determination could be mild, and ignore such uncertainties
in our final estimate.

\befs
\centering
\includegraphics[scale=1.3]{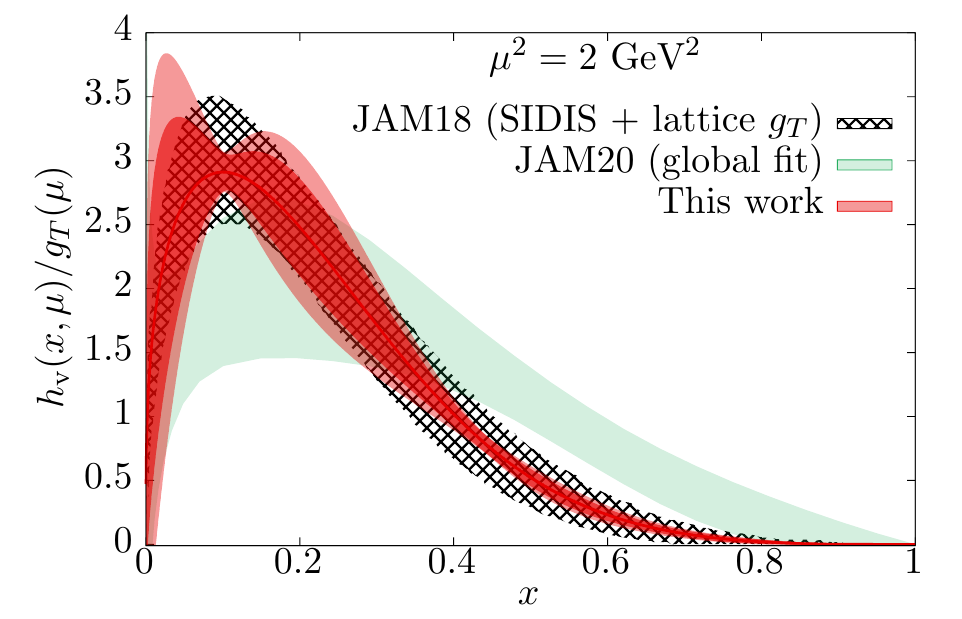}
\includegraphics[scale=1.3]{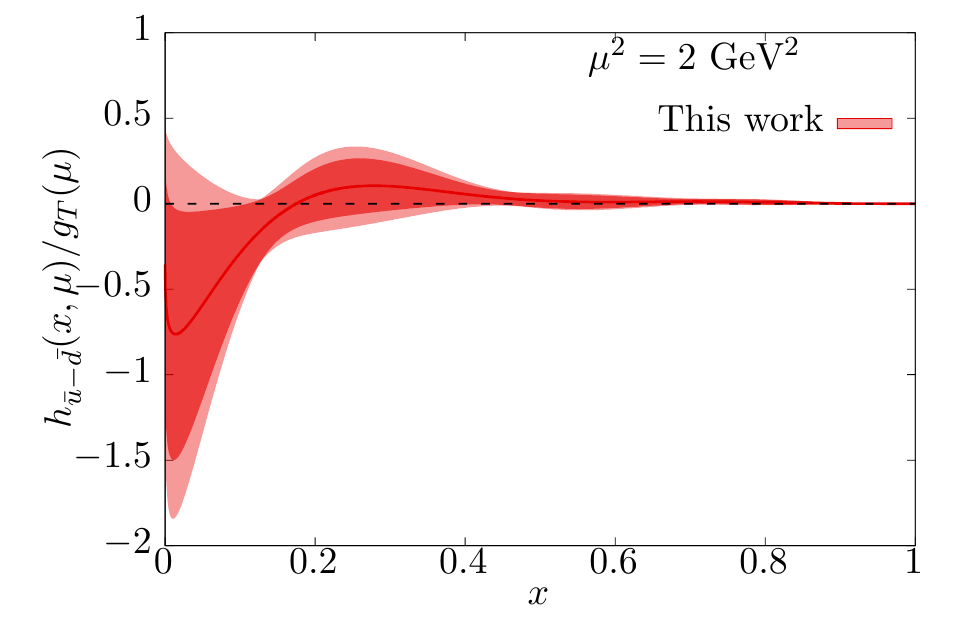}
\caption{
    Our lattice determination of the valence transversity distribution
    $h_{\rm v}(x,\mu)/g_T(\mu)$ using the pseudo-distribution
    approach is shown on the top panel, and the non-singlet antiquark
    transversity distribution $h_{\bar u - \bar d}(x,\mu)/g_T(\mu)$
    is shown on the bottom panel. The factorization scale used is
    $\mu=\sqrt{2}$ GeV for both the cases. In the two panels, the
    inner red band includes only the statistical error and the outer
    red band includes statistical and systematical errors in the
    PDF reconstruction. For the valence distribution, comparison
    is made with the previous phenomenological determinations 
    using SIDIS and lattice $g_T$ (JAM18)~\cite{Lin:2017stx}, shown using a patterned band, 
    and with the recently updated global fit analysis (JAM20)~\cite{Cammarota:2020qcw} of the single transverse spin 
    asymmetry data (but, without including lattice $g_T$), shown as a green band.
    The
    non-singlet antiquark distribution is consistent with an isospin
    symmetric intrinsic sea at all $x$.
}
\eefs{pdffinal}

In \fgn{pdffinal}, we present our final estimates of the $\msbar$
transversity PDFs at $\mu=\sqrt{2}$ GeV including the statistical and
systematic uncertainties. Our transversity PDF determination is
normalized with respect to $g_T(\mu)$ at $\mu=\sqrt{2}$ GeV, as in the
rest of the paper.  In the top panel, we show the valence transversity
PDF, $h_{\rm v}(x)=h_{-}(x)$ normalized by $g_T(\mu)$.
In the bottom panel, we show the non-singlet antiquark distribution
given by, $h_{\bar u - \bar d}(x) = \left[
h_{+}(x) - h_{-}(x)\right]/2$ normalized by
$g_T(\mu)$.  The outer red bands in both the panels include
both the statistical and systematic errors, whereas the inner red bands
include only the statistical error. In the top panel, we 
have compared our estimate for the valence transversity PDF with the 
expectations from fits to the experimental data. For this we used
two estimates from the Jefferson Angular Momentum 
Collaboration (JAM) based on two different fitting 
strategies as well as the processes that were considered.
First, we take the result presented in Ref.~\cite{Lin:2017stx}
where the analysis was based on fits to the single-transverse spin
asymmetry in pion production from deutron and proton targets, and 
further constrained by the lattice QCD input for the value of $g_T$.
We refer to this estimate as JAM18 in \fgn{pdffinal} and show it as
a black patterned band. Second, we take the recent updated result~\cite{Cammarota:2020qcw} from the JAM collaboration,
which considered single-transverse spin asymmetries in pion production via 
semi-inclusive $e^+e^-$ annihilation and $pp$ collisions in addition to the 
SIDIS data, but excluding the lattice input for $g_T$. We refer to this 
estimate as JAM20 in \fgn{pdffinal} and show it as the green band. 
In both cases, we have normalized them to the values of $g_T$ in 
their calculations, namely $g_T=1.01(6)$ for JAM18 and 
$g_T=0.86(12)$ for JAM20. While we see an overall agreement of 
our lattice estimate for the valence transversity PDF with the 
two phenomenological estimates, the very close agreement of our
result with JAM18 result is 
apparent. The source of the difference~\cite{nabuocomment} between the two phenomenological 
determinations, JAM18 and JAM20,
is likely to arise from the inclusion of single spin asymmetry 
data from $pp$ collisions from the RHIC experiment that results in a softer 
approach to zero as $x\to 1$, whereas the SIDIS data alone has a tendency for a 
harder fall near $x\to 1$. Since the experimental data are not
currently very precise to make a distinction between the two behaviors,
we expect our lattice determination, that has an inclination 
towards JAM18 result, could have an impact in the global fit
determinations in the near future.
However, we need to
immediately point the reader to the caveats that unlike the global
fit determination of the physical nucleon, our determination is at
a heavier-than-physical pion mass and at a fixed lattice spacing.
An effect of heavier pion mass could be through the trace terms to
the leading-twist OPE which we indirectly accounted for by introducing the nuisance fit terms proportional to $|z_3|^2$ in our fits and found
to be negligible. Another effect could be in changing the intrinsic
transversity PDF of the nucleon itself --- if this effect is found
to be small in the future computations at smaller pion masses, then
the overall agreement with the phenomenological determinations
would be remarkable.
We should also remark that while our lattice estimate could suffer from effects of heavier pion mass, 
our work is entirely within the collinear framework, whereas the 
global fits have to include chiral-odd TMD PDFs in order to extract the collinear transversity PDF.
In hindsight, the main non-vanishing contributions for $h_{\rm v}(x)$ 
coming from the intermediate $0.1 < x < 1$ is perhaps helping the lattice
determination due to a reduced small-$x$ uncertainty, unlike for
the case of the unpolarized valence PDFs. In the bottom panel of
\fgn{pdffinal}, we find that the non-singlet antiquark transversity 
PDF, which measures the difference between $\bar u$ and $\bar d$ in the
intrinsic sea of the nucleon,
vanishes at all $x$ within the uncertainties --- for $x<0.15$,
there is a slight excess of $\bar d$ compared to $\bar u$ if we
focus only on the central value, but these effects are statistically
insignificant. It should be noted that in the global fit
analyses of transversity PDF, a symmetric intrinsic sea is assumed from the start, whereas, our
result suggests that the intrinsic sea is indeed symmetric 
without any such prior assumptions.

\section{Conclusions}
\label{sec:conc}

We presented the formalism for the pseudo-distribution approach to
perturbatively match the renormalization group independent ratios
to the $\msbar$ transversity PDF. As a consequence, we were able
to separate the computation of transversity PDF into two independent
computations, namely, one for reconstructing the normalized quantity
$h(x)/g_T$ using the pseudo-distribution approach, and another for
finding $g_T$ to set the overall normalization that can be achieved
by well-known local operator methods.  In this paper, we presented
our computation of $h(x)/g_T$ and deferred $g_T$ to its dedicated
computation in the future.  We performed our analyses using the
nucleon matrix elements for the transversity pseudo-ITD obtained
by using the phased distillation
approach~\cite{Egerer:2021ymv,Egerer:2020hnc}, which forms an
important novel strategy followed in this paper.  We justified the
robustness of the excited-state extrapolations required to obtain
the matrix elements using the consistency between fits associated
with a spectral-decomposition and the summation method.
Through an application of the perturbative matching to capture the
Ioffe-time dependence at different fixed quark-antiquark separations,
$z_3$, we showed how to use the lattice data to directly infer the
presence of lattice spacing corrections, and to a lesser extent,
the higher-twist effects that presumably cancel in the RGI ratio.
The above steps formed the back-bone for our reconstruction of the
full $x$-dependent normalized transversity PDF $h(x)/g_T$ at
$\mu=\sqrt{2}$ GeV.

We used parametrized functional forms of  $h(x)/g_T$ in order to
overcome the inverse-problem associated with this approach.  First,
we reconstructed the transversity distribution
by employing a phenomenological
functional form of PDFs commonly used in  global fits (see
\eqn{jamansatz}) that is known to describe the cross-sections data
over a wide range of $x$ and $Q^2$. Using such a reconstructed
transversity PDF as our Bayesian prior, we used an expansion of
$h(x)/g_T$ in terms of a complete basis spanned by
Jacobi polynomials~\cite{Karpie:2021pap} in order to allow for more
flexibility in the PDF reconstruction. 
This strategy helped us remove any residual model
dependence as well as partially answered the question of whether a
more complex functional form could in principle change our conclusions.
We presented our final results in \fgn{pdffinal} for the valence
transversity PDF, $h_{\rm v}(x,\mu)/g_T$, and for the isovector
antiquark transversity PDF, $h_{\bar u - \bar d}(x,\mu)/g_T$. We
found a good agreement between our estimate of the valence
transversity PDF with the global fit analysis~\cite{Lin:2017stx} based on SIDIS and constraint from lattice $g_T$,
whereas we found only an overall agreement within larger statistical errors present in the recent
global fit analysis of single spin asymmetry data without any lattice input.
For the isovector antiquark PDF,
which is the difference between the $u$ and $d$ antiquark
distributions that are present in the
intrinsic sea (that is, not those radiated from the gluons),  
we found the resulting antiquark asymmetry to be consistent with zero at
all values of $x$.

The good agreement between our result for the valence transversity
PDF using pseudo-distribution approach is quite encouraging, given
comparable statistical errors in our estimate with that obtained
in the global fits.  Therefore, the lattice computations using
perturbative matching approaches are ideal for constraining the
transversity PDF in the lack of abundance of DIS cross-section data
sensitive to nucleon transversity. The path forward using this
approach is quite clear based on the results already presented in
this work.  The foremost, and also computationally the most
challenging, is to extend this computation to finer lattice spacings
to reduce the $a/|z_3|$ type short-distance lattice correction to
DGLAP (seen in \fgn{opewope}); such a correction will always be
present at $z_3$ of few lattice spacings however small the lattice
spacing becomes, but the idea would be to restrict our analysis for
physical distance $z_3 > z_3^{\rm min}$ for short-enough $z_3^{\rm
min}$ so as to ideally not add any corrections to our analysis and
rely only on the continuum DGLAP evolution.  Second, the good
comparison of our estimate in this work with the global-fit result
comes with the caveat that our computation was performed at a
heavier-than-physical pion mass of 358 MeV.  Therefore, it is
important to demonstrate that the observation holds as we reduce
the pion mass towards the physical point. Based on observations for
the unpolarized PDF~\cite{Joo:2020spy}, one could guess that the
effect of pion mass on the intrinsic quark structure of the nucleon
is not large. Third, we would like to fold in the estimates of
the tensor charge $g_T$ directly from the lattice to find $h(x,\mu)$
rather than the ratio $h(x,\mu)/g_T$ as in this work.  Finally, it would be 
interesting to use our estimated valence PDF as part of the global fit for
transversity PDF, such as those explored in Refs.~\cite{Bringewatt:2020ixn,DelDebbio:2020rgv,Cichy:2019ebf,Constantinou:2020hdm}.

\section*{Acknowledgments}
We thank Nobuo Sato and Wally Melnitchouk for discussions on the JAM determinations of the transversity PDF.
We would like to thank all the members of the HadStruc collaboration for fruitful and stimulating exchanges. 
This work is supported by Jefferson
Science Associates, LLC under U.S. DOE Contract \#DE-AC05-06OR23177.
KO was supported in part by U.S.  DOE grant \mbox{
  \#DE-FG02-04ER41302} and in part by the Center for Nuclear Femtography grants \#C2-2020-FEMT-006, \#C2019-FEMT-002-05.
    AR and WM are supported in part by U.S. DOE Grant
\mbox{\#DE-FG02-97ER41028. } NK and RSS are supported  in part by U.S. DOE Grant \mbox{\#DE-FG02-04ER41302}.
CE was supported in
part by the U.S. Department of Energy under Contract
No. DEFG02-04ER41302, a Department of Energy Office
of Science Graduate Student Research fellowship, through
the U.S. Department of Energy, Office of Science, Office of
Workforce Development for Teachers and Scientists, Office
of Science Graduate Student Research (SCGSR) program,
and a Jefferson Science Associates graduate fellowship.
The SCGSR program is administered by the Oak Ridge
Institute for Science and Education (ORISE) for the DOE.
ORISE is managed by ORAU under Contract No. DE-
SC0014664.
We would also like to thank the Texas Advanced Computing Center (TACC) at the University of Texas at Austin for providing HPC resources
on Frontera~\cite{frontera} that have contributed to the results in this paper. 
 We acknowledge the facilities of the USQCD Collaboration used for this research in part, which are funded by the Office of Science of the U.S. Department of Energy. This work was performed in part using computing
facilities at the College of William and Mary which were provided by
contributions from the National Science Foundation (MRI grant
PHY-1626177), and the Commonwealth of Virginia Equipment Trust Fund.
The authors acknowledge William \& Mary Research Computing for providing computational resources and/or technical support that have contributed to the results reported within this paper. This work used the Extreme Science and Engineering Discovery Environment (XSEDE), which is supported by National Science Foundation grant number ACI-1548562~\cite{xsede}.
 In addition, this work used resources at
NERSC, a DOE Office of Science User Facility supported by the Office
of Science of the U.S. Department of Energy under Contract
\#DE-AC02-05CH11231, as well as resources of the Oak Ridge Leadership Computing Facility at the Oak Ridge National Laboratory, which is supported by the Office of Science of the U.S. Department of Energy under Contract No. \mbox{\#DE-AC05-00OR22725}. In addition, this work was made possible using results obtained  at NERSC, a DOE Office of Science User Facility supported by the Office of Science of the U.S. Department of Energy under Contract \mbox{\#DE-AC02-05CH11231}, as well as resources of the Oak Ridge Leadership Computing Facility (ALCC and INCITE) at the Oak Ridge National Laboratory, which is supported by the Office of Science of the U.S. Department of Energy under Contract No. \mbox{\#DE-AC05-00OR22725}.  The software libraries used on these machines were Chroma~\cite{Edwards:2004sx}, QUDA ~\cite{Clark:2009wm,Babich:2010mu}, QDP-JIT~\cite{Winter:2014dka} and QPhiX~\cite{Joo:2013lwm,optimising} developed with  
 support from the U.S. Department of Energy, Office of Science, Office of Advanced Scientific Computing Research and Office of Nuclear Physics, Scientific Discovery through Advanced Computing (SciDAC) program, and of the U.S. Department of Energy Exascale Computing Project.

We acknowledge PRACE (Partnership for Advanced Computing in Europe) for awarding us access to the high performance computing system Marconi100 at CINECA (Consorzio Interuniversitario per il Calcolo Automatico dell’Italia Nord-orientale) under the grant Pra21-5389.
Results were obtained also by using Piz Daint at Centro Svizzero di Calcolo Scientifico (CSCS), via the project with id s994. We thank the staff of CSCS for access to the computational resources and for their constant support. This work also benefited from access to the Jean Zay supercomputer at the Institute for Development and Resources in Intensive Scientific Computing (IDRIS) in Orsay, France under project A0080511504.

\bibliography{pap.bib}

\begin{thebibliography}{120}%
\makeatletter
\providecommand \@ifxundefined [1]{%
 \@ifx{#1\undefined}
}%
\providecommand \@ifnum [1]{%
 \ifnum #1\expandafter \@firstoftwo
 \else \expandafter \@secondoftwo
 \fi
}%
\providecommand \@ifx [1]{%
 \ifx #1\expandafter \@firstoftwo
 \else \expandafter \@secondoftwo
 \fi
}%
\providecommand \natexlab [1]{#1}%
\providecommand \enquote  [1]{``#1''}%
\providecommand \bibnamefont  [1]{#1}%
\providecommand \bibfnamefont [1]{#1}%
\providecommand \citenamefont [1]{#1}%
\providecommand \href@noop [0]{\@secondoftwo}%
\providecommand \href [0]{\begingroup \@sanitize@url \@href}%
\providecommand \@href[1]{\@@startlink{#1}\@@href}%
\providecommand \@@href[1]{\endgroup#1\@@endlink}%
\providecommand \@sanitize@url [0]{\catcode `\\12\catcode `\$12\catcode
  `\&12\catcode `\#12\catcode `\^12\catcode `\_12\catcode `\%12\relax}%
\providecommand \@@startlink[1]{}%
\providecommand \@@endlink[0]{}%
\providecommand \url  [0]{\begingroup\@sanitize@url \@url }%
\providecommand \@url [1]{\endgroup\@href {#1}{\urlprefix }}%
\providecommand \urlprefix  [0]{URL }%
\providecommand \Eprint [0]{\href }%
\providecommand \doibase [0]{http://dx.doi.org/}%
\providecommand \selectlanguage [0]{\@gobble}%
\providecommand \bibinfo  [0]{\@secondoftwo}%
\providecommand \bibfield  [0]{\@secondoftwo}%
\providecommand \translation [1]{[#1]}%
\providecommand \BibitemOpen [0]{}%
\providecommand \bibitemStop [0]{}%
\providecommand \bibitemNoStop [0]{.\EOS\space}%
\providecommand \EOS [0]{\spacefactor3000\relax}%
\providecommand \BibitemShut  [1]{\csname bibitem#1\endcsname}%
\let\auto@bib@innerbib\@empty
\bibitem [{\citenamefont {Accardi}\ \emph
  {et~al.}(2016{\natexlab{a}})\citenamefont {Accardi} \emph
  {et~al.}}]{Accardi:2012qut}%
  \BibitemOpen
  \bibfield  {author} {\bibinfo {author} {\bibfnamefont {A.}~\bibnamefont
  {Accardi}} \emph {et~al.},\ }\href {\doibase 10.1140/epja/i2016-16268-9}
  {\bibfield  {journal} {\bibinfo  {journal} {Eur. Phys. J. A}\ }\textbf
  {\bibinfo {volume} {52}},\ \bibinfo {pages} {268} (\bibinfo {year}
  {2016}{\natexlab{a}})},\ \Eprint {http://arxiv.org/abs/1212.1701}
  {arXiv:1212.1701 [nucl-ex]} \BibitemShut {NoStop}%
\bibitem [{\citenamefont {Dudek}\ \emph {et~al.}(2012)\citenamefont {Dudek}
  \emph {et~al.}}]{Dudek:2012vr}%
  \BibitemOpen
  \bibfield  {author} {\bibinfo {author} {\bibfnamefont {J.}~\bibnamefont
  {Dudek}} \emph {et~al.},\ }\href {\doibase 10.1140/epja/i2012-12187-1}
  {\bibfield  {journal} {\bibinfo  {journal} {Eur. Phys. J. A}\ }\textbf
  {\bibinfo {volume} {48}},\ \bibinfo {pages} {187} (\bibinfo {year} {2012})},\
  \Eprint {http://arxiv.org/abs/1208.1244} {arXiv:1208.1244 [hep-ex]}
  \BibitemShut {NoStop}%
\bibitem [{\citenamefont {Chen}\ \emph {et~al.}(2014)\citenamefont {Chen},
  \citenamefont {Gao}, \citenamefont {Hemmick}, \citenamefont {Meziani},\ and\
  \citenamefont {Souder}}]{Chen:2014psa}%
  \BibitemOpen
  \bibfield  {author} {\bibinfo {author} {\bibfnamefont {J.~P.}\ \bibnamefont
  {Chen}}, \bibinfo {author} {\bibfnamefont {H.}~\bibnamefont {Gao}}, \bibinfo
  {author} {\bibfnamefont {T.~K.}\ \bibnamefont {Hemmick}}, \bibinfo {author}
  {\bibfnamefont {Z.~E.}\ \bibnamefont {Meziani}}, \ and\ \bibinfo {author}
  {\bibfnamefont {P.~A.}\ \bibnamefont {Souder}} (\bibinfo {collaboration}
  {SoLID}),\ }\href@noop {} {\  (\bibinfo {year} {2014})},\ \Eprint
  {http://arxiv.org/abs/1409.7741} {arXiv:1409.7741 [nucl-ex]} \BibitemShut
  {NoStop}%
\bibitem [{\citenamefont {Harland-Lang}\ \emph {et~al.}(2015)\citenamefont
  {Harland-Lang}, \citenamefont {Martin}, \citenamefont {Motylinski},\ and\
  \citenamefont {Thorne}}]{Harland-Lang:2014zoa}%
  \BibitemOpen
  \bibfield  {author} {\bibinfo {author} {\bibfnamefont {L.~A.}\ \bibnamefont
  {Harland-Lang}}, \bibinfo {author} {\bibfnamefont {A.~D.}\ \bibnamefont
  {Martin}}, \bibinfo {author} {\bibfnamefont {P.}~\bibnamefont {Motylinski}},
  \ and\ \bibinfo {author} {\bibfnamefont {R.~S.}\ \bibnamefont {Thorne}},\
  }\href {\doibase 10.1140/epjc/s10052-015-3397-6} {\bibfield  {journal}
  {\bibinfo  {journal} {Eur. Phys. J. C}\ }\textbf {\bibinfo {volume} {75}},\
  \bibinfo {pages} {204} (\bibinfo {year} {2015})},\ \Eprint
  {http://arxiv.org/abs/1412.3989} {arXiv:1412.3989 [hep-ph]} \BibitemShut
  {NoStop}%
\bibitem [{\citenamefont {Dulat}\ \emph {et~al.}(2016)\citenamefont {Dulat},
  \citenamefont {Hou}, \citenamefont {Gao}, \citenamefont {Guzzi},
  \citenamefont {Huston}, \citenamefont {Nadolsky}, \citenamefont {Pumplin},
  \citenamefont {Schmidt}, \citenamefont {Stump},\ and\ \citenamefont
  {Yuan}}]{Dulat:2015mca}%
  \BibitemOpen
  \bibfield  {author} {\bibinfo {author} {\bibfnamefont {S.}~\bibnamefont
  {Dulat}}, \bibinfo {author} {\bibfnamefont {T.-J.}\ \bibnamefont {Hou}},
  \bibinfo {author} {\bibfnamefont {J.}~\bibnamefont {Gao}}, \bibinfo {author}
  {\bibfnamefont {M.}~\bibnamefont {Guzzi}}, \bibinfo {author} {\bibfnamefont
  {J.}~\bibnamefont {Huston}}, \bibinfo {author} {\bibfnamefont
  {P.}~\bibnamefont {Nadolsky}}, \bibinfo {author} {\bibfnamefont
  {J.}~\bibnamefont {Pumplin}}, \bibinfo {author} {\bibfnamefont
  {C.}~\bibnamefont {Schmidt}}, \bibinfo {author} {\bibfnamefont
  {D.}~\bibnamefont {Stump}}, \ and\ \bibinfo {author} {\bibfnamefont {C.~P.}\
  \bibnamefont {Yuan}},\ }\href {\doibase 10.1103/PhysRevD.93.033006}
  {\bibfield  {journal} {\bibinfo  {journal} {Phys. Rev. D}\ }\textbf {\bibinfo
  {volume} {93}},\ \bibinfo {pages} {033006} (\bibinfo {year} {2016})},\
  \Eprint {http://arxiv.org/abs/1506.07443} {arXiv:1506.07443 [hep-ph]}
  \BibitemShut {NoStop}%
\bibitem [{\citenamefont {Accardi}\ \emph
  {et~al.}(2016{\natexlab{b}})\citenamefont {Accardi}, \citenamefont {Brady},
  \citenamefont {Melnitchouk}, \citenamefont {Owens},\ and\ \citenamefont
  {Sato}}]{Accardi:2016qay}%
  \BibitemOpen
  \bibfield  {author} {\bibinfo {author} {\bibfnamefont {A.}~\bibnamefont
  {Accardi}}, \bibinfo {author} {\bibfnamefont {L.~T.}\ \bibnamefont {Brady}},
  \bibinfo {author} {\bibfnamefont {W.}~\bibnamefont {Melnitchouk}}, \bibinfo
  {author} {\bibfnamefont {J.~F.}\ \bibnamefont {Owens}}, \ and\ \bibinfo
  {author} {\bibfnamefont {N.}~\bibnamefont {Sato}},\ }\href {\doibase
  10.1103/PhysRevD.93.114017} {\bibfield  {journal} {\bibinfo  {journal} {Phys.
  Rev. D}\ }\textbf {\bibinfo {volume} {93}},\ \bibinfo {pages} {114017}
  (\bibinfo {year} {2016}{\natexlab{b}})},\ \Eprint
  {http://arxiv.org/abs/1602.03154} {arXiv:1602.03154 [hep-ph]} \BibitemShut
  {NoStop}%
\bibitem [{\citenamefont {Ball}\ \emph {et~al.}(2017)\citenamefont {Ball} \emph
  {et~al.}}]{NNPDF:2017mvq}%
  \BibitemOpen
  \bibfield  {author} {\bibinfo {author} {\bibfnamefont {R.~D.}\ \bibnamefont
  {Ball}} \emph {et~al.} (\bibinfo {collaboration} {NNPDF}),\ }\href {\doibase
  10.1140/epjc/s10052-017-5199-5} {\bibfield  {journal} {\bibinfo  {journal}
  {Eur. Phys. J. C}\ }\textbf {\bibinfo {volume} {77}},\ \bibinfo {pages} {663}
  (\bibinfo {year} {2017})},\ \Eprint {http://arxiv.org/abs/1706.00428}
  {arXiv:1706.00428 [hep-ph]} \BibitemShut {NoStop}%
\bibitem [{\citenamefont {Accardi}\ \emph
  {et~al.}(2016{\natexlab{c}})\citenamefont {Accardi} \emph
  {et~al.}}]{Accardi:2016ndt}%
  \BibitemOpen
  \bibfield  {author} {\bibinfo {author} {\bibfnamefont {A.}~\bibnamefont
  {Accardi}} \emph {et~al.},\ }\href {\doibase 10.1140/epjc/s10052-016-4285-4}
  {\bibfield  {journal} {\bibinfo  {journal} {Eur. Phys. J. C}\ }\textbf
  {\bibinfo {volume} {76}},\ \bibinfo {pages} {471} (\bibinfo {year}
  {2016}{\natexlab{c}})},\ \Eprint {http://arxiv.org/abs/1603.08906}
  {arXiv:1603.08906 [hep-ph]} \BibitemShut {NoStop}%
\bibitem [{\citenamefont {Ralston}\ and\ \citenamefont
  {Soper}(1979)}]{Ralston:1979ys}%
  \BibitemOpen
  \bibfield  {author} {\bibinfo {author} {\bibfnamefont {J.~P.}\ \bibnamefont
  {Ralston}}\ and\ \bibinfo {author} {\bibfnamefont {D.~E.}\ \bibnamefont
  {Soper}},\ }\href {\doibase 10.1016/0550-3213(79)90082-8} {\bibfield
  {journal} {\bibinfo  {journal} {Nucl. Phys. B}\ }\textbf {\bibinfo {volume}
  {152}},\ \bibinfo {pages} {109} (\bibinfo {year} {1979})}\BibitemShut
  {NoStop}%
\bibitem [{\citenamefont {Artru}\ and\ \citenamefont
  {Mekhfi}(1990)}]{Artru:1989zv}%
  \BibitemOpen
  \bibfield  {author} {\bibinfo {author} {\bibfnamefont {X.}~\bibnamefont
  {Artru}}\ and\ \bibinfo {author} {\bibfnamefont {M.}~\bibnamefont {Mekhfi}},\
  }\href {\doibase 10.1007/BF01556280} {\bibfield  {journal} {\bibinfo
  {journal} {Z. Phys. C}\ }\textbf {\bibinfo {volume} {45}},\ \bibinfo {pages}
  {669} (\bibinfo {year} {1990})}\BibitemShut {NoStop}%
\bibitem [{\citenamefont {Cortes}\ \emph {et~al.}(1992)\citenamefont {Cortes},
  \citenamefont {Pire},\ and\ \citenamefont {Ralston}}]{Cortes:1991ja}%
  \BibitemOpen
  \bibfield  {author} {\bibinfo {author} {\bibfnamefont {J.~L.}\ \bibnamefont
  {Cortes}}, \bibinfo {author} {\bibfnamefont {B.}~\bibnamefont {Pire}}, \ and\
  \bibinfo {author} {\bibfnamefont {J.~P.}\ \bibnamefont {Ralston}},\ }\href
  {\doibase 10.1007/BF01565099} {\bibfield  {journal} {\bibinfo  {journal} {Z.
  Phys. C}\ }\textbf {\bibinfo {volume} {55}},\ \bibinfo {pages} {409}
  (\bibinfo {year} {1992})}\BibitemShut {NoStop}%
\bibitem [{\citenamefont {Jaffe}\ and\ \citenamefont
  {Ji}(1991)}]{Jaffe:1991kp}%
  \BibitemOpen
  \bibfield  {author} {\bibinfo {author} {\bibfnamefont {R.~L.}\ \bibnamefont
  {Jaffe}}\ and\ \bibinfo {author} {\bibfnamefont {X.-D.}\ \bibnamefont {Ji}},\
  }\href {\doibase 10.1103/PhysRevLett.67.552} {\bibfield  {journal} {\bibinfo
  {journal} {Phys. Rev. Lett.}\ }\textbf {\bibinfo {volume} {67}},\ \bibinfo
  {pages} {552} (\bibinfo {year} {1991})}\BibitemShut {NoStop}%
\bibitem [{\citenamefont {Ji}(1992)}]{Ji:1992ev}%
  \BibitemOpen
  \bibfield  {author} {\bibinfo {author} {\bibfnamefont {X.-D.}\ \bibnamefont
  {Ji}},\ }\href {\doibase 10.1016/0370-2693(92)91939-7} {\bibfield  {journal}
  {\bibinfo  {journal} {Phys. Lett. B}\ }\textbf {\bibinfo {volume} {284}},\
  \bibinfo {pages} {137} (\bibinfo {year} {1992})}\BibitemShut {NoStop}%
\bibitem [{\citenamefont {Anselmino}\ \emph {et~al.}(2007)\citenamefont
  {Anselmino}, \citenamefont {Boglione}, \citenamefont {D'Alesio},
  \citenamefont {Kotzinian}, \citenamefont {Murgia}, \citenamefont {Prokudin},\
  and\ \citenamefont {Turk}}]{Anselmino:2007fs}%
  \BibitemOpen
  \bibfield  {author} {\bibinfo {author} {\bibfnamefont {M.}~\bibnamefont
  {Anselmino}}, \bibinfo {author} {\bibfnamefont {M.}~\bibnamefont {Boglione}},
  \bibinfo {author} {\bibfnamefont {U.}~\bibnamefont {D'Alesio}}, \bibinfo
  {author} {\bibfnamefont {A.}~\bibnamefont {Kotzinian}}, \bibinfo {author}
  {\bibfnamefont {F.}~\bibnamefont {Murgia}}, \bibinfo {author} {\bibfnamefont
  {A.}~\bibnamefont {Prokudin}}, \ and\ \bibinfo {author} {\bibfnamefont
  {C.}~\bibnamefont {Turk}},\ }\href {\doibase 10.1103/PhysRevD.75.054032}
  {\bibfield  {journal} {\bibinfo  {journal} {Phys. Rev. D}\ }\textbf {\bibinfo
  {volume} {75}},\ \bibinfo {pages} {054032} (\bibinfo {year} {2007})},\
  \Eprint {http://arxiv.org/abs/hep-ph/0701006} {arXiv:hep-ph/0701006}
  \BibitemShut {NoStop}%
\bibitem [{\citenamefont {Airapetian}\ \emph {et~al.}(2005)\citenamefont
  {Airapetian} \emph {et~al.}}]{HERMES:2004mhh}%
  \BibitemOpen
  \bibfield  {author} {\bibinfo {author} {\bibfnamefont {A.}~\bibnamefont
  {Airapetian}} \emph {et~al.} (\bibinfo {collaboration} {HERMES}),\ }\href
  {\doibase 10.1103/PhysRevLett.94.012002} {\bibfield  {journal} {\bibinfo
  {journal} {Phys. Rev. Lett.}\ }\textbf {\bibinfo {volume} {94}},\ \bibinfo
  {pages} {012002} (\bibinfo {year} {2005})},\ \Eprint
  {http://arxiv.org/abs/hep-ex/0408013} {arXiv:hep-ex/0408013} \BibitemShut
  {NoStop}%
\bibitem [{\citenamefont {Ageev}\ \emph {et~al.}(2007)\citenamefont {Ageev}
  \emph {et~al.}}]{COMPASS:2006mkl}%
  \BibitemOpen
  \bibfield  {author} {\bibinfo {author} {\bibfnamefont {E.~S.}\ \bibnamefont
  {Ageev}} \emph {et~al.} (\bibinfo {collaboration} {COMPASS}),\ }\href
  {\doibase 10.1016/j.nuclphysb.2006.10.027} {\bibfield  {journal} {\bibinfo
  {journal} {Nucl. Phys. B}\ }\textbf {\bibinfo {volume} {765}},\ \bibinfo
  {pages} {31} (\bibinfo {year} {2007})},\ \Eprint
  {http://arxiv.org/abs/hep-ex/0610068} {arXiv:hep-ex/0610068} \BibitemShut
  {NoStop}%
\bibitem [{\citenamefont {Abe}\ \emph {et~al.}(2006)\citenamefont {Abe} \emph
  {et~al.}}]{Belle:2005dmx}%
  \BibitemOpen
  \bibfield  {author} {\bibinfo {author} {\bibfnamefont {K.}~\bibnamefont
  {Abe}} \emph {et~al.} (\bibinfo {collaboration} {Belle}),\ }\href {\doibase
  10.1103/PhysRevLett.96.232002} {\bibfield  {journal} {\bibinfo  {journal}
  {Phys. Rev. Lett.}\ }\textbf {\bibinfo {volume} {96}},\ \bibinfo {pages}
  {232002} (\bibinfo {year} {2006})},\ \Eprint
  {http://arxiv.org/abs/hep-ex/0507063} {arXiv:hep-ex/0507063} \BibitemShut
  {NoStop}%
\bibitem [{\citenamefont {Radici}\ and\ \citenamefont
  {Bacchetta}(2018)}]{Radici:2018iag}%
  \BibitemOpen
  \bibfield  {author} {\bibinfo {author} {\bibfnamefont {M.}~\bibnamefont
  {Radici}}\ and\ \bibinfo {author} {\bibfnamefont {A.}~\bibnamefont
  {Bacchetta}},\ }\href {\doibase 10.1103/PhysRevLett.120.192001} {\bibfield
  {journal} {\bibinfo  {journal} {Phys. Rev. Lett.}\ }\textbf {\bibinfo
  {volume} {120}},\ \bibinfo {pages} {192001} (\bibinfo {year} {2018})},\
  \Eprint {http://arxiv.org/abs/1802.05212} {arXiv:1802.05212 [hep-ph]}
  \BibitemShut {NoStop}%
\bibitem [{\citenamefont {Bacchetta}\ \emph {et~al.}(2011)\citenamefont
  {Bacchetta}, \citenamefont {Courtoy},\ and\ \citenamefont
  {Radici}}]{Bacchetta:2011ip}%
  \BibitemOpen
  \bibfield  {author} {\bibinfo {author} {\bibfnamefont {A.}~\bibnamefont
  {Bacchetta}}, \bibinfo {author} {\bibfnamefont {A.}~\bibnamefont {Courtoy}},
  \ and\ \bibinfo {author} {\bibfnamefont {M.}~\bibnamefont {Radici}},\ }\href
  {\doibase 10.1103/PhysRevLett.107.012001} {\bibfield  {journal} {\bibinfo
  {journal} {Phys. Rev. Lett.}\ }\textbf {\bibinfo {volume} {107}},\ \bibinfo
  {pages} {012001} (\bibinfo {year} {2011})},\ \Eprint
  {http://arxiv.org/abs/1104.3855} {arXiv:1104.3855 [hep-ph]} \BibitemShut
  {NoStop}%
\bibitem [{\citenamefont {Benel}\ \emph {et~al.}(2020)\citenamefont {Benel},
  \citenamefont {Courtoy},\ and\ \citenamefont
  {Ferro-Hernandez}}]{Benel:2019mcq}%
  \BibitemOpen
  \bibfield  {author} {\bibinfo {author} {\bibfnamefont {J.}~\bibnamefont
  {Benel}}, \bibinfo {author} {\bibfnamefont {A.}~\bibnamefont {Courtoy}}, \
  and\ \bibinfo {author} {\bibfnamefont {R.}~\bibnamefont {Ferro-Hernandez}},\
  }\href {\doibase 10.1140/epjc/s10052-020-8039-y} {\bibfield  {journal}
  {\bibinfo  {journal} {Eur. Phys. J. C}\ }\textbf {\bibinfo {volume} {80}},\
  \bibinfo {pages} {465} (\bibinfo {year} {2020})},\ \Eprint
  {http://arxiv.org/abs/1912.03289} {arXiv:1912.03289 [hep-ph]} \BibitemShut
  {NoStop}%
\bibitem [{\citenamefont {Cammarota}\ \emph {et~al.}(2020)\citenamefont
  {Cammarota}, \citenamefont {Gamberg}, \citenamefont {Kang}, \citenamefont
  {Miller}, \citenamefont {Pitonyak}, \citenamefont {Prokudin}, \citenamefont
  {Rogers},\ and\ \citenamefont {Sato}}]{Cammarota:2020qcw}%
  \BibitemOpen
  \bibfield  {author} {\bibinfo {author} {\bibfnamefont {J.}~\bibnamefont
  {Cammarota}}, \bibinfo {author} {\bibfnamefont {L.}~\bibnamefont {Gamberg}},
  \bibinfo {author} {\bibfnamefont {Z.-B.}\ \bibnamefont {Kang}}, \bibinfo
  {author} {\bibfnamefont {J.~A.}\ \bibnamefont {Miller}}, \bibinfo {author}
  {\bibfnamefont {D.}~\bibnamefont {Pitonyak}}, \bibinfo {author}
  {\bibfnamefont {A.}~\bibnamefont {Prokudin}}, \bibinfo {author}
  {\bibfnamefont {T.~C.}\ \bibnamefont {Rogers}}, \ and\ \bibinfo {author}
  {\bibfnamefont {N.}~\bibnamefont {Sato}} (\bibinfo {collaboration} {Jefferson
  Lab Angular Momentum}),\ }\href {\doibase 10.1103/PhysRevD.102.054002}
  {\bibfield  {journal} {\bibinfo  {journal} {Phys. Rev. D}\ }\textbf {\bibinfo
  {volume} {102}},\ \bibinfo {pages} {054002} (\bibinfo {year} {2020})},\
  \Eprint {http://arxiv.org/abs/2002.08384} {arXiv:2002.08384 [hep-ph]}
  \BibitemShut {NoStop}%
\bibitem [{\citenamefont {Ji}(2013)}]{Ji:2013dva}%
  \BibitemOpen
  \bibfield  {author} {\bibinfo {author} {\bibfnamefont {X.}~\bibnamefont
  {Ji}},\ }\href {\doibase 10.1103/PhysRevLett.110.262002} {\bibfield
  {journal} {\bibinfo  {journal} {Phys. Rev. Lett.}\ }\textbf {\bibinfo
  {volume} {110}},\ \bibinfo {pages} {262002} (\bibinfo {year} {2013})},\
  \Eprint {http://arxiv.org/abs/1305.1539} {arXiv:1305.1539 [hep-ph]}
  \BibitemShut {NoStop}%
\bibitem [{\citenamefont {Ji}(2014)}]{Ji:2014gla}%
  \BibitemOpen
  \bibfield  {author} {\bibinfo {author} {\bibfnamefont {X.}~\bibnamefont
  {Ji}},\ }\href {\doibase 10.1007/s11433-014-5492-3} {\bibfield  {journal}
  {\bibinfo  {journal} {Sci. China Phys. Mech. Astron.}\ }\textbf {\bibinfo
  {volume} {57}},\ \bibinfo {pages} {1407} (\bibinfo {year} {2014})},\ \Eprint
  {http://arxiv.org/abs/1404.6680} {arXiv:1404.6680 [hep-ph]} \BibitemShut
  {NoStop}%
\bibitem [{\citenamefont {Radyushkin}(2017)}]{Radyushkin:2017cyf}%
  \BibitemOpen
  \bibfield  {author} {\bibinfo {author} {\bibfnamefont {A.~V.}\ \bibnamefont
  {Radyushkin}},\ }\href {\doibase 10.1103/PhysRevD.96.034025} {\bibfield
  {journal} {\bibinfo  {journal} {Phys. Rev. D}\ }\textbf {\bibinfo {volume}
  {96}},\ \bibinfo {pages} {034025} (\bibinfo {year} {2017})},\ \Eprint
  {http://arxiv.org/abs/1705.01488} {arXiv:1705.01488 [hep-ph]} \BibitemShut
  {NoStop}%
\bibitem [{\citenamefont {Orginos}\ \emph {et~al.}(2017)\citenamefont
  {Orginos}, \citenamefont {Radyushkin}, \citenamefont {Karpie},\ and\
  \citenamefont {Zafeiropoulos}}]{Orginos:2017kos}%
  \BibitemOpen
  \bibfield  {author} {\bibinfo {author} {\bibfnamefont {K.}~\bibnamefont
  {Orginos}}, \bibinfo {author} {\bibfnamefont {A.}~\bibnamefont {Radyushkin}},
  \bibinfo {author} {\bibfnamefont {J.}~\bibnamefont {Karpie}}, \ and\ \bibinfo
  {author} {\bibfnamefont {S.}~\bibnamefont {Zafeiropoulos}},\ }\href {\doibase
  10.1103/PhysRevD.96.094503} {\bibfield  {journal} {\bibinfo  {journal} {Phys.
  Rev. D}\ }\textbf {\bibinfo {volume} {96}},\ \bibinfo {pages} {094503}
  (\bibinfo {year} {2017})},\ \Eprint {http://arxiv.org/abs/1706.05373}
  {arXiv:1706.05373 [hep-ph]} \BibitemShut {NoStop}%
\bibitem [{\citenamefont {Ma}\ and\ \citenamefont
  {Qiu}(2018{\natexlab{a}})}]{Ma:2014jla}%
  \BibitemOpen
  \bibfield  {author} {\bibinfo {author} {\bibfnamefont {Y.-Q.}\ \bibnamefont
  {Ma}}\ and\ \bibinfo {author} {\bibfnamefont {J.-W.}\ \bibnamefont {Qiu}},\
  }\href {\doibase 10.1103/PhysRevD.98.074021} {\bibfield  {journal} {\bibinfo
  {journal} {Phys. Rev. D}\ }\textbf {\bibinfo {volume} {98}},\ \bibinfo
  {pages} {074021} (\bibinfo {year} {2018}{\natexlab{a}})},\ \Eprint
  {http://arxiv.org/abs/1404.6860} {arXiv:1404.6860 [hep-ph]} \BibitemShut
  {NoStop}%
\bibitem [{\citenamefont {Ma}\ and\ \citenamefont
  {Qiu}(2018{\natexlab{b}})}]{Ma:2017pxb}%
  \BibitemOpen
  \bibfield  {author} {\bibinfo {author} {\bibfnamefont {Y.-Q.}\ \bibnamefont
  {Ma}}\ and\ \bibinfo {author} {\bibfnamefont {J.-W.}\ \bibnamefont {Qiu}},\
  }\href {\doibase 10.1103/PhysRevLett.120.022003} {\bibfield  {journal}
  {\bibinfo  {journal} {Phys. Rev. Lett.}\ }\textbf {\bibinfo {volume} {120}},\
  \bibinfo {pages} {022003} (\bibinfo {year} {2018}{\natexlab{b}})},\ \Eprint
  {http://arxiv.org/abs/1709.03018} {arXiv:1709.03018 [hep-ph]} \BibitemShut
  {NoStop}%
\bibitem [{\citenamefont {Braun}\ and\ \citenamefont
  {M\"uller}(2008)}]{Braun:2007wv}%
  \BibitemOpen
  \bibfield  {author} {\bibinfo {author} {\bibfnamefont {V.}~\bibnamefont
  {Braun}}\ and\ \bibinfo {author} {\bibfnamefont {D.}~\bibnamefont
  {M\"uller}},\ }\href {\doibase 10.1140/epjc/s10052-008-0608-4} {\bibfield
  {journal} {\bibinfo  {journal} {Eur. Phys. J. C}\ }\textbf {\bibinfo {volume}
  {55}},\ \bibinfo {pages} {349} (\bibinfo {year} {2008})},\ \Eprint
  {http://arxiv.org/abs/0709.1348} {arXiv:0709.1348 [hep-ph]} \BibitemShut
  {NoStop}%
\bibitem [{\citenamefont {Sufian}\ \emph {et~al.}(2019)\citenamefont {Sufian},
  \citenamefont {Karpie}, \citenamefont {Egerer}, \citenamefont {Orginos},
  \citenamefont {Qiu},\ and\ \citenamefont {Richards}}]{Sufian:2019bol}%
  \BibitemOpen
  \bibfield  {author} {\bibinfo {author} {\bibfnamefont {R.~S.}\ \bibnamefont
  {Sufian}}, \bibinfo {author} {\bibfnamefont {J.}~\bibnamefont {Karpie}},
  \bibinfo {author} {\bibfnamefont {C.}~\bibnamefont {Egerer}}, \bibinfo
  {author} {\bibfnamefont {K.}~\bibnamefont {Orginos}}, \bibinfo {author}
  {\bibfnamefont {J.-W.}\ \bibnamefont {Qiu}}, \ and\ \bibinfo {author}
  {\bibfnamefont {D.~G.}\ \bibnamefont {Richards}},\ }\href {\doibase
  10.1103/PhysRevD.99.074507} {\bibfield  {journal} {\bibinfo  {journal} {Phys.
  Rev. D}\ }\textbf {\bibinfo {volume} {99}},\ \bibinfo {pages} {074507}
  (\bibinfo {year} {2019})},\ \Eprint {http://arxiv.org/abs/1901.03921}
  {arXiv:1901.03921 [hep-lat]} \BibitemShut {NoStop}%
\bibitem [{\citenamefont {Sufian}\ \emph {et~al.}(2020)\citenamefont {Sufian},
  \citenamefont {Egerer}, \citenamefont {Karpie}, \citenamefont {Edwards},
  \citenamefont {Jo\'o}, \citenamefont {Ma}, \citenamefont {Orginos},
  \citenamefont {Qiu},\ and\ \citenamefont {Richards}}]{Sufian:2020vzb}%
  \BibitemOpen
  \bibfield  {author} {\bibinfo {author} {\bibfnamefont {R.~S.}\ \bibnamefont
  {Sufian}}, \bibinfo {author} {\bibfnamefont {C.}~\bibnamefont {Egerer}},
  \bibinfo {author} {\bibfnamefont {J.}~\bibnamefont {Karpie}}, \bibinfo
  {author} {\bibfnamefont {R.~G.}\ \bibnamefont {Edwards}}, \bibinfo {author}
  {\bibfnamefont {B.}~\bibnamefont {Jo\'o}}, \bibinfo {author} {\bibfnamefont
  {Y.-Q.}\ \bibnamefont {Ma}}, \bibinfo {author} {\bibfnamefont
  {K.}~\bibnamefont {Orginos}}, \bibinfo {author} {\bibfnamefont {J.-W.}\
  \bibnamefont {Qiu}}, \ and\ \bibinfo {author} {\bibfnamefont {D.~G.}\
  \bibnamefont {Richards}},\ }\href {\doibase 10.1103/PhysRevD.102.054508}
  {\bibfield  {journal} {\bibinfo  {journal} {Phys. Rev. D}\ }\textbf {\bibinfo
  {volume} {102}},\ \bibinfo {pages} {054508} (\bibinfo {year} {2020})},\
  \Eprint {http://arxiv.org/abs/2001.04960} {arXiv:2001.04960 [hep-lat]}
  \BibitemShut {NoStop}%
\bibitem [{\citenamefont {Martinelli}\ and\ \citenamefont
  {Sachrajda}(1987)}]{Martinelli:1987zd}%
  \BibitemOpen
  \bibfield  {author} {\bibinfo {author} {\bibfnamefont {G.}~\bibnamefont
  {Martinelli}}\ and\ \bibinfo {author} {\bibfnamefont {C.~T.}\ \bibnamefont
  {Sachrajda}},\ }\href {\doibase 10.1016/0370-2693(87)90601-0} {\bibfield
  {journal} {\bibinfo  {journal} {Phys. Lett. B}\ }\textbf {\bibinfo {volume}
  {196}},\ \bibinfo {pages} {184} (\bibinfo {year} {1987})}\BibitemShut
  {NoStop}%
\bibitem [{\citenamefont {Liu}\ and\ \citenamefont {Dong}(1994)}]{Liu:1993cv}%
  \BibitemOpen
  \bibfield  {author} {\bibinfo {author} {\bibfnamefont {K.-F.}\ \bibnamefont
  {Liu}}\ and\ \bibinfo {author} {\bibfnamefont {S.-J.}\ \bibnamefont {Dong}},\
  }\href {\doibase 10.1103/PhysRevLett.72.1790} {\bibfield  {journal} {\bibinfo
   {journal} {Phys. Rev. Lett.}\ }\textbf {\bibinfo {volume} {72}},\ \bibinfo
  {pages} {1790} (\bibinfo {year} {1994})},\ \Eprint
  {http://arxiv.org/abs/hep-ph/9306299} {arXiv:hep-ph/9306299} \BibitemShut
  {NoStop}%
\bibitem [{\citenamefont {Chambers}\ \emph {et~al.}(2017)\citenamefont
  {Chambers}, \citenamefont {Horsley}, \citenamefont {Nakamura}, \citenamefont
  {Perlt}, \citenamefont {Rakow}, \citenamefont {Schierholz}, \citenamefont
  {Schiller}, \citenamefont {Somfleth}, \citenamefont {Young},\ and\
  \citenamefont {Zanotti}}]{Chambers:2017dov}%
  \BibitemOpen
  \bibfield  {author} {\bibinfo {author} {\bibfnamefont {A.~J.}\ \bibnamefont
  {Chambers}}, \bibinfo {author} {\bibfnamefont {R.}~\bibnamefont {Horsley}},
  \bibinfo {author} {\bibfnamefont {Y.}~\bibnamefont {Nakamura}}, \bibinfo
  {author} {\bibfnamefont {H.}~\bibnamefont {Perlt}}, \bibinfo {author}
  {\bibfnamefont {P.~E.~L.}\ \bibnamefont {Rakow}}, \bibinfo {author}
  {\bibfnamefont {G.}~\bibnamefont {Schierholz}}, \bibinfo {author}
  {\bibfnamefont {A.}~\bibnamefont {Schiller}}, \bibinfo {author}
  {\bibfnamefont {K.}~\bibnamefont {Somfleth}}, \bibinfo {author}
  {\bibfnamefont {R.~D.}\ \bibnamefont {Young}}, \ and\ \bibinfo {author}
  {\bibfnamefont {J.~M.}\ \bibnamefont {Zanotti}},\ }\href {\doibase
  10.1103/PhysRevLett.118.242001} {\bibfield  {journal} {\bibinfo  {journal}
  {Phys. Rev. Lett.}\ }\textbf {\bibinfo {volume} {118}},\ \bibinfo {pages}
  {242001} (\bibinfo {year} {2017})},\ \Eprint
  {http://arxiv.org/abs/1703.01153} {arXiv:1703.01153 [hep-lat]} \BibitemShut
  {NoStop}%
\bibitem [{\citenamefont {Detmold}\ and\ \citenamefont
  {Lin}(2006)}]{Detmold:2005gg}%
  \BibitemOpen
  \bibfield  {author} {\bibinfo {author} {\bibfnamefont {W.}~\bibnamefont
  {Detmold}}\ and\ \bibinfo {author} {\bibfnamefont {C.~J.~D.}\ \bibnamefont
  {Lin}},\ }\href {\doibase 10.1103/PhysRevD.73.014501} {\bibfield  {journal}
  {\bibinfo  {journal} {Phys. Rev. D}\ }\textbf {\bibinfo {volume} {73}},\
  \bibinfo {pages} {014501} (\bibinfo {year} {2006})},\ \Eprint
  {http://arxiv.org/abs/hep-lat/0507007} {arXiv:hep-lat/0507007} \BibitemShut
  {NoStop}%
\bibitem [{\citenamefont {Detmold}\ \emph {et~al.}(2021)\citenamefont
  {Detmold}, \citenamefont {Grebe}, \citenamefont {Kanamori}, \citenamefont
  {Lin}, \citenamefont {Perry},\ and\ \citenamefont {Zhao}}]{Detmold:2021uru}%
  \BibitemOpen
  \bibfield  {author} {\bibinfo {author} {\bibfnamefont {W.}~\bibnamefont
  {Detmold}}, \bibinfo {author} {\bibfnamefont {A.~V.}\ \bibnamefont {Grebe}},
  \bibinfo {author} {\bibfnamefont {I.}~\bibnamefont {Kanamori}}, \bibinfo
  {author} {\bibfnamefont {C.~J.~D.}\ \bibnamefont {Lin}}, \bibinfo {author}
  {\bibfnamefont {R.~J.}\ \bibnamefont {Perry}}, \ and\ \bibinfo {author}
  {\bibfnamefont {Y.}~\bibnamefont {Zhao}},\ }\href@noop {} {\  (\bibinfo
  {year} {2021})},\ \Eprint {http://arxiv.org/abs/2103.09529} {arXiv:2103.09529
  [hep-lat]} \BibitemShut {NoStop}%
\bibitem [{\citenamefont {Ji}\ \emph {et~al.}(2021)\citenamefont {Ji},
  \citenamefont {Liu}, \citenamefont {Liu}, \citenamefont {Zhang},\ and\
  \citenamefont {Zhao}}]{Ji:2020ect}%
  \BibitemOpen
  \bibfield  {author} {\bibinfo {author} {\bibfnamefont {X.}~\bibnamefont
  {Ji}}, \bibinfo {author} {\bibfnamefont {Y.-S.}\ \bibnamefont {Liu}},
  \bibinfo {author} {\bibfnamefont {Y.}~\bibnamefont {Liu}}, \bibinfo {author}
  {\bibfnamefont {J.-H.}\ \bibnamefont {Zhang}}, \ and\ \bibinfo {author}
  {\bibfnamefont {Y.}~\bibnamefont {Zhao}},\ }\href {\doibase
  10.1103/RevModPhys.93.035005} {\bibfield  {journal} {\bibinfo  {journal}
  {Rev. Mod. Phys.}\ }\textbf {\bibinfo {volume} {93}},\ \bibinfo {pages}
  {035005} (\bibinfo {year} {2021})},\ \Eprint
  {http://arxiv.org/abs/2004.03543} {arXiv:2004.03543 [hep-ph]} \BibitemShut
  {NoStop}%
\bibitem [{\citenamefont {Radyushkin}(2020)}]{Radyushkin:2019mye}%
  \BibitemOpen
  \bibfield  {author} {\bibinfo {author} {\bibfnamefont {A.~V.}\ \bibnamefont
  {Radyushkin}},\ }\href {\doibase 10.1142/S0217751X20300021} {\bibfield
  {journal} {\bibinfo  {journal} {Int. J. Mod. Phys. A}\ }\textbf {\bibinfo
  {volume} {35}},\ \bibinfo {pages} {2030002} (\bibinfo {year} {2020})},\
  \Eprint {http://arxiv.org/abs/1912.04244} {arXiv:1912.04244 [hep-ph]}
  \BibitemShut {NoStop}%
\bibitem [{\citenamefont {Cichy}\ and\ \citenamefont
  {Constantinou}(2019)}]{Cichy:2018mum}%
  \BibitemOpen
  \bibfield  {author} {\bibinfo {author} {\bibfnamefont {K.}~\bibnamefont
  {Cichy}}\ and\ \bibinfo {author} {\bibfnamefont {M.}~\bibnamefont
  {Constantinou}},\ }\href {\doibase 10.1155/2019/3036904} {\bibfield
  {journal} {\bibinfo  {journal} {Adv. High Energy Phys.}\ }\textbf {\bibinfo
  {volume} {2019}},\ \bibinfo {pages} {3036904} (\bibinfo {year} {2019})},\
  \Eprint {http://arxiv.org/abs/1811.07248} {arXiv:1811.07248 [hep-lat]}
  \BibitemShut {NoStop}%
\bibitem [{\citenamefont {Monahan}(2018)}]{Monahan:2018euv}%
  \BibitemOpen
  \bibfield  {author} {\bibinfo {author} {\bibfnamefont {C.}~\bibnamefont
  {Monahan}},\ }\href {\doibase 10.22323/1.334.0018} {\bibfield  {journal}
  {\bibinfo  {journal} {PoS}\ }\textbf {\bibinfo {volume} {LATTICE2018}},\
  \bibinfo {pages} {018} (\bibinfo {year} {2018})},\ \Eprint
  {http://arxiv.org/abs/1811.00678} {arXiv:1811.00678 [hep-lat]} \BibitemShut
  {NoStop}%
\bibitem [{\citenamefont {Cichy}(2021)}]{Cichy:2021lih}%
  \BibitemOpen
  \bibfield  {author} {\bibinfo {author} {\bibfnamefont {K.}~\bibnamefont
  {Cichy}},\ }in\ \href@noop {} {\emph {\bibinfo {booktitle} {{38th
  International Symposium on Lattice Field Theory}}}}\ (\bibinfo {year}
  {2021})\ \Eprint {http://arxiv.org/abs/2110.07440} {arXiv:2110.07440
  [hep-lat]} \BibitemShut {NoStop}%
\bibitem [{\citenamefont {Egerer}\ \emph
  {et~al.}(2021{\natexlab{a}})\citenamefont {Egerer}, \citenamefont {Edwards},
  \citenamefont {Kallidonis}, \citenamefont {Orginos}, \citenamefont
  {Radyushkin}, \citenamefont {Richards}, \citenamefont {Romero},\ and\
  \citenamefont {Zafeiropoulos}}]{Egerer:2021ymv}%
  \BibitemOpen
  \bibfield  {author} {\bibinfo {author} {\bibfnamefont {C.}~\bibnamefont
  {Egerer}}, \bibinfo {author} {\bibfnamefont {R.~G.}\ \bibnamefont {Edwards}},
  \bibinfo {author} {\bibfnamefont {C.}~\bibnamefont {Kallidonis}}, \bibinfo
  {author} {\bibfnamefont {K.}~\bibnamefont {Orginos}}, \bibinfo {author}
  {\bibfnamefont {A.~V.}\ \bibnamefont {Radyushkin}}, \bibinfo {author}
  {\bibfnamefont {D.~G.}\ \bibnamefont {Richards}}, \bibinfo {author}
  {\bibfnamefont {E.}~\bibnamefont {Romero}}, \ and\ \bibinfo {author}
  {\bibfnamefont {S.}~\bibnamefont {Zafeiropoulos}},\ }\href@noop {} {\
  (\bibinfo {year} {2021}{\natexlab{a}})},\ \Eprint
  {http://arxiv.org/abs/2107.05199} {arXiv:2107.05199 [hep-lat]} \BibitemShut
  {NoStop}%
\bibitem [{\citenamefont {Karpie}\ \emph {et~al.}(2021)\citenamefont {Karpie},
  \citenamefont {Orginos}, \citenamefont {Radyushkin},\ and\ \citenamefont
  {Zafeiropoulos}}]{Karpie:2021pap}%
  \BibitemOpen
  \bibfield  {author} {\bibinfo {author} {\bibfnamefont {J.}~\bibnamefont
  {Karpie}}, \bibinfo {author} {\bibfnamefont {K.}~\bibnamefont {Orginos}},
  \bibinfo {author} {\bibfnamefont {A.}~\bibnamefont {Radyushkin}}, \ and\
  \bibinfo {author} {\bibfnamefont {S.}~\bibnamefont {Zafeiropoulos}},\
  }\href@noop {} {\  (\bibinfo {year} {2021})},\ \Eprint
  {http://arxiv.org/abs/2105.13313} {arXiv:2105.13313 [hep-lat]} \BibitemShut
  {NoStop}%
\bibitem [{\citenamefont {Jo\'o}\ \emph {et~al.}(2020)\citenamefont {Jo\'o},
  \citenamefont {Karpie}, \citenamefont {Orginos}, \citenamefont {Radyushkin},
  \citenamefont {Richards},\ and\ \citenamefont {Zafeiropoulos}}]{Joo:2020spy}%
  \BibitemOpen
  \bibfield  {author} {\bibinfo {author} {\bibfnamefont {B.}~\bibnamefont
  {Jo\'o}}, \bibinfo {author} {\bibfnamefont {J.}~\bibnamefont {Karpie}},
  \bibinfo {author} {\bibfnamefont {K.}~\bibnamefont {Orginos}}, \bibinfo
  {author} {\bibfnamefont {A.~V.}\ \bibnamefont {Radyushkin}}, \bibinfo
  {author} {\bibfnamefont {D.~G.}\ \bibnamefont {Richards}}, \ and\ \bibinfo
  {author} {\bibfnamefont {S.}~\bibnamefont {Zafeiropoulos}},\ }\href {\doibase
  10.1103/PhysRevLett.125.232003} {\bibfield  {journal} {\bibinfo  {journal}
  {Phys. Rev. Lett.}\ }\textbf {\bibinfo {volume} {125}},\ \bibinfo {pages}
  {232003} (\bibinfo {year} {2020})},\ \Eprint
  {http://arxiv.org/abs/2004.01687} {arXiv:2004.01687 [hep-lat]} \BibitemShut
  {NoStop}%
\bibitem [{\citenamefont {Jo\'o}\ \emph {et~al.}(2019)\citenamefont {Jo\'o},
  \citenamefont {Karpie}, \citenamefont {Orginos}, \citenamefont {Radyushkin},
  \citenamefont {Richards},\ and\ \citenamefont {Zafeiropoulos}}]{Joo:2019jct}%
  \BibitemOpen
  \bibfield  {author} {\bibinfo {author} {\bibfnamefont {B.}~\bibnamefont
  {Jo\'o}}, \bibinfo {author} {\bibfnamefont {J.}~\bibnamefont {Karpie}},
  \bibinfo {author} {\bibfnamefont {K.}~\bibnamefont {Orginos}}, \bibinfo
  {author} {\bibfnamefont {A.}~\bibnamefont {Radyushkin}}, \bibinfo {author}
  {\bibfnamefont {D.}~\bibnamefont {Richards}}, \ and\ \bibinfo {author}
  {\bibfnamefont {S.}~\bibnamefont {Zafeiropoulos}},\ }\href {\doibase
  10.1007/JHEP12(2019)081} {\bibfield  {journal} {\bibinfo  {journal} {JHEP}\
  }\textbf {\bibinfo {volume} {12}},\ \bibinfo {pages} {081} (\bibinfo {year}
  {2019})},\ \Eprint {http://arxiv.org/abs/1908.09771} {arXiv:1908.09771
  [hep-lat]} \BibitemShut {NoStop}%
\bibitem [{\citenamefont {Bhat}\ \emph {et~al.}(2021)\citenamefont {Bhat},
  \citenamefont {Cichy}, \citenamefont {Constantinou},\ and\ \citenamefont
  {Scapellato}}]{Bhat:2020ktg}%
  \BibitemOpen
  \bibfield  {author} {\bibinfo {author} {\bibfnamefont {M.}~\bibnamefont
  {Bhat}}, \bibinfo {author} {\bibfnamefont {K.}~\bibnamefont {Cichy}},
  \bibinfo {author} {\bibfnamefont {M.}~\bibnamefont {Constantinou}}, \ and\
  \bibinfo {author} {\bibfnamefont {A.}~\bibnamefont {Scapellato}},\ }\href
  {\doibase 10.1103/PhysRevD.103.034510} {\bibfield  {journal} {\bibinfo
  {journal} {Phys. Rev. D}\ }\textbf {\bibinfo {volume} {103}},\ \bibinfo
  {pages} {034510} (\bibinfo {year} {2021})},\ \Eprint
  {http://arxiv.org/abs/2005.02102} {arXiv:2005.02102 [hep-lat]} \BibitemShut
  {NoStop}%
\bibitem [{\citenamefont {Fan}\ \emph {et~al.}(2020)\citenamefont {Fan},
  \citenamefont {Gao}, \citenamefont {Li}, \citenamefont {Lin}, \citenamefont
  {Karthik}, \citenamefont {Mukherjee}, \citenamefont {Petreczky},
  \citenamefont {Syritsyn}, \citenamefont {Yang},\ and\ \citenamefont
  {Zhang}}]{Fan:2020nzz}%
  \BibitemOpen
  \bibfield  {author} {\bibinfo {author} {\bibfnamefont {Z.}~\bibnamefont
  {Fan}}, \bibinfo {author} {\bibfnamefont {X.}~\bibnamefont {Gao}}, \bibinfo
  {author} {\bibfnamefont {R.}~\bibnamefont {Li}}, \bibinfo {author}
  {\bibfnamefont {H.-W.}\ \bibnamefont {Lin}}, \bibinfo {author} {\bibfnamefont
  {N.}~\bibnamefont {Karthik}}, \bibinfo {author} {\bibfnamefont
  {S.}~\bibnamefont {Mukherjee}}, \bibinfo {author} {\bibfnamefont
  {P.}~\bibnamefont {Petreczky}}, \bibinfo {author} {\bibfnamefont
  {S.}~\bibnamefont {Syritsyn}}, \bibinfo {author} {\bibfnamefont {Y.-B.}\
  \bibnamefont {Yang}}, \ and\ \bibinfo {author} {\bibfnamefont
  {R.}~\bibnamefont {Zhang}},\ }\href {\doibase 10.1103/PhysRevD.102.074504}
  {\bibfield  {journal} {\bibinfo  {journal} {Phys. Rev. D}\ }\textbf {\bibinfo
  {volume} {102}},\ \bibinfo {pages} {074504} (\bibinfo {year} {2020})},\
  \Eprint {http://arxiv.org/abs/2005.12015} {arXiv:2005.12015 [hep-lat]}
  \BibitemShut {NoStop}%
\bibitem [{\citenamefont {Alexandrou}\ \emph
  {et~al.}(2021{\natexlab{a}})\citenamefont {Alexandrou}, \citenamefont
  {Cichy}, \citenamefont {Constantinou}, \citenamefont {Green}, \citenamefont
  {Hadjiyiannakou}, \citenamefont {Jansen}, \citenamefont {Manigrasso},
  \citenamefont {Scapellato},\ and\ \citenamefont
  {Steffens}}]{Alexandrou:2020qtt}%
  \BibitemOpen
  \bibfield  {author} {\bibinfo {author} {\bibfnamefont {C.}~\bibnamefont
  {Alexandrou}}, \bibinfo {author} {\bibfnamefont {K.}~\bibnamefont {Cichy}},
  \bibinfo {author} {\bibfnamefont {M.}~\bibnamefont {Constantinou}}, \bibinfo
  {author} {\bibfnamefont {J.~R.}\ \bibnamefont {Green}}, \bibinfo {author}
  {\bibfnamefont {K.}~\bibnamefont {Hadjiyiannakou}}, \bibinfo {author}
  {\bibfnamefont {K.}~\bibnamefont {Jansen}}, \bibinfo {author} {\bibfnamefont
  {F.}~\bibnamefont {Manigrasso}}, \bibinfo {author} {\bibfnamefont
  {A.}~\bibnamefont {Scapellato}}, \ and\ \bibinfo {author} {\bibfnamefont
  {F.}~\bibnamefont {Steffens}},\ }\href {\doibase 10.1103/PhysRevD.103.094512}
  {\bibfield  {journal} {\bibinfo  {journal} {Phys. Rev. D}\ }\textbf {\bibinfo
  {volume} {103}},\ \bibinfo {pages} {094512} (\bibinfo {year}
  {2021}{\natexlab{a}})},\ \Eprint {http://arxiv.org/abs/2011.00964}
  {arXiv:2011.00964 [hep-lat]} \BibitemShut {NoStop}%
\bibitem [{\citenamefont {Alexandrou}\ \emph
  {et~al.}(2018{\natexlab{a}})\citenamefont {Alexandrou}, \citenamefont
  {Cichy}, \citenamefont {Constantinou}, \citenamefont {Jansen}, \citenamefont
  {Scapellato},\ and\ \citenamefont {Steffens}}]{Alexandrou:2018pbm}%
  \BibitemOpen
  \bibfield  {author} {\bibinfo {author} {\bibfnamefont {C.}~\bibnamefont
  {Alexandrou}}, \bibinfo {author} {\bibfnamefont {K.}~\bibnamefont {Cichy}},
  \bibinfo {author} {\bibfnamefont {M.}~\bibnamefont {Constantinou}}, \bibinfo
  {author} {\bibfnamefont {K.}~\bibnamefont {Jansen}}, \bibinfo {author}
  {\bibfnamefont {A.}~\bibnamefont {Scapellato}}, \ and\ \bibinfo {author}
  {\bibfnamefont {F.}~\bibnamefont {Steffens}},\ }\href {\doibase
  10.1103/PhysRevLett.121.112001} {\bibfield  {journal} {\bibinfo  {journal}
  {Phys. Rev. Lett.}\ }\textbf {\bibinfo {volume} {121}},\ \bibinfo {pages}
  {112001} (\bibinfo {year} {2018}{\natexlab{a}})},\ \Eprint
  {http://arxiv.org/abs/1803.02685} {arXiv:1803.02685 [hep-lat]} \BibitemShut
  {NoStop}%
\bibitem [{\citenamefont {Lin}\ \emph {et~al.}(2020)\citenamefont {Lin},
  \citenamefont {Chen},\ and\ \citenamefont {Zhang}}]{Lin:2020fsj}%
  \BibitemOpen
  \bibfield  {author} {\bibinfo {author} {\bibfnamefont {H.-W.}\ \bibnamefont
  {Lin}}, \bibinfo {author} {\bibfnamefont {J.-W.}\ \bibnamefont {Chen}}, \
  and\ \bibinfo {author} {\bibfnamefont {R.}~\bibnamefont {Zhang}},\
  }\href@noop {} {\  (\bibinfo {year} {2020})},\ \Eprint
  {http://arxiv.org/abs/2011.14971} {arXiv:2011.14971 [hep-lat]} \BibitemShut
  {NoStop}%
\bibitem [{\citenamefont {Izubuchi}\ \emph {et~al.}(2019)\citenamefont
  {Izubuchi}, \citenamefont {Jin}, \citenamefont {Kallidonis}, \citenamefont
  {Karthik}, \citenamefont {Mukherjee}, \citenamefont {Petreczky},
  \citenamefont {Shugert},\ and\ \citenamefont {Syritsyn}}]{Izubuchi:2019lyk}%
  \BibitemOpen
  \bibfield  {author} {\bibinfo {author} {\bibfnamefont {T.}~\bibnamefont
  {Izubuchi}}, \bibinfo {author} {\bibfnamefont {L.}~\bibnamefont {Jin}},
  \bibinfo {author} {\bibfnamefont {C.}~\bibnamefont {Kallidonis}}, \bibinfo
  {author} {\bibfnamefont {N.}~\bibnamefont {Karthik}}, \bibinfo {author}
  {\bibfnamefont {S.}~\bibnamefont {Mukherjee}}, \bibinfo {author}
  {\bibfnamefont {P.}~\bibnamefont {Petreczky}}, \bibinfo {author}
  {\bibfnamefont {C.}~\bibnamefont {Shugert}}, \ and\ \bibinfo {author}
  {\bibfnamefont {S.}~\bibnamefont {Syritsyn}},\ }\href {\doibase
  10.1103/PhysRevD.100.034516} {\bibfield  {journal} {\bibinfo  {journal}
  {Phys. Rev. D}\ }\textbf {\bibinfo {volume} {100}},\ \bibinfo {pages}
  {034516} (\bibinfo {year} {2019})},\ \Eprint
  {http://arxiv.org/abs/1905.06349} {arXiv:1905.06349 [hep-lat]} \BibitemShut
  {NoStop}%
\bibitem [{\citenamefont {Gao}\ \emph {et~al.}(2020)\citenamefont {Gao},
  \citenamefont {Jin}, \citenamefont {Kallidonis}, \citenamefont {Karthik},
  \citenamefont {Mukherjee}, \citenamefont {Petreczky}, \citenamefont
  {Shugert}, \citenamefont {Syritsyn},\ and\ \citenamefont
  {Zhao}}]{Gao:2020ito}%
  \BibitemOpen
  \bibfield  {author} {\bibinfo {author} {\bibfnamefont {X.}~\bibnamefont
  {Gao}}, \bibinfo {author} {\bibfnamefont {L.}~\bibnamefont {Jin}}, \bibinfo
  {author} {\bibfnamefont {C.}~\bibnamefont {Kallidonis}}, \bibinfo {author}
  {\bibfnamefont {N.}~\bibnamefont {Karthik}}, \bibinfo {author} {\bibfnamefont
  {S.}~\bibnamefont {Mukherjee}}, \bibinfo {author} {\bibfnamefont
  {P.}~\bibnamefont {Petreczky}}, \bibinfo {author} {\bibfnamefont
  {C.}~\bibnamefont {Shugert}}, \bibinfo {author} {\bibfnamefont
  {S.}~\bibnamefont {Syritsyn}}, \ and\ \bibinfo {author} {\bibfnamefont
  {Y.}~\bibnamefont {Zhao}},\ }\href {\doibase 10.1103/PhysRevD.102.094513}
  {\bibfield  {journal} {\bibinfo  {journal} {Phys. Rev. D}\ }\textbf {\bibinfo
  {volume} {102}},\ \bibinfo {pages} {094513} (\bibinfo {year} {2020})},\
  \Eprint {http://arxiv.org/abs/2007.06590} {arXiv:2007.06590 [hep-lat]}
  \BibitemShut {NoStop}%
\bibitem [{\citenamefont {Lin}\ \emph {et~al.}(2021)\citenamefont {Lin},
  \citenamefont {Chen}, \citenamefont {Fan}, \citenamefont {Zhang},\ and\
  \citenamefont {Zhang}}]{Lin:2020ssv}%
  \BibitemOpen
  \bibfield  {author} {\bibinfo {author} {\bibfnamefont {H.-W.}\ \bibnamefont
  {Lin}}, \bibinfo {author} {\bibfnamefont {J.-W.}\ \bibnamefont {Chen}},
  \bibinfo {author} {\bibfnamefont {Z.}~\bibnamefont {Fan}}, \bibinfo {author}
  {\bibfnamefont {J.-H.}\ \bibnamefont {Zhang}}, \ and\ \bibinfo {author}
  {\bibfnamefont {R.}~\bibnamefont {Zhang}},\ }\href {\doibase
  10.1103/PhysRevD.103.014516} {\bibfield  {journal} {\bibinfo  {journal}
  {Phys. Rev. D}\ }\textbf {\bibinfo {volume} {103}},\ \bibinfo {pages}
  {014516} (\bibinfo {year} {2021})},\ \Eprint
  {http://arxiv.org/abs/2003.14128} {arXiv:2003.14128 [hep-lat]} \BibitemShut
  {NoStop}%
\bibitem [{\citenamefont {Alexandrou}\ \emph
  {et~al.}(2020{\natexlab{a}})\citenamefont {Alexandrou}, \citenamefont
  {Cichy}, \citenamefont {Constantinou}, \citenamefont {Hadjiyiannakou},
  \citenamefont {Jansen}, \citenamefont {Scapellato},\ and\ \citenamefont
  {Steffens}}]{Alexandrou:2020zbe}%
  \BibitemOpen
  \bibfield  {author} {\bibinfo {author} {\bibfnamefont {C.}~\bibnamefont
  {Alexandrou}}, \bibinfo {author} {\bibfnamefont {K.}~\bibnamefont {Cichy}},
  \bibinfo {author} {\bibfnamefont {M.}~\bibnamefont {Constantinou}}, \bibinfo
  {author} {\bibfnamefont {K.}~\bibnamefont {Hadjiyiannakou}}, \bibinfo
  {author} {\bibfnamefont {K.}~\bibnamefont {Jansen}}, \bibinfo {author}
  {\bibfnamefont {A.}~\bibnamefont {Scapellato}}, \ and\ \bibinfo {author}
  {\bibfnamefont {F.}~\bibnamefont {Steffens}},\ }\href {\doibase
  10.1103/PhysRevLett.125.262001} {\bibfield  {journal} {\bibinfo  {journal}
  {Phys. Rev. Lett.}\ }\textbf {\bibinfo {volume} {125}},\ \bibinfo {pages}
  {262001} (\bibinfo {year} {2020}{\natexlab{a}})},\ \Eprint
  {http://arxiv.org/abs/2008.10573} {arXiv:2008.10573 [hep-lat]} \BibitemShut
  {NoStop}%
\bibitem [{\citenamefont {Lin}(2020)}]{Lin:2020rxa}%
  \BibitemOpen
  \bibfield  {author} {\bibinfo {author} {\bibfnamefont {H.-W.}\ \bibnamefont
  {Lin}},\ }\href@noop {} {\  (\bibinfo {year} {2020})},\ \Eprint
  {http://arxiv.org/abs/2008.12474} {arXiv:2008.12474 [hep-ph]} \BibitemShut
  {NoStop}%
\bibitem [{\citenamefont {Chen}\ \emph
  {et~al.}(2020{\natexlab{a}})\citenamefont {Chen}, \citenamefont {Lin},\ and\
  \citenamefont {Zhang}}]{Chen:2019lcm}%
  \BibitemOpen
  \bibfield  {author} {\bibinfo {author} {\bibfnamefont {J.-W.}\ \bibnamefont
  {Chen}}, \bibinfo {author} {\bibfnamefont {H.-W.}\ \bibnamefont {Lin}}, \
  and\ \bibinfo {author} {\bibfnamefont {J.-H.}\ \bibnamefont {Zhang}},\ }\href
  {\doibase 10.1016/j.nuclphysb.2020.114940} {\bibfield  {journal} {\bibinfo
  {journal} {Nucl. Phys. B}\ }\textbf {\bibinfo {volume} {952}},\ \bibinfo
  {pages} {114940} (\bibinfo {year} {2020}{\natexlab{a}})},\ \Eprint
  {http://arxiv.org/abs/1904.12376} {arXiv:1904.12376 [hep-lat]} \BibitemShut
  {NoStop}%
\bibitem [{\citenamefont {Khan}\ \emph
  {et~al.}(2021{\natexlab{a}})\citenamefont {Khan} \emph
  {et~al.}}]{HadStruc:2021wmh}%
  \BibitemOpen
  \bibfield  {author} {\bibinfo {author} {\bibfnamefont {T.}~\bibnamefont
  {Khan}} \emph {et~al.} (\bibinfo {collaboration} {HadStruc}),\ }\href@noop {}
  {\  (\bibinfo {year} {2021}{\natexlab{a}})},\ \Eprint
  {http://arxiv.org/abs/2107.08960} {arXiv:2107.08960 [hep-lat]} \BibitemShut
  {NoStop}%
\bibitem [{\citenamefont {Fan}\ \emph {et~al.}(2018)\citenamefont {Fan},
  \citenamefont {Yang}, \citenamefont {Anthony}, \citenamefont {Lin},\ and\
  \citenamefont {Liu}}]{Fan:2018dxu}%
  \BibitemOpen
  \bibfield  {author} {\bibinfo {author} {\bibfnamefont {Z.-Y.}\ \bibnamefont
  {Fan}}, \bibinfo {author} {\bibfnamefont {Y.-B.}\ \bibnamefont {Yang}},
  \bibinfo {author} {\bibfnamefont {A.}~\bibnamefont {Anthony}}, \bibinfo
  {author} {\bibfnamefont {H.-W.}\ \bibnamefont {Lin}}, \ and\ \bibinfo
  {author} {\bibfnamefont {K.-F.}\ \bibnamefont {Liu}},\ }\href {\doibase
  10.1103/PhysRevLett.121.242001} {\bibfield  {journal} {\bibinfo  {journal}
  {Phys. Rev. Lett.}\ }\textbf {\bibinfo {volume} {121}},\ \bibinfo {pages}
  {242001} (\bibinfo {year} {2018})},\ \Eprint
  {http://arxiv.org/abs/1808.02077} {arXiv:1808.02077 [hep-lat]} \BibitemShut
  {NoStop}%
\bibitem [{\citenamefont {Fan}\ \emph {et~al.}(2021)\citenamefont {Fan},
  \citenamefont {Zhang},\ and\ \citenamefont {Lin}}]{Fan:2020cpa}%
  \BibitemOpen
  \bibfield  {author} {\bibinfo {author} {\bibfnamefont {Z.}~\bibnamefont
  {Fan}}, \bibinfo {author} {\bibfnamefont {R.}~\bibnamefont {Zhang}}, \ and\
  \bibinfo {author} {\bibfnamefont {H.-W.}\ \bibnamefont {Lin}},\ }\href
  {\doibase 10.1142/S0217751X21500809} {\bibfield  {journal} {\bibinfo
  {journal} {Int. J. Mod. Phys. A}\ }\textbf {\bibinfo {volume} {36}},\
  \bibinfo {pages} {2150080} (\bibinfo {year} {2021})},\ \Eprint
  {http://arxiv.org/abs/2007.16113} {arXiv:2007.16113 [hep-lat]} \BibitemShut
  {NoStop}%
\bibitem [{\citenamefont {Fan}\ and\ \citenamefont {Lin}(2021)}]{Fan:2021bcr}%
  \BibitemOpen
  \bibfield  {author} {\bibinfo {author} {\bibfnamefont {Z.}~\bibnamefont
  {Fan}}\ and\ \bibinfo {author} {\bibfnamefont {H.-W.}\ \bibnamefont {Lin}},\
  }\href@noop {} {\  (\bibinfo {year} {2021})},\ \Eprint
  {http://arxiv.org/abs/2104.06372} {arXiv:2104.06372 [hep-lat]} \BibitemShut
  {NoStop}%
\bibitem [{\citenamefont {Bhattacharya}\ \emph {et~al.}(2016)\citenamefont
  {Bhattacharya}, \citenamefont {Cirigliano}, \citenamefont {Cohen},
  \citenamefont {Gupta}, \citenamefont {Lin},\ and\ \citenamefont
  {Yoon}}]{Bhattacharya:2016zcn}%
  \BibitemOpen
  \bibfield  {author} {\bibinfo {author} {\bibfnamefont {T.}~\bibnamefont
  {Bhattacharya}}, \bibinfo {author} {\bibfnamefont {V.}~\bibnamefont
  {Cirigliano}}, \bibinfo {author} {\bibfnamefont {S.}~\bibnamefont {Cohen}},
  \bibinfo {author} {\bibfnamefont {R.}~\bibnamefont {Gupta}}, \bibinfo
  {author} {\bibfnamefont {H.-W.}\ \bibnamefont {Lin}}, \ and\ \bibinfo
  {author} {\bibfnamefont {B.}~\bibnamefont {Yoon}},\ }\href {\doibase
  10.1103/PhysRevD.94.054508} {\bibfield  {journal} {\bibinfo  {journal} {Phys.
  Rev. D}\ }\textbf {\bibinfo {volume} {94}},\ \bibinfo {pages} {054508}
  (\bibinfo {year} {2016})},\ \Eprint {http://arxiv.org/abs/1606.07049}
  {arXiv:1606.07049 [hep-lat]} \BibitemShut {NoStop}%
\bibitem [{\citenamefont {Bhattacharya}\ \emph {et~al.}(2015)\citenamefont
  {Bhattacharya}, \citenamefont {Cirigliano}, \citenamefont {Cohen},
  \citenamefont {Gupta}, \citenamefont {Joseph}, \citenamefont {Lin},\ and\
  \citenamefont {Yoon}}]{Bhattacharya:2015wna}%
  \BibitemOpen
  \bibfield  {author} {\bibinfo {author} {\bibfnamefont {T.}~\bibnamefont
  {Bhattacharya}}, \bibinfo {author} {\bibfnamefont {V.}~\bibnamefont
  {Cirigliano}}, \bibinfo {author} {\bibfnamefont {S.}~\bibnamefont {Cohen}},
  \bibinfo {author} {\bibfnamefont {R.}~\bibnamefont {Gupta}}, \bibinfo
  {author} {\bibfnamefont {A.}~\bibnamefont {Joseph}}, \bibinfo {author}
  {\bibfnamefont {H.-W.}\ \bibnamefont {Lin}}, \ and\ \bibinfo {author}
  {\bibfnamefont {B.}~\bibnamefont {Yoon}} (\bibinfo {collaboration} {PNDME}),\
  }\href {\doibase 10.1103/PhysRevD.92.094511} {\bibfield  {journal} {\bibinfo
  {journal} {Phys. Rev. D}\ }\textbf {\bibinfo {volume} {92}},\ \bibinfo
  {pages} {094511} (\bibinfo {year} {2015})},\ \Eprint
  {http://arxiv.org/abs/1506.06411} {arXiv:1506.06411 [hep-lat]} \BibitemShut
  {NoStop}%
\bibitem [{\citenamefont {Green}\ \emph {et~al.}(2012)\citenamefont {Green},
  \citenamefont {Negele}, \citenamefont {Pochinsky}, \citenamefont {Syritsyn},
  \citenamefont {Engelhardt},\ and\ \citenamefont {Krieg}}]{Green:2012ej}%
  \BibitemOpen
  \bibfield  {author} {\bibinfo {author} {\bibfnamefont {J.~R.}\ \bibnamefont
  {Green}}, \bibinfo {author} {\bibfnamefont {J.~W.}\ \bibnamefont {Negele}},
  \bibinfo {author} {\bibfnamefont {A.~V.}\ \bibnamefont {Pochinsky}}, \bibinfo
  {author} {\bibfnamefont {S.~N.}\ \bibnamefont {Syritsyn}}, \bibinfo {author}
  {\bibfnamefont {M.}~\bibnamefont {Engelhardt}}, \ and\ \bibinfo {author}
  {\bibfnamefont {S.}~\bibnamefont {Krieg}},\ }\href {\doibase
  10.1103/PhysRevD.86.114509} {\bibfield  {journal} {\bibinfo  {journal} {Phys.
  Rev. D}\ }\textbf {\bibinfo {volume} {86}},\ \bibinfo {pages} {114509}
  (\bibinfo {year} {2012})},\ \Eprint {http://arxiv.org/abs/1206.4527}
  {arXiv:1206.4527 [hep-lat]} \BibitemShut {NoStop}%
\bibitem [{\citenamefont {Aoki}\ \emph {et~al.}(2010)\citenamefont {Aoki},
  \citenamefont {Blum}, \citenamefont {Lin}, \citenamefont {Ohta},
  \citenamefont {Sasaki}, \citenamefont {Tweedie}, \citenamefont {Zanotti},\
  and\ \citenamefont {Yamazaki}}]{Aoki:2010xg}%
  \BibitemOpen
  \bibfield  {author} {\bibinfo {author} {\bibfnamefont {Y.}~\bibnamefont
  {Aoki}}, \bibinfo {author} {\bibfnamefont {T.}~\bibnamefont {Blum}}, \bibinfo
  {author} {\bibfnamefont {H.-W.}\ \bibnamefont {Lin}}, \bibinfo {author}
  {\bibfnamefont {S.}~\bibnamefont {Ohta}}, \bibinfo {author} {\bibfnamefont
  {S.}~\bibnamefont {Sasaki}}, \bibinfo {author} {\bibfnamefont
  {R.}~\bibnamefont {Tweedie}}, \bibinfo {author} {\bibfnamefont
  {J.}~\bibnamefont {Zanotti}}, \ and\ \bibinfo {author} {\bibfnamefont
  {T.}~\bibnamefont {Yamazaki}},\ }\href {\doibase 10.1103/PhysRevD.82.014501}
  {\bibfield  {journal} {\bibinfo  {journal} {Phys. Rev. D}\ }\textbf {\bibinfo
  {volume} {82}},\ \bibinfo {pages} {014501} (\bibinfo {year} {2010})},\
  \Eprint {http://arxiv.org/abs/1003.3387} {arXiv:1003.3387 [hep-lat]}
  \BibitemShut {NoStop}%
\bibitem [{\citenamefont {Abdel-Rehim}\ \emph {et~al.}(2015)\citenamefont
  {Abdel-Rehim} \emph {et~al.}}]{Abdel-Rehim:2015owa}%
  \BibitemOpen
  \bibfield  {author} {\bibinfo {author} {\bibfnamefont {A.}~\bibnamefont
  {Abdel-Rehim}} \emph {et~al.},\ }\href {\doibase 10.1103/PhysRevD.92.114513}
  {\bibfield  {journal} {\bibinfo  {journal} {Phys. Rev. D}\ }\textbf {\bibinfo
  {volume} {92}},\ \bibinfo {pages} {114513} (\bibinfo {year} {2015})},\
  \bibinfo {note} {[Erratum: Phys.Rev.D 93, 039904 (2016)]},\ \Eprint
  {http://arxiv.org/abs/1507.04936} {arXiv:1507.04936 [hep-lat]} \BibitemShut
  {NoStop}%
\bibitem [{\citenamefont {Bali}\ \emph {et~al.}(2015)\citenamefont {Bali},
  \citenamefont {Collins}, \citenamefont {Gl\"assle}, \citenamefont
  {G\"ockeler}, \citenamefont {Najjar}, \citenamefont {R\"odl}, \citenamefont
  {Sch\"afer}, \citenamefont {Schiel}, \citenamefont {S\"oldner},\ and\
  \citenamefont {Sternbeck}}]{Bali:2014nma}%
  \BibitemOpen
  \bibfield  {author} {\bibinfo {author} {\bibfnamefont {G.~S.}\ \bibnamefont
  {Bali}}, \bibinfo {author} {\bibfnamefont {S.}~\bibnamefont {Collins}},
  \bibinfo {author} {\bibfnamefont {B.}~\bibnamefont {Gl\"assle}}, \bibinfo
  {author} {\bibfnamefont {M.}~\bibnamefont {G\"ockeler}}, \bibinfo {author}
  {\bibfnamefont {J.}~\bibnamefont {Najjar}}, \bibinfo {author} {\bibfnamefont
  {R.~H.}\ \bibnamefont {R\"odl}}, \bibinfo {author} {\bibfnamefont
  {A.}~\bibnamefont {Sch\"afer}}, \bibinfo {author} {\bibfnamefont {R.~W.}\
  \bibnamefont {Schiel}}, \bibinfo {author} {\bibfnamefont {W.}~\bibnamefont
  {S\"oldner}}, \ and\ \bibinfo {author} {\bibfnamefont {A.}~\bibnamefont
  {Sternbeck}},\ }\href {\doibase 10.1103/PhysRevD.91.054501} {\bibfield
  {journal} {\bibinfo  {journal} {Phys. Rev. D}\ }\textbf {\bibinfo {volume}
  {91}},\ \bibinfo {pages} {054501} (\bibinfo {year} {2015})},\ \Eprint
  {http://arxiv.org/abs/1412.7336} {arXiv:1412.7336 [hep-lat]} \BibitemShut
  {NoStop}%
\bibitem [{\citenamefont {Yamazaki}\ \emph {et~al.}(2008)\citenamefont
  {Yamazaki}, \citenamefont {Aoki}, \citenamefont {Blum}, \citenamefont {Lin},
  \citenamefont {Lin}, \citenamefont {Ohta}, \citenamefont {Sasaki},
  \citenamefont {Tweedie},\ and\ \citenamefont {Zanotti}}]{Yamazaki:2008py}%
  \BibitemOpen
  \bibfield  {author} {\bibinfo {author} {\bibfnamefont {T.}~\bibnamefont
  {Yamazaki}}, \bibinfo {author} {\bibfnamefont {Y.}~\bibnamefont {Aoki}},
  \bibinfo {author} {\bibfnamefont {T.}~\bibnamefont {Blum}}, \bibinfo {author}
  {\bibfnamefont {H.~W.}\ \bibnamefont {Lin}}, \bibinfo {author} {\bibfnamefont
  {M.~F.}\ \bibnamefont {Lin}}, \bibinfo {author} {\bibfnamefont
  {S.}~\bibnamefont {Ohta}}, \bibinfo {author} {\bibfnamefont {S.}~\bibnamefont
  {Sasaki}}, \bibinfo {author} {\bibfnamefont {R.~J.}\ \bibnamefont {Tweedie}},
  \ and\ \bibinfo {author} {\bibfnamefont {J.~M.}\ \bibnamefont {Zanotti}}
  (\bibinfo {collaboration} {RBC+UKQCD}),\ }\href {\doibase
  10.1103/PhysRevLett.100.171602} {\bibfield  {journal} {\bibinfo  {journal}
  {Phys. Rev. Lett.}\ }\textbf {\bibinfo {volume} {100}},\ \bibinfo {pages}
  {171602} (\bibinfo {year} {2008})},\ \Eprint {http://arxiv.org/abs/0801.4016}
  {arXiv:0801.4016 [hep-lat]} \BibitemShut {NoStop}%
\bibitem [{\citenamefont {Mondal}\ \emph
  {et~al.}(2020{\natexlab{a}})\citenamefont {Mondal}, \citenamefont {Gupta},
  \citenamefont {Park}, \citenamefont {Yoon}, \citenamefont {Bhattacharya},
  \citenamefont {Jo\'o},\ and\ \citenamefont {Winter}}]{Mondal:2020ela}%
  \BibitemOpen
  \bibfield  {author} {\bibinfo {author} {\bibfnamefont {S.}~\bibnamefont
  {Mondal}}, \bibinfo {author} {\bibfnamefont {R.}~\bibnamefont {Gupta}},
  \bibinfo {author} {\bibfnamefont {S.}~\bibnamefont {Park}}, \bibinfo {author}
  {\bibfnamefont {B.}~\bibnamefont {Yoon}}, \bibinfo {author} {\bibfnamefont
  {T.}~\bibnamefont {Bhattacharya}}, \bibinfo {author} {\bibfnamefont
  {B.}~\bibnamefont {Jo\'o}}, \ and\ \bibinfo {author} {\bibfnamefont
  {F.}~\bibnamefont {Winter}} (\bibinfo {collaboration} {Nucleon Matrix
  Elements (NME)}),\ }\href {\doibase 10.1007/JHEP04(2021)044} {\bibfield
  {journal} {\bibinfo  {journal} {JHEP}\ }\textbf {\bibinfo {volume} {21}},\
  \bibinfo {pages} {004} (\bibinfo {year} {2020}{\natexlab{a}})},\ \Eprint
  {http://arxiv.org/abs/2011.12787} {arXiv:2011.12787 [hep-lat]} \BibitemShut
  {NoStop}%
\bibitem [{\citenamefont {Mondal}\ \emph
  {et~al.}(2020{\natexlab{b}})\citenamefont {Mondal}, \citenamefont {Gupta},
  \citenamefont {Park}, \citenamefont {Yoon}, \citenamefont {Bhattacharya},\
  and\ \citenamefont {Lin}}]{Mondal:2020cmt}%
  \BibitemOpen
  \bibfield  {author} {\bibinfo {author} {\bibfnamefont {S.}~\bibnamefont
  {Mondal}}, \bibinfo {author} {\bibfnamefont {R.}~\bibnamefont {Gupta}},
  \bibinfo {author} {\bibfnamefont {S.}~\bibnamefont {Park}}, \bibinfo {author}
  {\bibfnamefont {B.}~\bibnamefont {Yoon}}, \bibinfo {author} {\bibfnamefont
  {T.}~\bibnamefont {Bhattacharya}}, \ and\ \bibinfo {author} {\bibfnamefont
  {H.-W.}\ \bibnamefont {Lin}},\ }\href {\doibase 10.1103/PhysRevD.102.054512}
  {\bibfield  {journal} {\bibinfo  {journal} {Phys. Rev. D}\ }\textbf {\bibinfo
  {volume} {102}},\ \bibinfo {pages} {054512} (\bibinfo {year}
  {2020}{\natexlab{b}})},\ \Eprint {http://arxiv.org/abs/2005.13779}
  {arXiv:2005.13779 [hep-lat]} \BibitemShut {NoStop}%
\bibitem [{\citenamefont {Alexandrou}\ \emph
  {et~al.}(2020{\natexlab{b}})\citenamefont {Alexandrou} \emph
  {et~al.}}]{Alexandrou:2019ali}%
  \BibitemOpen
  \bibfield  {author} {\bibinfo {author} {\bibfnamefont {C.}~\bibnamefont
  {Alexandrou}} \emph {et~al.},\ }\href {\doibase 10.1103/PhysRevD.101.034519}
  {\bibfield  {journal} {\bibinfo  {journal} {Phys. Rev. D}\ }\textbf {\bibinfo
  {volume} {101}},\ \bibinfo {pages} {034519} (\bibinfo {year}
  {2020}{\natexlab{b}})},\ \Eprint {http://arxiv.org/abs/1908.10706}
  {arXiv:1908.10706 [hep-lat]} \BibitemShut {NoStop}%
\bibitem [{\citenamefont {Harris}\ \emph {et~al.}(2019)\citenamefont {Harris},
  \citenamefont {von Hippel}, \citenamefont {Junnarkar}, \citenamefont {Meyer},
  \citenamefont {Ottnad}, \citenamefont {Wilhelm}, \citenamefont {Wittig},\
  and\ \citenamefont {Wrang}}]{Harris:2019bih}%
  \BibitemOpen
  \bibfield  {author} {\bibinfo {author} {\bibfnamefont {T.}~\bibnamefont
  {Harris}}, \bibinfo {author} {\bibfnamefont {G.}~\bibnamefont {von Hippel}},
  \bibinfo {author} {\bibfnamefont {P.}~\bibnamefont {Junnarkar}}, \bibinfo
  {author} {\bibfnamefont {H.~B.}\ \bibnamefont {Meyer}}, \bibinfo {author}
  {\bibfnamefont {K.}~\bibnamefont {Ottnad}}, \bibinfo {author} {\bibfnamefont
  {J.}~\bibnamefont {Wilhelm}}, \bibinfo {author} {\bibfnamefont
  {H.}~\bibnamefont {Wittig}}, \ and\ \bibinfo {author} {\bibfnamefont
  {L.}~\bibnamefont {Wrang}},\ }\href {\doibase 10.1103/PhysRevD.100.034513}
  {\bibfield  {journal} {\bibinfo  {journal} {Phys. Rev. D}\ }\textbf {\bibinfo
  {volume} {100}},\ \bibinfo {pages} {034513} (\bibinfo {year} {2019})},\
  \Eprint {http://arxiv.org/abs/1905.01291} {arXiv:1905.01291 [hep-lat]}
  \BibitemShut {NoStop}%
\bibitem [{\citenamefont {Bali}\ \emph {et~al.}(2019)\citenamefont {Bali},
  \citenamefont {Collins}, \citenamefont {G\"ockeler}, \citenamefont {R\"odl},
  \citenamefont {Sch\"afer},\ and\ \citenamefont {Sternbeck}}]{Bali:2018zgl}%
  \BibitemOpen
  \bibfield  {author} {\bibinfo {author} {\bibfnamefont {G.~S.}\ \bibnamefont
  {Bali}}, \bibinfo {author} {\bibfnamefont {S.}~\bibnamefont {Collins}},
  \bibinfo {author} {\bibfnamefont {M.}~\bibnamefont {G\"ockeler}}, \bibinfo
  {author} {\bibfnamefont {R.}~\bibnamefont {R\"odl}}, \bibinfo {author}
  {\bibfnamefont {A.}~\bibnamefont {Sch\"afer}}, \ and\ \bibinfo {author}
  {\bibfnamefont {A.}~\bibnamefont {Sternbeck}},\ }\href {\doibase
  10.1103/PhysRevD.100.014507} {\bibfield  {journal} {\bibinfo  {journal}
  {Phys. Rev. D}\ }\textbf {\bibinfo {volume} {100}},\ \bibinfo {pages}
  {014507} (\bibinfo {year} {2019})},\ \Eprint
  {http://arxiv.org/abs/1812.08256} {arXiv:1812.08256 [hep-lat]} \BibitemShut
  {NoStop}%
\bibitem [{\citenamefont {Lin}\ \emph {et~al.}(2018{\natexlab{a}})\citenamefont
  {Lin}, \citenamefont {Melnitchouk}, \citenamefont {Prokudin}, \citenamefont
  {Sato},\ and\ \citenamefont {Shows}}]{Lin:2017stx}%
  \BibitemOpen
  \bibfield  {author} {\bibinfo {author} {\bibfnamefont {H.-W.}\ \bibnamefont
  {Lin}}, \bibinfo {author} {\bibfnamefont {W.}~\bibnamefont {Melnitchouk}},
  \bibinfo {author} {\bibfnamefont {A.}~\bibnamefont {Prokudin}}, \bibinfo
  {author} {\bibfnamefont {N.}~\bibnamefont {Sato}}, \ and\ \bibinfo {author}
  {\bibfnamefont {H.}~\bibnamefont {Shows}},\ }\href {\doibase
  10.1103/PhysRevLett.120.152502} {\bibfield  {journal} {\bibinfo  {journal}
  {Phys. Rev. Lett.}\ }\textbf {\bibinfo {volume} {120}},\ \bibinfo {pages}
  {152502} (\bibinfo {year} {2018}{\natexlab{a}})},\ \Eprint
  {http://arxiv.org/abs/1710.09858} {arXiv:1710.09858 [hep-ph]} \BibitemShut
  {NoStop}%
\bibitem [{\citenamefont {Liu}\ \emph {et~al.}(2018)\citenamefont {Liu},
  \citenamefont {Chen}, \citenamefont {Jin}, \citenamefont {Li}, \citenamefont
  {Lin}, \citenamefont {Yang}, \citenamefont {Zhang},\ and\ \citenamefont
  {Zhao}}]{Liu:2018hxv}%
  \BibitemOpen
  \bibfield  {author} {\bibinfo {author} {\bibfnamefont {Y.-S.}\ \bibnamefont
  {Liu}}, \bibinfo {author} {\bibfnamefont {J.-W.}\ \bibnamefont {Chen}},
  \bibinfo {author} {\bibfnamefont {L.}~\bibnamefont {Jin}}, \bibinfo {author}
  {\bibfnamefont {R.}~\bibnamefont {Li}}, \bibinfo {author} {\bibfnamefont
  {H.-W.}\ \bibnamefont {Lin}}, \bibinfo {author} {\bibfnamefont {Y.-B.}\
  \bibnamefont {Yang}}, \bibinfo {author} {\bibfnamefont {J.-H.}\ \bibnamefont
  {Zhang}}, \ and\ \bibinfo {author} {\bibfnamefont {Y.}~\bibnamefont {Zhao}},\
  }\href@noop {} {\  (\bibinfo {year} {2018})},\ \Eprint
  {http://arxiv.org/abs/1810.05043} {arXiv:1810.05043 [hep-lat]} \BibitemShut
  {NoStop}%
\bibitem [{\citenamefont {Alexandrou}\ \emph
  {et~al.}(2018{\natexlab{b}})\citenamefont {Alexandrou}, \citenamefont
  {Cichy}, \citenamefont {Constantinou}, \citenamefont {Jansen}, \citenamefont
  {Scapellato},\ and\ \citenamefont {Steffens}}]{Alexandrou:2018eet}%
  \BibitemOpen
  \bibfield  {author} {\bibinfo {author} {\bibfnamefont {C.}~\bibnamefont
  {Alexandrou}}, \bibinfo {author} {\bibfnamefont {K.}~\bibnamefont {Cichy}},
  \bibinfo {author} {\bibfnamefont {M.}~\bibnamefont {Constantinou}}, \bibinfo
  {author} {\bibfnamefont {K.}~\bibnamefont {Jansen}}, \bibinfo {author}
  {\bibfnamefont {A.}~\bibnamefont {Scapellato}}, \ and\ \bibinfo {author}
  {\bibfnamefont {F.}~\bibnamefont {Steffens}},\ }\href {\doibase
  10.1103/PhysRevD.98.091503} {\bibfield  {journal} {\bibinfo  {journal} {Phys.
  Rev. D}\ }\textbf {\bibinfo {volume} {98}},\ \bibinfo {pages} {091503}
  (\bibinfo {year} {2018}{\natexlab{b}})},\ \Eprint
  {http://arxiv.org/abs/1807.00232} {arXiv:1807.00232 [hep-lat]} \BibitemShut
  {NoStop}%
\bibitem [{\citenamefont {Chen}\ \emph {et~al.}(2016)\citenamefont {Chen},
  \citenamefont {Cohen}, \citenamefont {Ji}, \citenamefont {Lin},\ and\
  \citenamefont {Zhang}}]{Chen:2016utp}%
  \BibitemOpen
  \bibfield  {author} {\bibinfo {author} {\bibfnamefont {J.-W.}\ \bibnamefont
  {Chen}}, \bibinfo {author} {\bibfnamefont {S.~D.}\ \bibnamefont {Cohen}},
  \bibinfo {author} {\bibfnamefont {X.}~\bibnamefont {Ji}}, \bibinfo {author}
  {\bibfnamefont {H.-W.}\ \bibnamefont {Lin}}, \ and\ \bibinfo {author}
  {\bibfnamefont {J.-H.}\ \bibnamefont {Zhang}},\ }\href {\doibase
  10.1016/j.nuclphysb.2016.07.033} {\bibfield  {journal} {\bibinfo  {journal}
  {Nucl. Phys. B}\ }\textbf {\bibinfo {volume} {911}},\ \bibinfo {pages} {246}
  (\bibinfo {year} {2016})},\ \Eprint {http://arxiv.org/abs/1603.06664}
  {arXiv:1603.06664 [hep-ph]} \BibitemShut {NoStop}%
\bibitem [{\citenamefont {Alexandrou}\ \emph
  {et~al.}(2021{\natexlab{b}})\citenamefont {Alexandrou}, \citenamefont
  {Cichy}, \citenamefont {Constantinou}, \citenamefont {Hadjiyiannakou},
  \citenamefont {Jansen}, \citenamefont {Scapellato},\ and\ \citenamefont
  {Steffens}}]{Alexandrou:2021bbo}%
  \BibitemOpen
  \bibfield  {author} {\bibinfo {author} {\bibfnamefont {C.}~\bibnamefont
  {Alexandrou}}, \bibinfo {author} {\bibfnamefont {K.}~\bibnamefont {Cichy}},
  \bibinfo {author} {\bibfnamefont {M.}~\bibnamefont {Constantinou}}, \bibinfo
  {author} {\bibfnamefont {K.}~\bibnamefont {Hadjiyiannakou}}, \bibinfo
  {author} {\bibfnamefont {K.}~\bibnamefont {Jansen}}, \bibinfo {author}
  {\bibfnamefont {A.}~\bibnamefont {Scapellato}}, \ and\ \bibinfo {author}
  {\bibfnamefont {F.}~\bibnamefont {Steffens}},\ }\href@noop {} {\  (\bibinfo
  {year} {2021}{\natexlab{b}})},\ \Eprint {http://arxiv.org/abs/2108.10789}
  {arXiv:2108.10789 [hep-lat]} \BibitemShut {NoStop}%
\bibitem [{\citenamefont {Lin}\ \emph {et~al.}(2018{\natexlab{b}})\citenamefont
  {Lin} \emph {et~al.}}]{Lin:2017snn}%
  \BibitemOpen
  \bibfield  {author} {\bibinfo {author} {\bibfnamefont {H.-W.}\ \bibnamefont
  {Lin}} \emph {et~al.},\ }\href {\doibase 10.1016/j.ppnp.2018.01.007}
  {\bibfield  {journal} {\bibinfo  {journal} {Prog. Part. Nucl. Phys.}\
  }\textbf {\bibinfo {volume} {100}},\ \bibinfo {pages} {107} (\bibinfo {year}
  {2018}{\natexlab{b}})},\ \Eprint {http://arxiv.org/abs/1711.07916}
  {arXiv:1711.07916 [hep-ph]} \BibitemShut {NoStop}%
\bibitem [{\citenamefont {Vogelsang}(1998)}]{Vogelsang:1997ak}%
  \BibitemOpen
  \bibfield  {author} {\bibinfo {author} {\bibfnamefont {W.}~\bibnamefont
  {Vogelsang}},\ }\href {\doibase 10.1103/PhysRevD.57.1886} {\bibfield
  {journal} {\bibinfo  {journal} {Phys. Rev. D}\ }\textbf {\bibinfo {volume}
  {57}},\ \bibinfo {pages} {1886} (\bibinfo {year} {1998})},\ \Eprint
  {http://arxiv.org/abs/hep-ph/9706511} {arXiv:hep-ph/9706511} \BibitemShut
  {NoStop}%
\bibitem [{\citenamefont {Musch}\ \emph {et~al.}(2011)\citenamefont {Musch},
  \citenamefont {Hagler}, \citenamefont {Negele},\ and\ \citenamefont
  {Schafer}}]{Musch:2010ka}%
  \BibitemOpen
  \bibfield  {author} {\bibinfo {author} {\bibfnamefont {B.~U.}\ \bibnamefont
  {Musch}}, \bibinfo {author} {\bibfnamefont {P.}~\bibnamefont {Hagler}},
  \bibinfo {author} {\bibfnamefont {J.~W.}\ \bibnamefont {Negele}}, \ and\
  \bibinfo {author} {\bibfnamefont {A.}~\bibnamefont {Schafer}},\ }\href
  {\doibase 10.1103/PhysRevD.83.094507} {\bibfield  {journal} {\bibinfo
  {journal} {Phys. Rev. D}\ }\textbf {\bibinfo {volume} {83}},\ \bibinfo
  {pages} {094507} (\bibinfo {year} {2011})},\ \Eprint
  {http://arxiv.org/abs/1011.1213} {arXiv:1011.1213 [hep-lat]} \BibitemShut
  {NoStop}%
\bibitem [{\citenamefont {Ioffe}(1969)}]{Ioffe:1969kf}%
  \BibitemOpen
  \bibfield  {author} {\bibinfo {author} {\bibfnamefont {B.~L.}\ \bibnamefont
  {Ioffe}},\ }\href {\doibase 10.1016/0370-2693(69)90415-8} {\bibfield
  {journal} {\bibinfo  {journal} {Phys. Lett. B}\ }\textbf {\bibinfo {volume}
  {30}},\ \bibinfo {pages} {123} (\bibinfo {year} {1969})}\BibitemShut
  {NoStop}%
\bibitem [{\citenamefont {Braun}\ \emph {et~al.}(1995)\citenamefont {Braun},
  \citenamefont {Gornicki},\ and\ \citenamefont {Mankiewicz}}]{Braun:1994jq}%
  \BibitemOpen
  \bibfield  {author} {\bibinfo {author} {\bibfnamefont {V.}~\bibnamefont
  {Braun}}, \bibinfo {author} {\bibfnamefont {P.}~\bibnamefont {Gornicki}}, \
  and\ \bibinfo {author} {\bibfnamefont {L.}~\bibnamefont {Mankiewicz}},\
  }\href {\doibase 10.1103/PhysRevD.51.6036} {\bibfield  {journal} {\bibinfo
  {journal} {Phys. Rev. D}\ }\textbf {\bibinfo {volume} {51}},\ \bibinfo
  {pages} {6036} (\bibinfo {year} {1995})},\ \Eprint
  {http://arxiv.org/abs/hep-ph/9410318} {arXiv:hep-ph/9410318} \BibitemShut
  {NoStop}%
\bibitem [{\citenamefont {Constantinou}\ and\ \citenamefont
  {Panagopoulos}(2017)}]{Constantinou:2017sej}%
  \BibitemOpen
  \bibfield  {author} {\bibinfo {author} {\bibfnamefont {M.}~\bibnamefont
  {Constantinou}}\ and\ \bibinfo {author} {\bibfnamefont {H.}~\bibnamefont
  {Panagopoulos}},\ }\href {\doibase 10.1103/PhysRevD.96.054506} {\bibfield
  {journal} {\bibinfo  {journal} {Phys. Rev. D}\ }\textbf {\bibinfo {volume}
  {96}},\ \bibinfo {pages} {054506} (\bibinfo {year} {2017})},\ \Eprint
  {http://arxiv.org/abs/1705.11193} {arXiv:1705.11193 [hep-lat]} \BibitemShut
  {NoStop}%
\bibitem [{\citenamefont {Ji}\ \emph {et~al.}(2018)\citenamefont {Ji},
  \citenamefont {Zhang},\ and\ \citenamefont {Zhao}}]{Ji:2017oey}%
  \BibitemOpen
  \bibfield  {author} {\bibinfo {author} {\bibfnamefont {X.}~\bibnamefont
  {Ji}}, \bibinfo {author} {\bibfnamefont {J.-H.}\ \bibnamefont {Zhang}}, \
  and\ \bibinfo {author} {\bibfnamefont {Y.}~\bibnamefont {Zhao}},\ }\href
  {\doibase 10.1103/PhysRevLett.120.112001} {\bibfield  {journal} {\bibinfo
  {journal} {Phys. Rev. Lett.}\ }\textbf {\bibinfo {volume} {120}},\ \bibinfo
  {pages} {112001} (\bibinfo {year} {2018})},\ \Eprint
  {http://arxiv.org/abs/1706.08962} {arXiv:1706.08962 [hep-ph]} \BibitemShut
  {NoStop}%
\bibitem [{\citenamefont {Ishikawa}\ \emph {et~al.}(2017)\citenamefont
  {Ishikawa}, \citenamefont {Ma}, \citenamefont {Qiu},\ and\ \citenamefont
  {Yoshida}}]{Ishikawa:2017faj}%
  \BibitemOpen
  \bibfield  {author} {\bibinfo {author} {\bibfnamefont {T.}~\bibnamefont
  {Ishikawa}}, \bibinfo {author} {\bibfnamefont {Y.-Q.}\ \bibnamefont {Ma}},
  \bibinfo {author} {\bibfnamefont {J.-W.}\ \bibnamefont {Qiu}}, \ and\
  \bibinfo {author} {\bibfnamefont {S.}~\bibnamefont {Yoshida}},\ }\href
  {\doibase 10.1103/PhysRevD.96.094019} {\bibfield  {journal} {\bibinfo
  {journal} {Phys. Rev. D}\ }\textbf {\bibinfo {volume} {96}},\ \bibinfo
  {pages} {094019} (\bibinfo {year} {2017})},\ \Eprint
  {http://arxiv.org/abs/1707.03107} {arXiv:1707.03107 [hep-ph]} \BibitemShut
  {NoStop}%
\bibitem [{\citenamefont {Green}\ \emph {et~al.}(2018)\citenamefont {Green},
  \citenamefont {Jansen},\ and\ \citenamefont {Steffens}}]{Green:2017xeu}%
  \BibitemOpen
  \bibfield  {author} {\bibinfo {author} {\bibfnamefont {J.}~\bibnamefont
  {Green}}, \bibinfo {author} {\bibfnamefont {K.}~\bibnamefont {Jansen}}, \
  and\ \bibinfo {author} {\bibfnamefont {F.}~\bibnamefont {Steffens}},\ }\href
  {\doibase 10.1103/PhysRevLett.121.022004} {\bibfield  {journal} {\bibinfo
  {journal} {Phys. Rev. Lett.}\ }\textbf {\bibinfo {volume} {121}},\ \bibinfo
  {pages} {022004} (\bibinfo {year} {2018})},\ \Eprint
  {http://arxiv.org/abs/1707.07152} {arXiv:1707.07152 [hep-lat]} \BibitemShut
  {NoStop}%
\bibitem [{\citenamefont {Balitsky}\ and\ \citenamefont
  {Braun}(1989)}]{Balitsky:1987bk}%
  \BibitemOpen
  \bibfield  {author} {\bibinfo {author} {\bibfnamefont {I.~I.}\ \bibnamefont
  {Balitsky}}\ and\ \bibinfo {author} {\bibfnamefont {V.~M.}\ \bibnamefont
  {Braun}},\ }\href {\doibase 10.1016/0550-3213(89)90168-5} {\bibfield
  {journal} {\bibinfo  {journal} {Nucl. Phys. B}\ }\textbf {\bibinfo {volume}
  {311}},\ \bibinfo {pages} {541} (\bibinfo {year} {1989})}\BibitemShut
  {NoStop}%
\bibitem [{\citenamefont {Braun}\ \emph {et~al.}(2021)\citenamefont {Braun},
  \citenamefont {Ji},\ and\ \citenamefont {Vladimirov}}]{Braun:2021gvv}%
  \BibitemOpen
  \bibfield  {author} {\bibinfo {author} {\bibfnamefont {V.~M.}\ \bibnamefont
  {Braun}}, \bibinfo {author} {\bibfnamefont {Y.}~\bibnamefont {Ji}}, \ and\
  \bibinfo {author} {\bibfnamefont {A.}~\bibnamefont {Vladimirov}},\ }\href
  {\doibase 10.1007/JHEP10(2021)087} {\bibfield  {journal} {\bibinfo  {journal}
  {JHEP}\ }\textbf {\bibinfo {volume} {10}},\ \bibinfo {pages} {087} (\bibinfo
  {year} {2021})},\ \Eprint {http://arxiv.org/abs/2108.03065} {arXiv:2108.03065
  [hep-ph]} \BibitemShut {NoStop}%
\bibitem [{\citenamefont {Izubuchi}\ \emph {et~al.}(2018)\citenamefont
  {Izubuchi}, \citenamefont {Ji}, \citenamefont {Jin}, \citenamefont
  {Stewart},\ and\ \citenamefont {Zhao}}]{Izubuchi:2018srq}%
  \BibitemOpen
  \bibfield  {author} {\bibinfo {author} {\bibfnamefont {T.}~\bibnamefont
  {Izubuchi}}, \bibinfo {author} {\bibfnamefont {X.}~\bibnamefont {Ji}},
  \bibinfo {author} {\bibfnamefont {L.}~\bibnamefont {Jin}}, \bibinfo {author}
  {\bibfnamefont {I.~W.}\ \bibnamefont {Stewart}}, \ and\ \bibinfo {author}
  {\bibfnamefont {Y.}~\bibnamefont {Zhao}},\ }\href {\doibase
  10.1103/PhysRevD.98.056004} {\bibfield  {journal} {\bibinfo  {journal} {Phys.
  Rev. D}\ }\textbf {\bibinfo {volume} {98}},\ \bibinfo {pages} {056004}
  (\bibinfo {year} {2018})},\ \Eprint {http://arxiv.org/abs/1801.03917}
  {arXiv:1801.03917 [hep-ph]} \BibitemShut {NoStop}%
\bibitem [{\citenamefont {Tanabashi}\ \emph {et~al.}(2018)\citenamefont
  {Tanabashi} \emph {et~al.}}]{Tanabashi:2018oca}%
  \BibitemOpen
  \bibfield  {author} {\bibinfo {author} {\bibfnamefont {M.}~\bibnamefont
  {Tanabashi}} \emph {et~al.} (\bibinfo {collaboration} {Particle Data
  Group}),\ }\href {\doibase 10.1103/PhysRevD.98.030001} {\bibfield  {journal}
  {\bibinfo  {journal} {Phys. Rev. D}\ }\textbf {\bibinfo {volume} {98}},\
  \bibinfo {pages} {030001} (\bibinfo {year} {2018})}\BibitemShut {NoStop}%
\bibitem [{\citenamefont {Edwards}\ \emph {et~al.}(2016)\citenamefont
  {Edwards}, \citenamefont {Jo\'o}, \citenamefont {Orginos}, \citenamefont
  {Richards},\ and\ \citenamefont {Winter}}]{edwardsetal}%
  \BibitemOpen
  \bibfield  {author} {\bibinfo {author} {\bibfnamefont {R.}~\bibnamefont
  {Edwards}}, \bibinfo {author} {\bibfnamefont {B.}~\bibnamefont {Jo\'o}},
  \bibinfo {author} {\bibfnamefont {K.}~\bibnamefont {Orginos}}, \bibinfo
  {author} {\bibfnamefont {D.}~\bibnamefont {Richards}}, \ and\ \bibinfo
  {author} {\bibfnamefont {F.}~\bibnamefont {Winter}},\ }\href@noop {}
  {\bibfield  {journal} {\bibinfo  {journal} {unpublished}\ } (\bibinfo {year}
  {2016})}\BibitemShut {NoStop}%
\bibitem [{\citenamefont {Yoon}\ \emph {et~al.}(2017)\citenamefont {Yoon} \emph
  {et~al.}}]{Yoon:2016jzj}%
  \BibitemOpen
  \bibfield  {author} {\bibinfo {author} {\bibfnamefont {B.}~\bibnamefont
  {Yoon}} \emph {et~al.},\ }\href {\doibase 10.1103/PhysRevD.95.074508}
  {\bibfield  {journal} {\bibinfo  {journal} {Phys. Rev. D}\ }\textbf {\bibinfo
  {volume} {95}},\ \bibinfo {pages} {074508} (\bibinfo {year} {2017})},\
  \Eprint {http://arxiv.org/abs/1611.07452} {arXiv:1611.07452 [hep-lat]}
  \BibitemShut {NoStop}%
\bibitem [{\citenamefont {Yoon}\ \emph {et~al.}(2016)\citenamefont {Yoon} \emph
  {et~al.}}]{Yoon:2016dij}%
  \BibitemOpen
  \bibfield  {author} {\bibinfo {author} {\bibfnamefont {B.}~\bibnamefont
  {Yoon}} \emph {et~al.},\ }\href {\doibase 10.1103/PhysRevD.93.114506}
  {\bibfield  {journal} {\bibinfo  {journal} {Phys. Rev. D}\ }\textbf {\bibinfo
  {volume} {93}},\ \bibinfo {pages} {114506} (\bibinfo {year} {2016})},\
  \Eprint {http://arxiv.org/abs/1602.07737} {arXiv:1602.07737 [hep-lat]}
  \BibitemShut {NoStop}%
\bibitem [{\citenamefont {Peardon}\ \emph {et~al.}(2009)\citenamefont
  {Peardon}, \citenamefont {Bulava}, \citenamefont {Foley}, \citenamefont
  {Morningstar}, \citenamefont {Dudek}, \citenamefont {Edwards}, \citenamefont
  {Joo}, \citenamefont {Lin}, \citenamefont {Richards},\ and\ \citenamefont
  {Juge}}]{HadronSpectrum:2009krc}%
  \BibitemOpen
  \bibfield  {author} {\bibinfo {author} {\bibfnamefont {M.}~\bibnamefont
  {Peardon}}, \bibinfo {author} {\bibfnamefont {J.}~\bibnamefont {Bulava}},
  \bibinfo {author} {\bibfnamefont {J.}~\bibnamefont {Foley}}, \bibinfo
  {author} {\bibfnamefont {C.}~\bibnamefont {Morningstar}}, \bibinfo {author}
  {\bibfnamefont {J.}~\bibnamefont {Dudek}}, \bibinfo {author} {\bibfnamefont
  {R.~G.}\ \bibnamefont {Edwards}}, \bibinfo {author} {\bibfnamefont
  {B.}~\bibnamefont {Joo}}, \bibinfo {author} {\bibfnamefont {H.-W.}\
  \bibnamefont {Lin}}, \bibinfo {author} {\bibfnamefont {D.~G.}\ \bibnamefont
  {Richards}}, \ and\ \bibinfo {author} {\bibfnamefont {K.~J.}\ \bibnamefont
  {Juge}} (\bibinfo {collaboration} {Hadron Spectrum}),\ }\href {\doibase
  10.1103/PhysRevD.80.054506} {\bibfield  {journal} {\bibinfo  {journal} {Phys.
  Rev. D}\ }\textbf {\bibinfo {volume} {80}},\ \bibinfo {pages} {054506}
  (\bibinfo {year} {2009})},\ \Eprint {http://arxiv.org/abs/0905.2160}
  {arXiv:0905.2160 [hep-lat]} \BibitemShut {NoStop}%
\bibitem [{\citenamefont {Egerer}\ \emph
  {et~al.}(2021{\natexlab{b}})\citenamefont {Egerer}, \citenamefont {Edwards},
  \citenamefont {Orginos},\ and\ \citenamefont {Richards}}]{Egerer:2020hnc}%
  \BibitemOpen
  \bibfield  {author} {\bibinfo {author} {\bibfnamefont {C.}~\bibnamefont
  {Egerer}}, \bibinfo {author} {\bibfnamefont {R.~G.}\ \bibnamefont {Edwards}},
  \bibinfo {author} {\bibfnamefont {K.}~\bibnamefont {Orginos}}, \ and\
  \bibinfo {author} {\bibfnamefont {D.~G.}\ \bibnamefont {Richards}},\ }\href
  {\doibase 10.1103/PhysRevD.103.034502} {\bibfield  {journal} {\bibinfo
  {journal} {Phys. Rev. D}\ }\textbf {\bibinfo {volume} {103}},\ \bibinfo
  {pages} {034502} (\bibinfo {year} {2021}{\natexlab{b}})},\ \Eprint
  {http://arxiv.org/abs/2009.10691} {arXiv:2009.10691 [hep-lat]} \BibitemShut
  {NoStop}%
\bibitem [{\citenamefont {Maiani}\ \emph {et~al.}(1987)\citenamefont {Maiani},
  \citenamefont {Martinelli}, \citenamefont {Paciello},\ and\ \citenamefont
  {Taglienti}}]{Maiani:1987by}%
  \BibitemOpen
  \bibfield  {author} {\bibinfo {author} {\bibfnamefont {L.}~\bibnamefont
  {Maiani}}, \bibinfo {author} {\bibfnamefont {G.}~\bibnamefont {Martinelli}},
  \bibinfo {author} {\bibfnamefont {M.~L.}\ \bibnamefont {Paciello}}, \ and\
  \bibinfo {author} {\bibfnamefont {B.}~\bibnamefont {Taglienti}},\ }\href
  {\doibase 10.1016/0550-3213(87)90078-2} {\bibfield  {journal} {\bibinfo
  {journal} {Nucl. Phys. B}\ }\textbf {\bibinfo {volume} {293}},\ \bibinfo
  {pages} {420} (\bibinfo {year} {1987})}\BibitemShut {NoStop}%
\bibitem [{\citenamefont {Capitani}\ \emph {et~al.}(2012)\citenamefont
  {Capitani}, \citenamefont {Della~Morte}, \citenamefont {von Hippel},
  \citenamefont {Jager}, \citenamefont {Juttner}, \citenamefont {Knippschild},
  \citenamefont {Meyer},\ and\ \citenamefont {Wittig}}]{Capitani:2012gj}%
  \BibitemOpen
  \bibfield  {author} {\bibinfo {author} {\bibfnamefont {S.}~\bibnamefont
  {Capitani}}, \bibinfo {author} {\bibfnamefont {M.}~\bibnamefont
  {Della~Morte}}, \bibinfo {author} {\bibfnamefont {G.}~\bibnamefont {von
  Hippel}}, \bibinfo {author} {\bibfnamefont {B.}~\bibnamefont {Jager}},
  \bibinfo {author} {\bibfnamefont {A.}~\bibnamefont {Juttner}}, \bibinfo
  {author} {\bibfnamefont {B.}~\bibnamefont {Knippschild}}, \bibinfo {author}
  {\bibfnamefont {H.~B.}\ \bibnamefont {Meyer}}, \ and\ \bibinfo {author}
  {\bibfnamefont {H.}~\bibnamefont {Wittig}},\ }\href {\doibase
  10.1103/PhysRevD.86.074502} {\bibfield  {journal} {\bibinfo  {journal} {Phys.
  Rev. D}\ }\textbf {\bibinfo {volume} {86}},\ \bibinfo {pages} {074502}
  (\bibinfo {year} {2012})},\ \Eprint {http://arxiv.org/abs/1205.0180}
  {arXiv:1205.0180 [hep-lat]} \BibitemShut {NoStop}%
\bibitem [{\citenamefont {Khan}\ \emph
  {et~al.}(2021{\natexlab{b}})\citenamefont {Khan}, \citenamefont {Richards},\
  and\ \citenamefont {Winter}}]{Khan:2020ahz}%
  \BibitemOpen
  \bibfield  {author} {\bibinfo {author} {\bibfnamefont {T.}~\bibnamefont
  {Khan}}, \bibinfo {author} {\bibfnamefont {D.}~\bibnamefont {Richards}}, \
  and\ \bibinfo {author} {\bibfnamefont {F.}~\bibnamefont {Winter}},\ }\href
  {\doibase 10.1103/PhysRevD.104.034503} {\bibfield  {journal} {\bibinfo
  {journal} {Phys. Rev. D}\ }\textbf {\bibinfo {volume} {104}},\ \bibinfo
  {pages} {034503} (\bibinfo {year} {2021}{\natexlab{b}})},\ \Eprint
  {http://arxiv.org/abs/2010.03052} {arXiv:2010.03052 [hep-lat]} \BibitemShut
  {NoStop}%
\bibitem [{\citenamefont {Karpie}\ \emph {et~al.}(2018)\citenamefont {Karpie},
  \citenamefont {Orginos},\ and\ \citenamefont
  {Zafeiropoulos}}]{Karpie:2018zaz}%
  \BibitemOpen
  \bibfield  {author} {\bibinfo {author} {\bibfnamefont {J.}~\bibnamefont
  {Karpie}}, \bibinfo {author} {\bibfnamefont {K.}~\bibnamefont {Orginos}}, \
  and\ \bibinfo {author} {\bibfnamefont {S.}~\bibnamefont {Zafeiropoulos}},\
  }\href {\doibase 10.1007/JHEP11(2018)178} {\bibfield  {journal} {\bibinfo
  {journal} {JHEP}\ }\textbf {\bibinfo {volume} {11}},\ \bibinfo {pages} {178}
  (\bibinfo {year} {2018})},\ \Eprint {http://arxiv.org/abs/1807.10933}
  {arXiv:1807.10933 [hep-lat]} \BibitemShut {NoStop}%
\bibitem [{\citenamefont {Braun}\ \emph {et~al.}(2019)\citenamefont {Braun},
  \citenamefont {Vladimirov},\ and\ \citenamefont {Zhang}}]{Braun:2018brg}%
  \BibitemOpen
  \bibfield  {author} {\bibinfo {author} {\bibfnamefont {V.~M.}\ \bibnamefont
  {Braun}}, \bibinfo {author} {\bibfnamefont {A.}~\bibnamefont {Vladimirov}}, \
  and\ \bibinfo {author} {\bibfnamefont {J.-H.}\ \bibnamefont {Zhang}},\ }\href
  {\doibase 10.1103/PhysRevD.99.014013} {\bibfield  {journal} {\bibinfo
  {journal} {Phys. Rev. D}\ }\textbf {\bibinfo {volume} {99}},\ \bibinfo
  {pages} {014013} (\bibinfo {year} {2019})},\ \Eprint
  {http://arxiv.org/abs/1810.00048} {arXiv:1810.00048 [hep-ph]} \BibitemShut
  {NoStop}%
\bibitem [{\citenamefont {Karpie}\ \emph {et~al.}(2019)\citenamefont {Karpie},
  \citenamefont {Orginos}, \citenamefont {Rothkopf},\ and\ \citenamefont
  {Zafeiropoulos}}]{Karpie:2019eiq}%
  \BibitemOpen
  \bibfield  {author} {\bibinfo {author} {\bibfnamefont {J.}~\bibnamefont
  {Karpie}}, \bibinfo {author} {\bibfnamefont {K.}~\bibnamefont {Orginos}},
  \bibinfo {author} {\bibfnamefont {A.}~\bibnamefont {Rothkopf}}, \ and\
  \bibinfo {author} {\bibfnamefont {S.}~\bibnamefont {Zafeiropoulos}},\ }\href
  {\doibase 10.1007/JHEP04(2019)057} {\bibfield  {journal} {\bibinfo  {journal}
  {JHEP}\ }\textbf {\bibinfo {volume} {04}},\ \bibinfo {pages} {057} (\bibinfo
  {year} {2019})},\ \Eprint {http://arxiv.org/abs/1901.05408} {arXiv:1901.05408
  [hep-lat]} \BibitemShut {NoStop}%
\bibitem [{\citenamefont {Golub}\ and\ \citenamefont
  {Pereyra}(1973)}]{doi:10.1137/0710036}%
  \BibitemOpen
  \bibfield  {author} {\bibinfo {author} {\bibfnamefont {G.~H.}\ \bibnamefont
  {Golub}}\ and\ \bibinfo {author} {\bibfnamefont {V.}~\bibnamefont
  {Pereyra}},\ }\href {\doibase 10.1137/0710036} {\bibfield  {journal}
  {\bibinfo  {journal} {SIAM Journal on Numerical Analysis}\ }\textbf {\bibinfo
  {volume} {10}},\ \bibinfo {pages} {413} (\bibinfo {year} {1973})},\ \Eprint
  {http://arxiv.org/abs/https://doi.org/10.1137/0710036}
  {https://doi.org/10.1137/0710036} \BibitemShut {NoStop}%
\bibitem [{\citenamefont {Symonds}\ and\ \citenamefont
  {Moussalli}(2011)}]{symonds2011brief}%
  \BibitemOpen
  \bibfield  {author} {\bibinfo {author} {\bibfnamefont {M.~R.}\ \bibnamefont
  {Symonds}}\ and\ \bibinfo {author} {\bibfnamefont {A.}~\bibnamefont
  {Moussalli}},\ }\href@noop {} {\bibfield  {journal} {\bibinfo  {journal}
  {Behavioral Ecology and Sociobiology}\ }\textbf {\bibinfo {volume} {65}},\
  \bibinfo {pages} {13} (\bibinfo {year} {2011})}\BibitemShut {NoStop}%
\bibitem [{\citenamefont {Chen}\ \emph
  {et~al.}(2020{\natexlab{b}})\citenamefont {Chen}, \citenamefont {Wang},\ and\
  \citenamefont {Zhu}}]{Chen:2020arf}%
  \BibitemOpen
  \bibfield  {author} {\bibinfo {author} {\bibfnamefont {L.-B.}\ \bibnamefont
  {Chen}}, \bibinfo {author} {\bibfnamefont {W.}~\bibnamefont {Wang}}, \ and\
  \bibinfo {author} {\bibfnamefont {R.}~\bibnamefont {Zhu}},\ }\href {\doibase
  10.1103/PhysRevD.102.011503} {\bibfield  {journal} {\bibinfo  {journal}
  {Phys. Rev. D}\ }\textbf {\bibinfo {volume} {102}},\ \bibinfo {pages}
  {011503} (\bibinfo {year} {2020}{\natexlab{b}})},\ \Eprint
  {http://arxiv.org/abs/2005.13757} {arXiv:2005.13757 [hep-ph]} \BibitemShut
  {NoStop}%
\bibitem [{\citenamefont {Chen}\ \emph
  {et~al.}(2020{\natexlab{c}})\citenamefont {Chen}, \citenamefont {Wang},\ and\
  \citenamefont {Zhu}}]{Chen:2020iqi}%
  \BibitemOpen
  \bibfield  {author} {\bibinfo {author} {\bibfnamefont {L.-B.}\ \bibnamefont
  {Chen}}, \bibinfo {author} {\bibfnamefont {W.}~\bibnamefont {Wang}}, \ and\
  \bibinfo {author} {\bibfnamefont {R.}~\bibnamefont {Zhu}},\ }\href {\doibase
  10.1007/JHEP10(2020)079} {\bibfield  {journal} {\bibinfo  {journal} {JHEP}\
  }\textbf {\bibinfo {volume} {10}},\ \bibinfo {pages} {079} (\bibinfo {year}
  {2020}{\natexlab{c}})},\ \Eprint {http://arxiv.org/abs/2006.10917}
  {arXiv:2006.10917 [hep-ph]} \BibitemShut {NoStop}%
\bibitem [{\citenamefont {Li}\ \emph {et~al.}(2021)\citenamefont {Li},
  \citenamefont {Ma},\ and\ \citenamefont {Qiu}}]{Li:2020xml}%
  \BibitemOpen
  \bibfield  {author} {\bibinfo {author} {\bibfnamefont {Z.-Y.}\ \bibnamefont
  {Li}}, \bibinfo {author} {\bibfnamefont {Y.-Q.}\ \bibnamefont {Ma}}, \ and\
  \bibinfo {author} {\bibfnamefont {J.-W.}\ \bibnamefont {Qiu}},\ }\href
  {\doibase 10.1103/PhysRevLett.126.072001} {\bibfield  {journal} {\bibinfo
  {journal} {Phys. Rev. Lett.}\ }\textbf {\bibinfo {volume} {126}},\ \bibinfo
  {pages} {072001} (\bibinfo {year} {2021})},\ \Eprint
  {http://arxiv.org/abs/2006.12370} {arXiv:2006.12370 [hep-ph]} \BibitemShut
  {NoStop}%
\bibitem [{\citenamefont {Gao}\ \emph {et~al.}(2021)\citenamefont {Gao},
  \citenamefont {Lee}, \citenamefont {Mukherjee}, \citenamefont {Shugert},\
  and\ \citenamefont {Zhao}}]{Gao:2021hxl}%
  \BibitemOpen
  \bibfield  {author} {\bibinfo {author} {\bibfnamefont {X.}~\bibnamefont
  {Gao}}, \bibinfo {author} {\bibfnamefont {K.}~\bibnamefont {Lee}}, \bibinfo
  {author} {\bibfnamefont {S.}~\bibnamefont {Mukherjee}}, \bibinfo {author}
  {\bibfnamefont {C.}~\bibnamefont {Shugert}}, \ and\ \bibinfo {author}
  {\bibfnamefont {Y.}~\bibnamefont {Zhao}},\ }\href {\doibase
  10.1103/PhysRevD.103.094504} {\bibfield  {journal} {\bibinfo  {journal}
  {Phys. Rev. D}\ }\textbf {\bibinfo {volume} {103}},\ \bibinfo {pages}
  {094504} (\bibinfo {year} {2021})},\ \Eprint
  {http://arxiv.org/abs/2102.01101} {arXiv:2102.01101 [hep-ph]} \BibitemShut
  {NoStop}%
\bibitem [{\citenamefont {Karthik}\ and\ \citenamefont
  {Sufian}(2021)}]{Karthik:2021sbj}%
  \BibitemOpen
  \bibfield  {author} {\bibinfo {author} {\bibfnamefont {N.}~\bibnamefont
  {Karthik}}\ and\ \bibinfo {author} {\bibfnamefont {R.~S.}\ \bibnamefont
  {Sufian}},\ }\href {\doibase 10.1103/PhysRevD.104.074506} {\bibfield
  {journal} {\bibinfo  {journal} {Phys. Rev. D}\ }\textbf {\bibinfo {volume}
  {104}},\ \bibinfo {pages} {074506} (\bibinfo {year} {2021})},\ \Eprint
  {http://arxiv.org/abs/2106.03875} {arXiv:2106.03875 [hep-lat]} \BibitemShut
  {NoStop}%
\bibitem [{\citenamefont {Sato}()}]{nabuocomment}%
  \BibitemOpen
  \bibfield  {author} {\bibinfo {author} {\bibfnamefont {N.}~\bibnamefont
  {Sato}},\ }\href@noop {} {\bibinfo  {journal} {personal communication}\
  }\BibitemShut {NoStop}%
\bibitem [{\citenamefont {Bringewatt}\ \emph {et~al.}(2021)\citenamefont
  {Bringewatt}, \citenamefont {Sato}, \citenamefont {Melnitchouk},
  \citenamefont {Qiu}, \citenamefont {Steffens},\ and\ \citenamefont
  {Constantinou}}]{Bringewatt:2020ixn}%
  \BibitemOpen
\bibfield  {journal} {  }\bibfield  {author} {\bibinfo {author} {\bibfnamefont
  {J.}~\bibnamefont {Bringewatt}}, \bibinfo {author} {\bibfnamefont
  {N.}~\bibnamefont {Sato}}, \bibinfo {author} {\bibfnamefont {W.}~\bibnamefont
  {Melnitchouk}}, \bibinfo {author} {\bibfnamefont {J.-W.}\ \bibnamefont
  {Qiu}}, \bibinfo {author} {\bibfnamefont {F.}~\bibnamefont {Steffens}}, \
  and\ \bibinfo {author} {\bibfnamefont {M.}~\bibnamefont {Constantinou}},\
  }\href {\doibase 10.1103/PhysRevD.103.016003} {\bibfield  {journal} {\bibinfo
   {journal} {Phys. Rev. D}\ }\textbf {\bibinfo {volume} {103}},\ \bibinfo
  {pages} {016003} (\bibinfo {year} {2021})},\ \Eprint
  {http://arxiv.org/abs/2010.00548} {arXiv:2010.00548 [hep-ph]} \BibitemShut
  {NoStop}%
\bibitem [{\citenamefont {Del~Debbio}\ \emph {et~al.}(2021)\citenamefont
  {Del~Debbio}, \citenamefont {Giani}, \citenamefont {Karpie}, \citenamefont
  {Orginos}, \citenamefont {Radyushkin},\ and\ \citenamefont
  {Zafeiropoulos}}]{DelDebbio:2020rgv}%
  \BibitemOpen
  \bibfield  {author} {\bibinfo {author} {\bibfnamefont {L.}~\bibnamefont
  {Del~Debbio}}, \bibinfo {author} {\bibfnamefont {T.}~\bibnamefont {Giani}},
  \bibinfo {author} {\bibfnamefont {J.}~\bibnamefont {Karpie}}, \bibinfo
  {author} {\bibfnamefont {K.}~\bibnamefont {Orginos}}, \bibinfo {author}
  {\bibfnamefont {A.}~\bibnamefont {Radyushkin}}, \ and\ \bibinfo {author}
  {\bibfnamefont {S.}~\bibnamefont {Zafeiropoulos}},\ }\href {\doibase
  10.1007/JHEP02(2021)138} {\bibfield  {journal} {\bibinfo  {journal} {JHEP}\
  }\textbf {\bibinfo {volume} {02}},\ \bibinfo {pages} {138} (\bibinfo {year}
  {2021})},\ \Eprint {http://arxiv.org/abs/2010.03996} {arXiv:2010.03996
  [hep-ph]} \BibitemShut {NoStop}%
\bibitem [{\citenamefont {Cichy}\ \emph {et~al.}(2019)\citenamefont {Cichy},
  \citenamefont {Del~Debbio},\ and\ \citenamefont {Giani}}]{Cichy:2019ebf}%
  \BibitemOpen
  \bibfield  {author} {\bibinfo {author} {\bibfnamefont {K.}~\bibnamefont
  {Cichy}}, \bibinfo {author} {\bibfnamefont {L.}~\bibnamefont {Del~Debbio}}, \
  and\ \bibinfo {author} {\bibfnamefont {T.}~\bibnamefont {Giani}},\ }\href
  {\doibase 10.1007/JHEP10(2019)137} {\bibfield  {journal} {\bibinfo  {journal}
  {JHEP}\ }\textbf {\bibinfo {volume} {10}},\ \bibinfo {pages} {137} (\bibinfo
  {year} {2019})},\ \Eprint {http://arxiv.org/abs/1907.06037} {arXiv:1907.06037
  [hep-ph]} \BibitemShut {NoStop}%
\bibitem [{\citenamefont {Constantinou}\ \emph {et~al.}(2021)\citenamefont
  {Constantinou} \emph {et~al.}}]{Constantinou:2020hdm}%
  \BibitemOpen
  \bibfield  {author} {\bibinfo {author} {\bibfnamefont {M.}~\bibnamefont
  {Constantinou}} \emph {et~al.},\ }\href {\doibase 10.1016/j.ppnp.2021.103908}
  {\bibfield  {journal} {\bibinfo  {journal} {Prog. Part. Nucl. Phys.}\
  }\textbf {\bibinfo {volume} {121}},\ \bibinfo {pages} {103908} (\bibinfo
  {year} {2021})},\ \Eprint {http://arxiv.org/abs/2006.08636} {arXiv:2006.08636
  [hep-ph]} \BibitemShut {NoStop}%
\bibitem [{\citenamefont {Stanzione}\ \emph {et~al.}(2020)\citenamefont
  {Stanzione}, \citenamefont {West}, \citenamefont {Evans}, \citenamefont
  {Minyard}, \citenamefont {Ghattas},\ and\ \citenamefont {Panda}}]{frontera}%
  \BibitemOpen
  \bibfield  {author} {\bibinfo {author} {\bibfnamefont {D.}~\bibnamefont
  {Stanzione}}, \bibinfo {author} {\bibfnamefont {J.}~\bibnamefont {West}},
  \bibinfo {author} {\bibfnamefont {R.~T.}\ \bibnamefont {Evans}}, \bibinfo
  {author} {\bibfnamefont {T.}~\bibnamefont {Minyard}}, \bibinfo {author}
  {\bibfnamefont {O.}~\bibnamefont {Ghattas}}, \ and\ \bibinfo {author}
  {\bibfnamefont {D.~K.}\ \bibnamefont {Panda}},\ }in\ \href {\doibase
  10.1145/3311790.3396656} {\emph {\bibinfo {booktitle} {Practice and
  Experience in Advanced Research Computing}}},\ \bibinfo {series and number}
  {PEARC '20}\ (\bibinfo  {publisher} {Association for Computing Machinery},\
  \bibinfo {address} {New York, NY, USA},\ \bibinfo {year} {2020})\ p.\
  \bibinfo {pages} {106–111}\BibitemShut {NoStop}%
\bibitem [{\citenamefont {Towns}\ \emph {et~al.}(2014)\citenamefont {Towns},
  \citenamefont {Cockerill}, \citenamefont {Dahan}, \citenamefont {Foster},
  \citenamefont {Gaither}, \citenamefont {Grimshaw}, \citenamefont {Hazlewood},
  \citenamefont {Lathrop}, \citenamefont {Lifka}, \citenamefont {Peterson},
  \citenamefont {Roskies}, \citenamefont {Scott},\ and\ \citenamefont
  {Wilkins-Diehr}}]{xsede}%
  \BibitemOpen
  \bibfield  {author} {\bibinfo {author} {\bibfnamefont {J.}~\bibnamefont
  {Towns}}, \bibinfo {author} {\bibfnamefont {T.}~\bibnamefont {Cockerill}},
  \bibinfo {author} {\bibfnamefont {M.}~\bibnamefont {Dahan}}, \bibinfo
  {author} {\bibfnamefont {I.}~\bibnamefont {Foster}}, \bibinfo {author}
  {\bibfnamefont {K.}~\bibnamefont {Gaither}}, \bibinfo {author} {\bibfnamefont
  {A.}~\bibnamefont {Grimshaw}}, \bibinfo {author} {\bibfnamefont
  {V.}~\bibnamefont {Hazlewood}}, \bibinfo {author} {\bibfnamefont
  {S.}~\bibnamefont {Lathrop}}, \bibinfo {author} {\bibfnamefont
  {D.}~\bibnamefont {Lifka}}, \bibinfo {author} {\bibfnamefont {G.~D.}\
  \bibnamefont {Peterson}}, \bibinfo {author} {\bibfnamefont {R.}~\bibnamefont
  {Roskies}}, \bibinfo {author} {\bibfnamefont {J.}~\bibnamefont {Scott}}, \
  and\ \bibinfo {author} {\bibfnamefont {N.}~\bibnamefont {Wilkins-Diehr}},\
  }\href {\doibase 10.1109/MCSE.2014.80} {\bibfield  {journal} {\bibinfo
  {journal} {Computing in Science \& Engineering}\ }\textbf {\bibinfo {volume}
  {16}},\ \bibinfo {pages} {62} (\bibinfo {year} {2014})}\BibitemShut {NoStop}%
\bibitem [{\citenamefont {Edwards}\ and\ \citenamefont
  {Joo}(2005)}]{Edwards:2004sx}%
  \BibitemOpen
  \bibfield  {author} {\bibinfo {author} {\bibfnamefont {R.~G.}\ \bibnamefont
  {Edwards}}\ and\ \bibinfo {author} {\bibfnamefont {B.}~\bibnamefont {Joo}}
  (\bibinfo {collaboration} {SciDAC, LHPC, UKQCD}),\ }\href {\doibase
  10.1016/j.nuclphysbps.2004.11.254} {\bibfield  {journal} {\bibinfo  {journal}
  {Nucl. Phys. B Proc. Suppl.}\ }\textbf {\bibinfo {volume} {140}},\ \bibinfo
  {pages} {832} (\bibinfo {year} {2005})},\ \Eprint
  {http://arxiv.org/abs/hep-lat/0409003} {arXiv:hep-lat/0409003} \BibitemShut
  {NoStop}%
\bibitem [{\citenamefont {Clark}\ \emph {et~al.}(2010)\citenamefont {Clark},
  \citenamefont {Babich}, \citenamefont {Barros}, \citenamefont {Brower},\ and\
  \citenamefont {Rebbi}}]{Clark:2009wm}%
  \BibitemOpen
  \bibfield  {author} {\bibinfo {author} {\bibfnamefont {M.~A.}\ \bibnamefont
  {Clark}}, \bibinfo {author} {\bibfnamefont {R.}~\bibnamefont {Babich}},
  \bibinfo {author} {\bibfnamefont {K.}~\bibnamefont {Barros}}, \bibinfo
  {author} {\bibfnamefont {R.~C.}\ \bibnamefont {Brower}}, \ and\ \bibinfo
  {author} {\bibfnamefont {C.}~\bibnamefont {Rebbi}},\ }\href {\doibase
  10.1016/j.cpc.2010.05.002} {\bibfield  {journal} {\bibinfo  {journal}
  {Comput. Phys. Commun.}\ }\textbf {\bibinfo {volume} {181}},\ \bibinfo
  {pages} {1517} (\bibinfo {year} {2010})},\ \Eprint
  {http://arxiv.org/abs/0911.3191} {arXiv:0911.3191 [hep-lat]} \BibitemShut
  {NoStop}%
\bibitem [{\citenamefont {Babich}\ \emph {et~al.}(2010)\citenamefont {Babich},
  \citenamefont {Clark},\ and\ \citenamefont {Joo}}]{Babich:2010mu}%
  \BibitemOpen
  \bibfield  {author} {\bibinfo {author} {\bibfnamefont {R.}~\bibnamefont
  {Babich}}, \bibinfo {author} {\bibfnamefont {M.~A.}\ \bibnamefont {Clark}}, \
  and\ \bibinfo {author} {\bibfnamefont {B.}~\bibnamefont {Joo}},\ }in\
  \href@noop {} {\emph {\bibinfo {booktitle} {{SC 10 (Supercomputing 2010)}}}}\
  (\bibinfo {year} {2010})\ \Eprint {http://arxiv.org/abs/1011.0024}
  {arXiv:1011.0024 [hep-lat]} \BibitemShut {NoStop}%
\bibitem [{\citenamefont {Winter}\ \emph {et~al.}(2014)\citenamefont {Winter},
  \citenamefont {Clark}, \citenamefont {Edwards},\ and\ \citenamefont
  {Jo\'o}}]{Winter:2014dka}%
  \BibitemOpen
  \bibfield  {author} {\bibinfo {author} {\bibfnamefont {F.~T.}\ \bibnamefont
  {Winter}}, \bibinfo {author} {\bibfnamefont {M.~A.}\ \bibnamefont {Clark}},
  \bibinfo {author} {\bibfnamefont {R.~G.}\ \bibnamefont {Edwards}}, \ and\
  \bibinfo {author} {\bibfnamefont {B.}~\bibnamefont {Jo\'o}},\ }in\ \href
  {\doibase 10.1109/IPDPS.2014.112} {\emph {\bibinfo {booktitle} {{28th IEEE
  International Parallel and Distributed Processing Symposium}}}}\ (\bibinfo
  {year} {2014})\ \Eprint {http://arxiv.org/abs/1408.5925} {arXiv:1408.5925
  [hep-lat]} \BibitemShut {NoStop}%
\bibitem [{\citenamefont {Jo\'o}\ \emph {et~al.}(2013)\citenamefont {Jo\'o},
  \citenamefont {Kalamkar}, \citenamefont {Vaidyanathan}, \citenamefont
  {Smelyanskiy}, \citenamefont {Pamnany}, \citenamefont {Lee}, \citenamefont
  {Dubey},\ and\ \citenamefont {Watson}}]{Joo:2013lwm}%
  \BibitemOpen
  \bibfield  {author} {\bibinfo {author} {\bibfnamefont {B.}~\bibnamefont
  {Jo\'o}}, \bibinfo {author} {\bibfnamefont {D.~D.}\ \bibnamefont {Kalamkar}},
  \bibinfo {author} {\bibfnamefont {K.}~\bibnamefont {Vaidyanathan}}, \bibinfo
  {author} {\bibfnamefont {M.}~\bibnamefont {Smelyanskiy}}, \bibinfo {author}
  {\bibfnamefont {K.}~\bibnamefont {Pamnany}}, \bibinfo {author} {\bibfnamefont
  {V.~W.}\ \bibnamefont {Lee}}, \bibinfo {author} {\bibfnamefont
  {P.}~\bibnamefont {Dubey}}, \ and\ \bibinfo {author} {\bibfnamefont
  {W.}~\bibnamefont {Watson}},\ }\href {\doibase 10.1007/978-3-642-38750-0_4}
  {\bibfield  {journal} {\bibinfo  {journal} {Lect. Notes Comput. Sci.}\
  }\textbf {\bibinfo {volume} {7905}},\ \bibinfo {pages} {40} (\bibinfo {year}
  {2013})}\BibitemShut {NoStop}%
\bibitem [{\citenamefont {Jo{\'o}}\ \emph {et~al.}(2016)\citenamefont
  {Jo{\'o}}, \citenamefont {Kalamkar}, \citenamefont {Kurth}, \citenamefont
  {Vaidyanathan},\ and\ \citenamefont {Walden}}]{optimising}%
  \BibitemOpen
  \bibfield  {author} {\bibinfo {author} {\bibfnamefont {B.}~\bibnamefont
  {Jo{\'o}}}, \bibinfo {author} {\bibfnamefont {D.~D.}\ \bibnamefont
  {Kalamkar}}, \bibinfo {author} {\bibfnamefont {T.}~\bibnamefont {Kurth}},
  \bibinfo {author} {\bibfnamefont {K.}~\bibnamefont {Vaidyanathan}}, \ and\
  \bibinfo {author} {\bibfnamefont {A.}~\bibnamefont {Walden}},\ }in\
  \href@noop {} {\emph {\bibinfo {booktitle} {High Performance Computing}}},\
  \bibinfo {editor} {edited by\ \bibinfo {editor} {\bibfnamefont
  {M.}~\bibnamefont {Taufer}}, \bibinfo {editor} {\bibfnamefont
  {B.}~\bibnamefont {Mohr}}, \ and\ \bibinfo {editor} {\bibfnamefont {J.~M.}\
  \bibnamefont {Kunkel}}}\ (\bibinfo  {publisher} {Springer International
  Publishing},\ \bibinfo {address} {Cham},\ \bibinfo {year} {2016})\ pp.\
  \bibinfo {pages} {415--427}\BibitemShut {NoStop}%
\end{thebibliography}%

\end{document}